\documentclass[12pt]{article}
\usepackage{graphicx} 
\def\laq{\raise 0.4ex\hbox{$<$}\kern -0.8em\lower 0.62
ex\hbox{$\sim$}}
\def\gaq{\raise 0.4ex\hbox{$>$}\kern -0.7em\lower 0.62
ex\hbox{$\sim$}}

\begin{document}

\begin{titlepage}
\begin{flushright}
CERN-PH-TH/2008-153
\end{flushright}
\vspace*{1cm}

\begin{center}
{\Large{\bf Thermal history of the plasma and high-frequency gravitons}}
\vskip2.cm
\large{Massimo Giovannini \footnote{e-mail address: massimo.giovannini@cern.ch}}
\vskip 1.cm
{\it   Department of Physics, Theory Division, CERN, 1211 Geneva 23, Switzerland}
\vskip 0.5cm
{\it  INFN, Section of Milan-Bicocca, 20126 Milan, Italy}
\vskip 1cm
\end{center}

\begin{abstract}
Possible deviations from a  radiation-dominated evolution, occurring prior to the synthesis of light nuclei, impacted on the spectral energy density of high-frequency gravitons. For a systematic scrutiny of this situation, the  $\Lambda$CDM paradigm must be complemented by (at least two) physical parameters describing, respectively, a threshold frequency and a slope. The supplementary frequency scale sets the lower border of a  high-frequency domain where the spectral energy grows with a slope which depends, 
predominantly, upon  the total sound speed of the plasma right after inflation. While the infra-red region of the graviton energy spectrum is nearly scale-invariant, 
the expected signals for typical frequencies larger than $0.01$ nHz 
are hereby analyzed in a  model-independent framework by requiring that the total 
sound speed of the post-inflationary plasma be smaller than the speed of light.  Current (e.g. low-frequency)  upper limits on the tensor power spectra (determined from the combined analysis of the three large-scale  data sets) are shown to be compatible with a detectable signal in the frequency range of wide-band interferometers. 
In the present context, the scrutiny of the early evolution of the sound speed of the plasma can then be mapped onto a reliable strategy of parameter extraction including
not only the well established cosmological observables but also the forthcoming data from wide band interferometers.
\end{abstract}
\end{titlepage}
\renewcommand{\theequation}{1.\arabic{equation}}
\setcounter{equation}{0}
\section{The general famework}
\label{sec1}
Cosmological observations rely on three pivotal data sets, i.e. the 
Cosmic Microwave Background (CMB) data, the determinations of the matter power spectrum from galaxy surveys and the supernova light curve observations.  The large-scale measurements are inextricably bound to the model used to interpret the data.  Consequently the three aforementioned data sets are jointly analyzed in terms of a standard scenario which is often dubbed $\Lambda$CDM paradigm, where $\Lambda$ qualifies the dark energy component and CDM denotes the cold dark matter component.
While all the current observations are based, directly or indirectly, on the electromagnetic spectrum, there is the hope, in the future, that the 
electromagnetic observations could be complemented by the analysis of the spectrum of the relic gravitons which have been produced 
both in the context of the $\Lambda$CDM paradigm as well as in other related contexts.  The problem is, therefore, twofold: on the one hand 
reliable estimates of the spectrum of the relic gravitons arising in the $\Lambda$CDM paradigm are needed. On the other 
hand it will be important to analyze other complementary scenarios. The purpose of the present paper is to address both issues 
in quantitative terms. The relic graviton background produced in the context of the $\Lambda$CDM paradigm is expected to be rather 
minute and undetectable by wide-band interferometers \cite{LIGO,VIRGO,TAMA,GEO} in one of their future realizations. There are, however, 
extensions of the $\Lambda$CDM scenario where the signal potentially detectable by wide-band interferometers is much 
larger than in the current paradigm.   

The recent WMAP 5-yr data \cite{WMAP51,WMAP52,WMAP53,WMAP54,WMAP55}  set quite 
stringent bounds on the amplitude of the relic graviton spectral energy density for typical frequency scales 
\footnote{Natural units $\hbar = c = k_{\mathrm{B}} =1$ will be consistently 
adopted all along the present investigation.}:
\begin{equation}
\nu_{\mathrm{p}} = \frac{k_{\mathrm{p}}}{2\pi}  = 3.092\times 10^{-18} \,\, \mathrm{Hz} \equiv 3.092 \,\,\mathrm{aHz},
\label{EQ1}
\end{equation}
where, according to the prefixes of the international system of units, $1\,\mathrm{aHz} = 10^{-18}$ Hz. The wavenumber $k_{\mathrm{p}}$ is 
also called sometimes pivot scale\footnote{The pivot wavenumber $k_{\mathrm{p}}$ 
corresponds to an effective multipole $\ell_{\mathrm{eff}} \simeq 30$.} since it is customary, in the experimental analyses of CMB data \cite{WMAP54,WMAP55}, to assign the amplitude of 
the scalar and tensor modes exactly at $k_{\mathrm{p}}$.  In the same units of Eq. (\ref{EQ1}) the typical frequency interval 
potentially accessible to the observations of the wide-band interferometers ranges between few Hz and $10$ kHz with a peak 
of sensitivity around $100$Hz.

The logic followed in the present investigation will be to compute as accurately as possible 
the relic graviton spectra in terms of the parameters of the putative $\Lambda$CDM 
paradigm. The CMB data will then be used to enforce the normalization of the spectra at $\nu_{\mathrm{p}}$. This will allow
for the estimate of the spectral energy density at the frequency explored by wide-band interferometers (i.e., approximately, $100$ Hz) not only 
in the case of the $\Lambda$CDM paradigm but also in the context of its extensions.  The latter extensions will be examined on the basis 
of their plausibility, e.g. the models which are already incompatible (or barely compatible) with CMB observations 
will not be analyzed and the attention will be focussed on those scenarios which are not ruled out by (current) large-scale observations 
and which may lead to a potentially large signal at the wide-band interferometer scale.
In the present introductory section, after a general discussion of the typical frequencies 
of the graviton spectrum, the present status of CMB observations and wide-band interferometers 
observations will be swiftly discussed.  Specific attention will be paid to those quantitative aspects 
which are germane to our theme, i.e. the stochastic backgrounds of relic gravitons. Some of the concepts introduced here 
will also be more specifically addressed in the forthcoming sections.  At the end of this introduction, the
 purposes of the present investigation will be more specifically outlined.

\subsection{Typical frequencies of the problem}
To compare frequencies it is mandatory 
to specify the background and the appropriate conventions 
on the normalization of the scale factor.
Consistently with the $\Lambda$CDM paradigm,  
the background geometry will be taken to be conformally flat, i.e. 
\begin{equation}
ds^2 = \overline{g}_{\mu\nu} dx^{\mu} dx^{\nu} \equiv a^2(\tau) [ d\tau^2 - d\vec{x}^2], 
\qquad \overline{g}_{\mu\nu} = a^2(\tau) \eta_{\mu\nu},
\label{EQ2}
\end{equation}
where $\eta_{\mu\nu}$ is the Minkowski metric with signature mostly minus, 
i.e. $(+, -, -, -)$.
The scale factor at the present time will be normalized to unity, i.e. $a_{0} = 1$. Within the latter
 convention, the comoving frequencies (or wavelengths) coincide, at the present time,
with the physical frequencies (or wavelengths). The (conformal) time derivative 
of the logarithm of the scale factor will be used throughout the script and it is defined as 
\begin{equation}
{\mathcal H} = \frac{a'}{a}  = \frac{d \ln{a}}{d\tau},
\label{T7a}
\end{equation}
Note that, in Eq. (\ref{T7a}), the prime denotes a derivation with respect to the conformal time coordinate $\tau$: this 
notation will be consistently enforced in the whole investigation.  The evolution of the background can be expressed in terms 
of ${\mathcal H}$ and ${\mathcal H}'$ and it is given by:
\begin{eqnarray}
&& 3 {\mathcal H}^2 = a^2 \ell_{\mathrm{P}}^2 \rho_{\mathrm{t}},
\label{FL1}\\
&& 2({\mathcal H}^2 - {\mathcal H}') =  a^2 \ell_{\mathrm{P}}^2 (\rho_{\mathrm{t}} + p_{\mathrm{t}}),
\label{FL2}\\
&& \rho_{\mathrm{t}}' + 3 {\mathcal H} (\rho_{\mathrm{t}} + p_{\mathrm{t}}) =0,
\label{FL3}
\end{eqnarray}
where $\rho_{\mathrm{t}}$ and $p_{\mathrm{t}}$ denote, respectively, the total energy density and the 
total pressure of the plasma. 

The frequency of Eq. (\ref{EQ1}) can be usefully compared with two other 
important frequencies, i.e. the frequency of matter-radiation equality (be it 
$\nu_{\mathrm{eq}}$) and the frequency of neutrino decoupling (which also coincides, in loose 
terms, with the Hubble radius at the onset of big bang nucleosynthesis). These 
two frequencies can then be written, respectively, as:
\begin{eqnarray}
\nu_{\mathrm{eq}} &=& \frac{k_{\mathrm{eq}}}{2 \pi} = 1.281 \times 10^{-17} \biggl(\frac{h_{0}^2 \Omega_{\mathrm{M}0}}{0.1326}\biggr) \biggl(\frac{h_{0}^2 \Omega_{\mathrm{R}0}}{4.15 \times 10^{-5}}\biggr)^{-1/2}\,\, \mathrm{Hz},
\label{EQ3}\\
\nu_{\mathrm{bbn}}&=& 
2.252\times 10^{-11} \biggl(\frac{g_{\rho}}{10.75}\biggr)^{1/4} \biggl(\frac{T_{\mathrm{bbn}}}{\,\,\mathrm{MeV}}\biggr) 
\biggl(\frac{h_{0}^2 \Omega_{\mathrm{R}0}}{4.15 \times 10^{-5}}\biggr)^{1/4}\,\,\mathrm{Hz}\simeq 0.01 \,\,\mathrm{nHz}.
\label{EQ4}
\end{eqnarray}
In Eqs. (\ref{EQ3}) and (\ref{EQ4}) $\Omega_{\mathrm{M}0}$ and $\Omega_{\mathrm{R}0}$ denote, respectively, the present critical fraction of matter and radiation with typical values drawn from the best fit to the WMAP 5-yr data alone and within 
the $\Lambda$CDM paradigm. In Eq. (\ref{EQ4}) $g_{\rho}$ denotes the effective number of relativistic degrees of freedom entering the total energy density of the plasma.  While $\nu_{\mathrm{eq}}$  is still close 
to the aHz, $\nu_{\mathrm{bbn}}$ is rather in the nHz range. 

The success of the CMB and BBN calculations implicitly 
demands that, after neutrino decoupling,
the Universe was already dominated by radiation. If we assume that 
the radiation dominates right at the end of inflation, then the maximal frequency of the graviton spectrum can be computed and it is given by
\begin{equation}
\nu_{\mathrm{max}}  = 0.346 \,\biggl(\frac{\epsilon}{0.01}\biggr)^{1/4} 
\biggl(\frac{{\mathcal A}_{\mathcal R}}{2.41 \times 10^{-9}}\biggr)^{1/4}
\biggl(\frac{h_{0}^2 \Omega_{\mathrm{R}0}}{4.15 \times 10^{-5}}\biggr)^{1/4} \,\mathrm{GHz},
\label{EQ5}
\end{equation}
where ${\mathcal A}_{\mathcal R}$ denotes the amplitude of the power spectrum 
of curvature perturbations evaluated at the pivot wavenumber $k_{\mathrm{p}}$.
Between $\nu_{\mathrm{bbn}}$ and $\nu_{\mathrm{max}}$ there are roughly 20 orders of magnitude in frequency.
In the $\Lambda$CDM scenario the relic graviton spectrum has, in this range, always the same slope.

\subsection{CMB data and relic gravitons}
As already mentioned, CMB experiments are sensitive to long wavelength gravitons 
with typical frequencies of the order of $\nu_{\mathrm{p}} \sim \mathrm{aHz}$ (see also Eq. (\ref{EQ1})).
The number of CMB parameters depends upon the specific model used to interpret (and fit) the data.
The $\Lambda$CDM scenario probably contains the fewest number of parameters 
required to have a consistent fit of CMB data. 

The $\Lambda$CDM parameters can be inferred from various experiments and, among them, a central 
role is played by  WMAP   \cite{WMAP51,WMAP52,WMAP53,WMAP54,WMAP55}  (see also \cite{WMAPfirst1,WMAPfirst2,WMAPfirst3} for first year data release and \cite{WMAPthird1,WMAPthird2} for the third year data release)
as well as other experiments (see, for instance, \cite{AC1} in connection with the 5-yr WMAP data release).
The TT, TE  and, partially EE angular power spectra\footnote{Following the custom the TT correlations will  simply denote the angular power spectra of the temperature autocorrelations. The TE and the EE power spectra  denote, respectively, the cross power spectrum between temperature and polarization and the polarization autocorrelations.} have been measured 
by the WMAP experiment. Other (i.e. non space-borne) experiments are now measuring polarization observables, in particular there are the 3-yr Dasi release \cite{dasi3yr},  the CAPMAP experiment \cite{CAPMAP}, the recent QUAD data \cite{Quad1,Quad2}, as well as various other experiments at different stages of development.  The TT, TE and EE power spectra are customarily analyzed in the light of the minimal $\Lambda$CDM scenario but also 
other models are possible and they include, for instance, the addition of spatial curvature (i.e. 
the open-$\Lambda$CDM), more general parametrizations for the equation of state of dark-energy and so 
on and so forth.

The combined analysis of the CMB data, of the 
large-scale structure data \cite{LSS1,LSS2} and of the supernova data \cite{SN1,SN2} 
can lead to quantitative upper limits on the possible contribution of the 
tensor modes to the initial conditions of the CMB temperature and polarization 
anisotropies. These upper limits 
can be phrased in terms of $r_{\mathrm{T}}$, i.e.  the ratio between the power 
spectrum of tensor fluctuations and the power spectrum of the scalar fluctuations evaluated at the pivot wavenumber  $k_{\mathrm{p}} = 0.002\,\,\mathrm{Mpc}^{-1}$.  In the minimal paradigm (i.e. the $\Lambda$CDM scenario) the tensor are not 
included in the fit. 

If the inflationary phase is driven by a single scalar degree of freedom 
and if the radiation dominance kicks in almost suddenly after inflation, 
  $r_{\mathrm{T}}$ not only determines the tensor amplitude but also, thanks 
to the algebra obeyed by the slow-roll parameters, the  slope of the tensor power spectrum, customarily denoted by 
$n_{\mathrm{T}}$. To lowest order in the slow-roll 
expansion, therefore, the tensor spectral index is slightly red and it is related to $r_{\mathrm{T}}$ (and to the slow-roll parameter) as $n_{\mathrm{T}} \simeq - r_{\mathrm{T}}/8 \simeq  - 2 \epsilon$, where 
$\epsilon = - \dot{H}/H^2$ measures the rate of decrease of the Hubble 
parameter during the inflationary epoch \footnote{The overdot will denote
throughout the paper a derivation with respect to the cosmic 
time coordinate $t$ while the prime will denote a derivation with respect 
to the conformal time coordinate $\tau$.}. Within the established set of conventions 
the scalar spectral index $n_{\mathrm{s}}$ 
is given by $n_{\mathrm{s}} = (1 - 6 \epsilon + 2 \overline{\eta})$ and it depends not only upon $\epsilon$ 
but also upon the second slow-roll parameter 
$\overline{\eta} = \overline{M}_{\mathrm{P}}^2 V_{,\varphi\varphi}/V$ (where 
$V$ is the inflaton potential, $V_{,\varphi\varphi}$ denotes the second derivative  of the potential with respect to the inflaton field and $\overline{M}_{\mathrm{P}} = 1/\sqrt{8\pi G}$).

Depending upon the specific data used in the analysis, the upper limits on $r_{\mathrm{T}}$ as well as the determination of the other cosmological parameters may change slightly. 
\begin{table}[!ht]
\begin{center}
\begin{tabular}{||l|c|c|c|c|c||}
\hline
Data & $r_{\mathrm{T}}$ &$n_{\mathrm{s}}$ &$\Omega_{\Lambda}$&$\Omega_{\mathrm{M}0}$&$k_{\mathrm{eq}} \mathrm{Mpc}$\\
\hline
 WMAP5 alone& $<0.43$ &$0.986 \pm 0.22$& $0.770_{-0.032}^{+0.033}$&$0.230_{-0.033}^{0.032}$ & $0.00936$ \\
WMAP5 + Acbar& $< 0.40  $& $0.985_{-0.020}^{0.019}$&$0.767 \pm 0.032$&$0.233\pm 0.032$ &$0.00944$\\
WMAP5+ LSS + SN &$<0.20$ &$0.968 \pm 0.015$&$0.725 \pm 0.015$&$0.275 \pm 0.015$&$0.00999$\\
WMAP5+ other CMB data & $<0.36$ & $0.979\pm 0.020$&$0.775\pm 0.032$&$0.225\pm 0.032$&$0.00922$\\
\hline
\end{tabular}
\caption{The values of $r_{\mathrm{T}}$ are reported as they have been estimated in the absence of any running of the (scalar) spectral index.}
\end{center}
\label{MOD1}
\end{table}
In Tab. \ref{MOD1} the upper limits on $r_{\mathrm{T}}$ are illustrated as they are determined from the combination of different 
data sets. For illustration the determined values of the scalar spectral index (i.e. $n_{\mathrm{s}}$), 
of the dark energy and dark matter fractions (i.e., respectively, $\Omega_{\Lambda}$ and $\Omega_{\mathrm{M}0}$),
and of the typical wavenumber of equality $k_{\mathrm{eq}}$ are also reported in the remaining  columns.  While different analyses can be performed, it is clear, by looking at Tab. \ref{MOD1} that the typical upper bounds on $r_{\mathrm{T}}$ range between, say, $0.2$ and $0.4$.
Slightly more stringent limits can also be obtained by adding supplementary assumptions. 
Within a conservative perspective, the tensor power spectra are, at least, ten times  smaller than the power spectra of curvature perturbations.
In the near future the Planck explorer satellite \cite{planck} might be able to set  more direct limits on $r_{\mathrm{T}}$ by measuring (hopefully) the BB angular power spectra \footnote{Forthcoming projects like Clover \cite{CLOVER}, Brain \cite{BRAIN}, Quiet \cite{QUIET} 
and Spider \cite{SPIDER} have polarization as specific target.}. 
The E-mode power spectra and the B-mode power spectra arise as two 
orthogonal combinations of the Stokes parameters which are frame-dependent 
(i.e. $Q$ and $U$). While the adiabatic mode leads naturally to the E-mode 
polarization, the only way of obtaining the B-mode (in the standard $\Lambda$CDM
paradigm) is through the contribution of the tensor modes. Consequently, a detection o the BB angular power 
spectra would be equivalent, in the $\Lambda$CDM framework, to a first determination of $r_{\mathrm{T}}$.
Having reviewed the essentials of CMB data and their connection with the relic graviton spectra, we can now move 
to higher frequencies and describe the status of the other devices which could shed light on the 
relic gravitons, i.e. the wide-band interferometers.

\subsection{Wide-band interferometers}

The wide-band interferometers operate in a window 
ranging from few Hz up to $10$ kHz. The available interferometers 
are Ligo \cite{LIGO}, Virgo \cite{VIRGO}, Tama \cite{TAMA} and Geo
\cite{GEO}. The sensitivity of a given pair of wide-band detectors to a stochastic background of relic gravitons depends upon 
the relative orientation of the instruments. The wideness of the band 
(important for the correlation among different instruments)
is not as large as $10$ kHz  but typically narrower and, in an optimistic perspective,  it could range up to $100$ Hz. The putative frequency of wide-band detectors will therefore be 
indicated as $\nu_{\mathrm{LV}}$, i.e. in loose terms, the Ligo/Virgo frequency. 
 There are daring projects of wide-band detectors in space like the Lisa \cite{LISA}, the BBO
\cite{BBO} and the Decigo \cite{DECIGO} projects. The common denominator of these three projects 
is that they are all space-borne missions and that they are all sensitive to frequencies 
smaller than the mHz. While wide-band interferometers are now operating 
and might even reach their advanced sensitivities along the incoming decade, 
the achievable sensitivities of space-borne interferometers are still on the 
edge of the achievable technologies.  Since $\nu_{\mathrm{bbn}} < \nu_{\mathrm{LV}} < \nu_{\mathrm{max}}$ the wide-band interferometers 
are an ideal instrument to investigate the relic graviton spectrum in the unknown 
territory where there are neither direct nor indirect tests on the thermal history 
of the plasma. The problem is that, as it will be carefully shown, the  spectral energy density of the relic gravitons produced within the $\Lambda$CDM model is quite minute and it is undetectable by interferometers even in their advanced version where the sensitivity is expected to improve by 5 or even 6 orders of magnitude in comparison with 
the present performances \cite{LIGOS1,LIGOS2,LIGOS3} (see also \cite{virgoligo} and \cite{auriga}). 
This impasse, as previously stressed, stems from the assumption that, right after inflation, the radiation-dominated 
evolution kicks in almost suddenly. At the moment, there are no evidences neither in favor of such a statement nor against it.  The main theme 
of the present investigation will be to reverse this problem. It will be argued that 
wide-band detectors, in their advanced version, will be certainly able to test 
definite deviations from a simplistic thermal history of the plasma, i.e. 
the one stipulating that, after inflation, the radiation was suddenly 
dominating the evolution.  

\subsection{Layout of the investigation}

In the present investigation the spectral energy density of the relic gravitons will be calculated 
first at small frequencies (compatible with the CMB observations) and then at 
higher frequencies (compatible with the operational window of wide-band interferometers). 
The latter calculation will be performed both n the case of the $\Lambda$CDM paradigm but also in those 
extensions which may lead to a large spectral energy density at the scale of the wide-band 
interferometers without violating any of the bounds stemming from CMB observations. 

The spectral energy density of the relic gravitons will be introduced in section \ref{sec2}.
Gravity is inherently a non-Abelian gauge theory there are potential ambiguities in defining 
univocally an energy-momentum (pseudo)-tensor for the relic gravitons: 
it will be shown that different choices of the energy-momentum 
pseudo-tensor lead to the same spectral energy density of the relic gravitons. The punch line 
of section \ref{sec2} will be that the spectral energy density 
 can be more accurately performed with numerical methods rather than resorting, as often done, 
 to semi-analytical estimates which amount to estimate first the power spectrum and then the 
 spectral energy density. 
 
 In section \ref{sec3}, using the numerical techniques introduced in section \ref{sec2}, the spectral energy density 
 of the relic gravitons will be computed in the case of the standard $\Lambda$CDM paradigm which leads
to nearly scale-invariant spectra. The signal arising in the context of the $\Lambda$CDM paradigm 
will be confronted with the sensitivity of wide-band interferometers.

After showing the compatibility of the new methods with the results of the nearly scale-invariant spectra, 
possible scaling violations in the spectral energy density will be discussed in section \ref{sec4}.
While in the $\Lambda$CDM paradigm the spectral energy density of the relic gravitons is nearly scale invariant it is plausible 
 to construct a class of models where the spectral energy density is fully compatible 
with the CMB and with the large-scale data at low frequencies while it is potentially detectable by wide-band interferometers. 
When we say that it is plausible this simply means that it is not forbidden by any of the current observational data.
The proposed extensions of the $\Lambda$CDM paradigm have also a physical interpretation (see section \ref{sec4}) 
since they naturally arise when the thermal history of the Universe deviates, for sufficiently early times, from the 
usual assumptions of the $\Lambda$CDM scenario which stipulates that, right after inflation, the Universe 
suddenly becomes dominated by radiation. 

The minimal realization of the ideas pursued in section \ref{sec4} is scrutinized in section \ref{sec5} and it is dubbed 
 T$\Lambda$CDM scenario (for tensor-$\Lambda$CDM). The T$\Lambda$CDM paradigm
consists of two supplementary parameters, i.e., in broad terms, a new pivotal frequency and a new spectral 
slope. The new frequency marks the onset of the high-frequency branch of the spectral energy density 
of the relic gravitons.  In section \ref{sec5} the T$\Lambda$CDM paradigm is compatible 
with the current bounds stemming not only from CMB and large-scale structure. It will also be required that 
the big-bang nucleosynthesis (BBN) as well as pulsar timing constraints are satisfied and this 
will allow to spell out quantitatively the restrictions on the two supplementary 
parameters characterizing the T$\Lambda$CDM scenario. 
\renewcommand{\theequation}{2.\arabic{equation}}
\setcounter{equation}{0}
\section{Basic Equations}
\label{sec2}
The basic technical tools required to pursue the present analysis will be hereby summarized. 
The first part of the present section (i.e. subsection \ref{sec2aa}) contains an introduction to the basic terminology and 
a swift derivation of the main equations. Subsections \ref{sec2a} and \ref{sec2b} contain the 
details of our numerical approach whose results will also be illustrated in various physically relevant examples (see subsection 
\ref{EXCALC}) and compared to the corresponding semi-analytical 
estimates (see subsection \ref{sec2c}). Finally, the exponential damping of the relic graviton spectrum will be numerically discussed 
in subsection \ref{sec2d}. As explained in the general layout of the investigation, all the considerations 
of the present section are bound to the $\Lambda$CDM paradigm so that the typical values 
of the cosmological parameters may be usefully drawn from Tab. \ref{MOD1}.
\subsection{Generalities}
\label{sec2aa}
In the $\Lambda$CDM paradigm the geometry is conformally flat  (see Eq. (\ref{EQ2})) and the corresponding 
 tensor fluctuations  are defined as
\begin{equation}
\delta_{\mathrm{t}}^{(1)} g_{ij} = - a^2(\tau) h_{ij}, \qquad \delta_{\mathrm{t}}^{(1)} g^{ij} = \frac{h^{ij}}{a^2}, \qquad \delta_{\mathrm{t}}^{(2)} g^{ij} = - \frac{h_{k}^{i} h^{kj}}{a^2},
\label{T1}
\end{equation}
where $h_{i}^{i} = \partial_{i} h^{i}_{j} =0$. The second order action obeyed by $h_{ij}$ can be written as
\begin{equation}
S_{\mathrm{GW}} = \frac{1}{8\ell_{\mathrm{P}}^2}\int d^{4} x \sqrt{- \overline{g}} g^{\mu\nu} \partial_{\mu} h_{ij} 
\partial_{\nu} h^{ij}, 
\label{T2}
\end{equation}
where $\ell_{\mathrm{P}}^2$ is defined as 
\footnote{Some authors 
include a $\sqrt{8\pi}$ is the definition of the reduced Planck mass (what we call $\overline{M}_{\mathrm{P}} = \ell_{\mathrm{P}}^{-1}$). This 
convention is per se harmless, however, it may be confusing in practice. In the present script the conventions expressed 
by Eq. (\ref{defPL}) will always be carefully followed.}:
\begin{equation}
\ell_{\mathrm{P}}^2 = 8\pi G = \frac{1}{\overline{M}_{\mathrm{P}}^2} = \frac{8\pi}{M_{\mathrm{P}}^2}.
\label{defPL}
\end{equation}
 Equation (\ref{T2}) is effectively equivalent to the sum of the  actions of two 
(scalar) degrees of freedom minimally coupled to the background geometry. To derive Eq. (\ref{T2}) the Einstein-Hilbert 
action must be perturbed to second order in the amplitude $h_{ij}$, i.e. 
\begin{equation}
\delta_{\mathrm{t}}^{(2)} S = - \frac{1}{16\pi G} \int d^{4} x [ \delta_{\mathrm{t}}^{(2)} \sqrt{-g}\, \overline{R} + 
\sqrt{-\overline{g}}\, \delta_{\mathrm{t}}^{(2)} R + \delta_{\mathrm{t}}^{(1)} \sqrt{-g} \,\,\delta_{\mathrm{t}}^{(1)}R].
\label{SI4a}
\end{equation}
To evaluate Eq. (\ref{SI4a}) in explicit terms it is necessary to compute the Ricci tensors 
both to first and second order, i.e. 
\begin{eqnarray}
\delta_{\rm t}^{(1)} R_{ij} &=& \frac{1}{2} [h_{ij}'' + 2 {\cal H} h_{ij}' - \nabla^2 h_{ij}] 
+ ({\cal H}' + 2 {\cal H}^2) h_{ij},
\label{rij1}\\
\delta_{\rm t}^{(2)} R_{00} &=& \frac{1}{4} h_{ij}' {h^{ij}}' - \frac{{\cal H}}{2} 
h_{ij} {h^{ij}}' + \frac{1}{2} h^{ij} \nabla^2 h_{ij},
\label{r00}\\
\delta_{\rm t}^{(2)} R_{ij} &=& \frac{1}{2} h^{k \ell} [ \partial_{k} \partial_{\ell} h_{ij} - 
\partial_{k} \partial_{j} h_{\ell i} - \partial_{k} \partial_{i} h_{j \ell}] 
\nonumber\\
&-& \frac{1}{2} \partial_{j} [ h^{k \ell} ( \partial_{\ell} h_{ik} - \partial_{k} h_{\ell i} - 
\partial_{i} h_{k\ell})] - \frac{{\cal H}}{2} h^{k\ell} h_{k \ell}' \delta_{ij}
\nonumber\\
&+& \frac{{\cal H}}{2} h^{\ell}_{j} h_{\ell i}' + \frac{{\cal H}}{2} h^{\ell}_{i} h_{\ell j}' 
- \frac{1}{4} {h^{k}_{j}}' h_{ik}' - \frac{{\cal H}}{2} {h^{k}_{j}}' h_{ik} - 
\frac{1}{4} {h^{k}_{i}}' h_{kj}' - \frac{{\cal H}}{2} {h^{k}_{i}}' h_{kj}
\nonumber\\
&-& \frac{1}{4} [ \partial_{i} h_{k}^{\ell} + \partial_{k} h^{\ell}_{i} - \partial^{\ell}h_{ik}]
[\partial_{\ell} h^{k}_{j} + \partial_{j} h^{k}_{\ell} - \partial^{k}h_{j \ell}].
\label{rij2}
\end{eqnarray}
The Ricci scalar is zero to first order in the tensor fluctuations, i.e. $\delta_{\rm t}^{(1)} R =0$. This is due to the traceless nature of these fluctuations.
To second-order, however, $\delta_{\rm t}^{(2)} R \neq 0$ and its form is:
\begin{eqnarray}
\delta_{\rm t}^{(2)} R &=& \frac{1}{a^2} \biggl\{ \frac{3}{4} h_{k \ell}' {h^{k\ell}}' + 
{\cal H} h_{k\ell}' h^{k\ell} + \frac{1}{2} h^{k\ell} \nabla^2 h_{k\ell} - 
\frac{1}{4}\partial_{i} h^{k\ell}  \partial^{i} h_{k\ell}\biggr\}
\nonumber\\
&+& \frac{1}{a^2} \biggl\{ - \frac{1}{2} \partial_{i}[ h^{k\ell}(\partial_{\ell} h^{i}_{k} 
- \partial_{k} h_{\ell}^{i} - \partial^{i} h_{k\ell})] 
\nonumber\\
&-& \frac{1}{4}[ \partial_{i} h^{\ell}_{k} \partial_{\ell} h^{k}_{i} - \partial_{i} h^{\ell}_{k}
\partial^{k} h_{i\ell} + \partial_{k} h^{\ell i} \partial_{\ell} h^{k}_{i} - \partial^{\ell} h_{ik}  
\partial^{i} h^{k}_{\ell} + \partial^{\ell} h_{ik} \partial^{k} h_{i \ell}]\biggr\}.
\label{r}
\end{eqnarray}
Using the results of Eqs. (\ref{rij1})--(\ref{r}) into Eq. (\ref{SI4a}) the second-order action for the tensor modes
of Eq. (\ref{T2}) can be obtained by getting rid of a number of total derivatives. 

According to Eq. (\ref{T1}), $h_{ij}$ carries 
two degrees of freedom associated with the two polarizations of the graviton in a Friedmann-Robertson-Walker (FRW)
space-time. Defining as $\hat{k} = \vec{k}/|\vec{k}|$ the direction along which a given tensor mode propagates, the two 
polarizations can be defined as 
 \begin{eqnarray}
\epsilon_{ij}^{\oplus}(\vec{k}) = ( \hat{m}_{i} \hat{m}_{j} - \hat{n}_{i} \hat{n}_{j}), \qquad \epsilon_{ij}^{\otimes}(\vec{k}) = (\hat{m}_{i} \hat{n}_{j} + \hat{m}_{j} \hat{n}_{i}),
\label{T4}
\end{eqnarray}
where $\hat{m}$ and $\hat{n}$ are two mutually orthogonal unit vectors which are also 
orthogonal to $\hat{k}$ (i.e. $\hat{m} \cdot \hat{n} = \hat{n} \cdot \hat{k} = \hat{m} \cdot \hat{k}=0$).
During the early stages of the $\Lambda$CDM model (i.e. during the inflationary phase)  $h_{ij}(\vec{x},\tau)$ can be expanded in terms of the appropriate 
creation and annihilation operators as:
\begin{equation}
\hat{h}_{ij}(\vec{x},\tau) = \frac{\sqrt{2} \ell_{\mathrm{P}}}{(2\pi)^{3/2}}\sum_{\lambda} \int \, d^{3} k \,\,\epsilon^{(\lambda)}_{ij}(\vec{k})\, [ F_{k,\lambda}(\tau) \hat{a}_{\vec{k}\,\lambda } e^{- i \vec{k} \cdot \vec{x}} + F^{*}_{k,\lambda}(\tau) \hat{a}_{\vec{k}\,\lambda }^{\dagger} e^{ i \vec{k} \cdot \vec{x}} ],
\label{T8}
\end{equation}
where the index $\lambda$ counts the two polarizations, i.e. $\lambda = \otimes, \oplus$;
 $k$ denotes the wavenumber and $F_{k\,\lambda}(\tau)$ is the (complex) mode function obeying 
\begin{eqnarray}
F_{k,\,\lambda}' &=& G_{k,\,\lambda},
\label{T9a}\\
G_{k,\, \lambda}' &=& -2 {\mathcal H} G_{k,\,\lambda} - k^2 F_{k,\,\lambda}.
\label{T9}
\end{eqnarray}
In Eq. (\ref{T8}) $[\hat{a}_{\vec{k},\lambda}, \hat{a}^{\dagger}_{\vec{p},\lambda'}] = \delta^{(3)}(\vec{k} - \vec{p})
\delta_{\lambda\lambda'}$.  The initial state $|0\rangle$ 
(annihilated by $\hat{a}_{\vec{k},\lambda}$) minimizes the tensor Hamiltonian when all the wavelengths of the field are shorter than the event horizon at the onset of the inflationary evolution.
 The main observables which are used to characterize the relic graviton background are the two-point function evaluated at 
equal times and the spectral energy density in critical units.  
The two-point function is defined as 
\begin{eqnarray}
\langle 0| \hat{h}_{ij}(\vec{x},\tau) \, \hat{h}_{ij}(\vec{y},\tau)|0 \rangle &=& \int_{0}^{\infty} d\ln{k} \,{\mathcal P}_{\mathrm{T}}(k,\tau)\, \frac{\sin{kr}}{kr},\qquad r = |\vec{x} - \vec{y}|,
\label{T10a}\\
{\mathcal P}_{\mathrm{T}}(k,\tau) &=& \frac{4 \ell_{\mathrm{P}}^2\,\, k^3}{\pi^2} |F_{k}(\tau)|^2.
\label{T10b}
\end{eqnarray}
The quantity ${\mathcal P}_{\mathrm{T}}(k,\tau)$ is, by definition,  the tensor power spectrum. 
Equations (\ref{T10a})--(\ref{T10b})  can be derived by recalling the following pair of relations: 
\begin{equation}
\epsilon_{ij}^{(\lambda)} \epsilon_{ij}^{(\lambda')} = 2 \delta_{\lambda\lambda'}, \qquad 
F_{k,\,\oplus}(\tau)= F_{k,\,\otimes}(\tau) \equiv F_{k}(\tau).
\label{T11a}
\end{equation}
Out of Eqs. (\ref{T10a})--(\ref{T10b}) it is sometimes practical to introduce the spectral amplitude $ S_{h}(\nu,\tau)$, namely, 
\begin{equation}
\langle 0| \hat{h}_{ij}(\vec{x},\tau) \, \hat{h}_{ij}(\vec{x},\tau)|0 \rangle = 
\int_{0}^{\infty}\, {\mathcal P}_{\mathrm{T}}(k,\tau) d\ln{k}  = 4 \int_{0}^{\infty} \nu S_{h}(\nu,\tau) d\ln{\nu},
\label{T11}
\end{equation}
where $k = 2 \pi \nu$. 
By definition, $\rho_{\mathrm{GW}}(\vec{x},\tau) = \langle 0| T_{0}^{0}(\vec{x},\tau)|0 \rangle$ 
where $|0\rangle$ is, again, the state annihilated by $\hat{a}_{k,\lambda}$ (see also Eq. (\ref{T8})) and where 
$T_{\mu}^{\nu}$ is the energy-momentum (pseudo)-tensor of the relic gravitons. 
The spectral energy density of the relic gravitons in critical units can then be computed from the 
expectation value of $T_{0}^{0}$
\footnote{The natural logarithms will be denoted by $\ln$ while the common logarithms 
will be denoted by $\log$.}
\begin{equation}
\Omega_{\mathrm{GW}}(k,\tau) = \frac{1}{\rho_{\mathrm{crit}}} \frac{d \rho_{\mathrm{GW}}}{d \ln{k}},\qquad 
\rho_{\mathrm{GW}} = \langle 0| T_{0}^{0}|0\rangle.
\label{T12}
\end{equation}
where $\rho_{\mathrm{crit}} = 3 H^2/\ell_{\mathrm{P}}^2$ is the critical energy density.
In FRW space-times the energy-momentum 
pseudo-tensor of the relic gravitons can be assigned in manners which are 
conceptually different but physically complementary.  This will be one of the topics discussed in subsections \ref{sec2a} and \ref{sec2b}.

\subsection{Transfer function for the amplitude}
\label{sec2a}
To connect the early moment of the normalization of the relic gravitons to the moment where 
the tensor modes of the geometry reenter the Hubble radius and  affect, in principle, terrestrial 
detectors the customary approach is to solve for the tensor mode function and to define 
the so-called amplitude transfer function. From the amplitude transfer function the spectral energy density 
(see e.g. Eq. (\ref{T12})) can be computed. What we propose here is to do the opposite, i.e. to compute, numerically and in one 
shot, the transfer function for the spectral energy density.  To show the equivalence (but also the inherent differences) of the two approaches,  the present subsection will be concerned with 
the transfer function for the amplitude. In the following subsection (i.e. subsection \ref{sec2b}) the transfer function 
of the spectral energy density will be more specifically discussed. 

During the inflationary phase, the tensor power 
spectrum can be easily computed by solving Eqs. (\ref{T9a}) and (\ref{T9}) in the slow-roll approximation
\begin{equation}
F_{k}(\tau) = \frac{{\mathcal N}}{a(\tau) 
\sqrt{2 k}} \sqrt{ - k \tau} H_{\nu}^{(1)}( - k\tau),\qquad 
{\mathcal N} = \sqrt{\frac{\pi}{2}} e^{ i\pi(\nu + 1/2)/2},\qquad \nu = \frac{3- \epsilon}{2( 1 -\epsilon)}.
\label{TA1}
\end{equation}
where $H_{\nu}^{(1)}(z) = J_{\nu}(z) + i Y_{\nu}(z)$ is the Hankel function 
of first kind \cite{abr,tric} and where $\epsilon = - \dot{H}/{H^2}$. 
To obtain the result of Eq. (\ref{TA1}) from Eqs. (\ref{T9a}) and (\ref{T9}) it is 
useful to bear in mind the following pair of identities 
\begin{equation}
{\mathcal H}^2 + {\mathcal H}^2 = a^2 H^2 ( 2 - \epsilon), \qquad  a H = - \frac{1}{\tau (1 - \epsilon)}.
\label{TA2aa}
\end{equation}
The second equality in Eq. (\ref{TA2aa}) can be simply deduced (after integration 
by parts) from the relation between cosmic and conformal times, i.e. 
$a(\tau) d\tau = dt$. Physically Eqs. (\ref{TA1}) and (\ref{TA2aa}) hold under the 
approximation that, at early times, the background geometry is of quasi-de Sitter type.  

By substituting Eq. (\ref{TA1}) 
into Eq. (\ref{T10b}) the standard expression of the tensor 
power spectrum  can be obtained. When the relevant modes exited the Hubble 
radius during inflation:
\begin{equation}
\overline{{\mathcal P}}_{\mathrm{T}}(k,\tau) = \ell_{\mathrm{P}}^2 H^2 \frac{2^{2 \nu}}{\pi^3} \Gamma^2(\nu) ( 1 - \epsilon)^{ 2 \nu -1} \biggl(\frac{k}{a H}\biggr)^{3 - 2 \nu}, \qquad 
\nu = \frac{3}{2} + \epsilon + {\mathcal O}(\epsilon^2),
\label{TA2}
\end{equation}
where the small argument limit of the Hankel functions has been taken and where the slow-roll approximation 
has been enforced, i.e. in formulas:
\begin{equation}
x = k\tau \simeq \frac{k}{{\mathcal H}} = \frac{k}{a H} \ll 1, \qquad \epsilon = -\frac{\dot{H}}{H^2} < 1.
\label{TA2aaa}
\end{equation}
The two approximations introduced in Eq. (\ref{TA2aaa}) will be often employed and, therefore, it is appropriate 
to spell out clearly their physical meaning. The first relation of Eq. (\ref{TA2aaa}) implies that $k \tau <1$ 
this means that the wave-numbers are, in practice, smaller than the Hubble rate. Conversely 
the corresponding wavelengths will be larger than the Hubble radius. The chain of equalities 
appearing in Eq. (\ref{TA2aaa}) can be easily understood since, by definition, ${\mathcal H} = a H$ and $H$ is 
the Hubble rate. In this long wavelength limit, as we shall see, the evolution of the tensor modes can be 
derived in semi-analytical terms and it corresponds to a tensor mode function $F_{k}(\tau)$ which 
is approximately constant. The latter statement holds if the geometry is of quasi-de Sitter type. The 
latter condition is verified if the second relation of Eq. (\ref{TA2aaa}), stipulating $\epsilon<1$, holds. 
The latter conditions is also dubbed slow-roll approximation and it allows to simplify the tensor power spectrum 
even further:
\begin{equation} 
\overline{{\mathcal P}}_{\mathrm{T}}(k) \simeq \frac{2}{3 \pi^2} \biggl(\frac{V}{\overline{M}_{\mathrm{P}}^4}\biggr)_{k \simeq a H} \simeq \frac{128}{3} 
\biggl(\frac{V}{M_{\mathrm{P}}^4}\biggr)_{k \simeq a H}.
\label{TA3}
\end{equation}
The spectral index defined from Eq. (\ref{TA3}) is nothing but 
\begin{equation}
n_{\mathrm{T}} = \frac{d \ln{\overline{{\mathcal P}}_{\mathrm{T}}}}{d \ln{k}} = - \frac{2\epsilon}{1 - \epsilon} = - 2\epsilon + {\mathcal O}(\epsilon^2).
\label{TA4}
\end{equation}
 where the second equality can be derived with the standard rules of the slow-roll algebra.  The spectral amplitude and  slope are then parametrized, for practical purposes, as
\begin{equation}
\overline{{\mathcal P}}_{\mathrm{T}}(k) = {\mathcal A}_{\mathrm{T}} 
\biggl(\frac{k}{k_{\mathrm{p}}}\biggr)^{n_{\mathrm{T}}}, \qquad k_{\mathrm{p}} = 0.002 \,\, \mathrm{Mpc}^{-1},
\label{TA5}
\end{equation}
where, by definition, ${\mathcal A}_{\mathrm{T}}$ is the amplitude of the tensor power spectrum evaluated at the pivot scale $k_{\mathrm{p}}$. The pivot wavenumber of Eq. (\ref{TA5}) is simply related to the pivot frequency defined in Eq. (\ref{EQ1}) as $\nu_{\mathrm{p}} = k_{\mathrm{p}}/(2\pi)$. 
 Bearing in mind that the power spectrum of curvature perturbations is given, in single field inflationary models, as 
\begin{equation}
\overline{{\mathcal P}}_{{\mathcal R}}(k) = \frac{8}{3}\biggl(\frac{V}{\epsilon\, M_{\mathrm{P}}^4}\biggr)_{k \simeq a H} = 
{\mathcal A}_{{\mathcal R}} \biggl(\frac{k}{k_{\mathrm{p}}}\biggr)^{n_{\mathrm{s}} -1},
\label{TA6}
\end{equation}
the ratio between the tensor and the scalar power spectra is simply given by 
\begin{equation}
r_{\mathrm{T}} = \frac{\overline{{\mathcal P}}_{\mathrm{T}}(k)}{\overline{{\mathcal P}}_{{\mathcal R}}(k)}= \frac{{\mathcal A}_{\mathrm{T}}}{{\mathcal A}_{{\mathcal R}}} =  16 \epsilon, 
\label{TA6a}
\end{equation}
Equation (\ref{TA6a}) implies, recalling Eq. (\ref{TA4}), that  $r_{\mathrm{T}} = - 8 n_{\mathrm{T}}$. 
 In Tab. \ref{MOD1} the values of  $r_{\mathrm{T}}$ have been reported as they can be 
estimated in few different analyses of the cosmological data sets. 

Equation (\ref{TA5}) correctly parametrizes the spectrum only when the relevant 
wavelengths are larger than the Hubble radius before matter-radiation equality.
To transfer the spectrum {\em inside} the Hubble radius the procedure 
is to integrate numerically Eqs. (\ref{T9a})--(\ref{T9}) (as well as Eqs. (\ref{FL1})--(\ref{FL3})) across the relevant transitions of the background geometry. While the geometry passes from inflation to radiation Eq. (\ref{TA5}) implies that 
the tensor mode function is constant while the relevant wavelengths are larger than the Hubble radius:
\begin{equation}
F_{k}(\tau) = A_{k} + B_{k} \int \frac{d\tau'}{a^2(\tau')}, \qquad \frac{k}{a H} \ll 1, \qquad |A_{k}|^2 = \frac{\pi^2}{4 \ell_{\mathrm{P}}^2 k^3} \overline{P}_{\mathrm{T}}(k).
\label{TA7}
\end{equation}
The term proportional to $B_{k}$ in Eq. (\ref{TA7}) leads to a decaying mode and 
$F_{k}(\tau)$ is therefore determined, for 
$|k\tau|\ll 1$, by the first term whose squared modulus coincides with the spectrum computed in Eq. (\ref{TA3}) 
and parametrized as in Eq. (\ref{TA5}).
The evolution of the background (i.e. Eqs. (\ref{FL1})--(\ref{FL3})) and of the tensor 
mode functions (i.e. Eqs. (\ref{T9a})--(\ref{T9})) should therefore be solved across the radiation matter transition and 
the usual approach is to compute the transfer function for the amplitude \cite{tur1} i.e. 
\begin{equation}
T_{h}(k) = \sqrt{\frac{\langle|F_{k}(\tau)|^2\rangle}{\langle|\overline{F}_{k}(\tau)|^2\rangle}}. 
\label{TA8}
\end{equation}
In Eq. (\ref{TA8}), $\overline{F}_{k}(\tau)$ denotes the approximate form of the mode function (holding during the 
matter-dominated phase); $F_{k}(\tau)$ denotes, instead, the solution obtained by fully numerical methods.  
As the wavelengths become shorter than the Hubble radius, $F_{k}(\tau)$ oscillates. Consequently,  
To get $T_{h}(k)$ the oscillations must be carefully averaged and this is the meaning of the averages appearing in 
 Eq. (\ref{TA8}). Hence, the calculation of $T_{h}(k)$ requires a careful matching over the phases between the numerical and the 
approximate (but analytical) solution.  Consider, indeed, one of the most important applications 
of the previous results, i.e. the radiation-matter transition. 
After matter-radiation equality, the scale factor is going, approximately, as $a(\tau) \simeq \tau^2$ and, therefore, the 
(approximate) solution of Eqs. (\ref{T9a})--(\ref{T9})  is given by 
\begin{equation}
\overline{F}_{k}(\tau) = \frac{3 j_{1}(k\tau)}{k\tau} A_{k}, \qquad j_{1}(k\tau) = \frac{\sin{k\tau}}{(k\tau)^2} - 
\frac{\cos{k\tau}}{(k\tau)}.
\label{TA9}
\end{equation}
which is constant for $k\tau < 1$.
\begin{figure}[!ht]
\centering
\includegraphics[height=6.2cm]{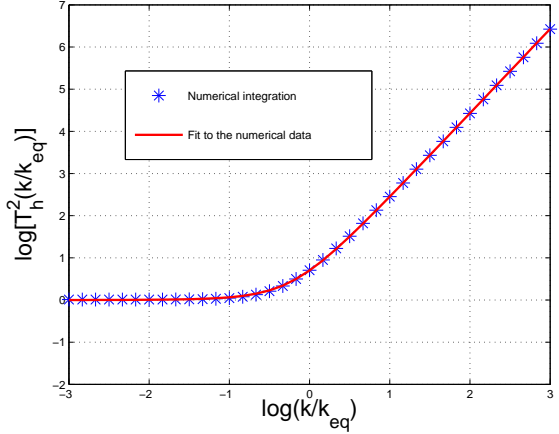}
\caption[a]{The starred points represent the numerical values of the amplitude transfer function of the amplitude  
across the matter-radiation transition. The logarithm (to base $10$) is reported on both axes. The full  line represents the 
numerical fit.}
\label{Figure1}      
\end{figure}
In Fig. \ref{Figure1} the result of the numerical integration is reported in terms of 
$T_{h}^2(k/k_{\mathrm{eq}})$.  In Fig. \ref{Figure1} the fit to the numerical points is also reported and it can be parametrized as:
\begin{equation}
T_{h}(k/k_{\mathrm{eq}}) = \sqrt{1 + c_{1} \biggl(\frac{k}{k_{\mathrm{eq}}}\biggr) + b_{1} \biggl(\frac{k}{k_{\mathrm{eq}}}\biggr)^2}.
\label{TA10}
\end{equation}
By applying the standard tools of the regression analysis  $c_{1}$ and $b_{1}$ can be determined as $c_{1}= 1.260$ and $b_{1}= 2.683$. The latter result agrees with the findings of  \cite{tur1} who obtain $ \overline{c}_{1}= 1.34$ and $\overline{b}_{1} = 2.50$. The value of $k_{\mathrm{eq}}$ can be 
obtained directly from the experimental data (see, for instance, last column of Tab. \ref{MOD1} implying $k_{\mathrm{eq}} \simeq 
{\mathcal O}(0.009)\, \mathrm{Mpc}^{-1}$).  For instance, the WMAP  5-yr data combined with the supernova data 
and with the large-scale structure data would give $k_{\mathrm{eq}} = 0.00999^{+0.00028}_{-0.00027}\,\, \mathrm{Mpc}^{-1}$.
It turns out that a rather good analytical estimate 
of $k_{\mathrm{eq}}$ can be presented as 
\begin{eqnarray}
k_{\mathrm{eq}} = 0.0082879 \,\, \biggl(\frac{h_{0}^2 \Omega_{\mathrm{M}0}}{0.1326}\biggr) \biggl(\frac{h_{0}^2 \Omega_{\mathrm{R}0}}{4.15 \times 10^{-5}}\biggr)^{-1/2}\,\, \mathrm{Mpc}^{-1}.
\label{TA11}
\end{eqnarray}
where the typical value selected for 
$h_{0}^2 \Omega_{\mathrm{R}0}$  is given by the sum of the photon component (i.e. 
$h_{0}^2 \Omega_{\gamma0} = 2.47 \times 10^{-5}$) and of the neutrino component (i.e. $h_{0}^2 \Omega_{\gamma0} = 1.68 \times 10^{-5}$): the neutrinos, consistently with the $\Lambda$CDM paradigm, are taken to be massless and their (present) 
kinetic temperature is just a factor $(4/11)^{1/3}$ smaller than the (present) photon temperature. From Eq. (\ref{TA11}) it is straightforward to estimate the equality 
frequency of Eq. (\ref{EQ3}).  

The analytical estimate stems from the observation that the exact solution of 
Eqs. (\ref{FL1})--(\ref{FL3}) for the matter-radiation transition can be given as $a(\tau) = a_{\mathrm{eq}}[ y^2 + 2 y]$ where $y= \tau/\tau_{1}$. The time-scale $\tau_{1} = \tau_{\mathrm{eq}}(\sqrt{2} +1)$ 
is related to the equality time $\tau_{\mathrm{eq}}$ which can be estimated as 
\begin{equation}
\tau_{\mathrm{eq}} = 
\frac{2 (\sqrt{2} -1)}{H_{0}} \frac{\sqrt{ \Omega_{\mathrm{R}0}}}{ \Omega_{\mathrm{M}0}} 
= 120.658 \biggl(\frac{h_{0}^2 \Omega_{\mathrm{M}0}}{0.1326}\biggr)^{-1} \biggl(\frac{h_{0}^2 \Omega_{\mathrm{R}0}}{4.15 \times 10^{-5}}\biggr)^{1/2}\,\,\mathrm{Mpc}.
\label{TA13}
\end{equation}
In the case of the WMAP 5-yr data combined with the supernova and large-scale structure data 
$h_{0}^2 \Omega_{\mathrm{M0}}= 0.1368^{0.0038}_{-0.0037}$.
Consequently, Eqs. (\ref{TA8}), (\ref{TA9}) and (\ref{TA10}) imply that  the spectrum of the tensor modes is given, at the present time, as 
\begin{equation}
{\mathcal P}_{\mathrm{T}}(k,\tau_{0}) = \frac{9 j_{1}^2(k\tau_{0})}{(k\tau_{0})^2} 
T^2_{h}(k/k_{\mathrm{eq}})\overline{{\mathcal P}}_{\mathrm{T}}(k).
\label{TA14}
\end{equation}
Within the standard approach, Eq. (\ref{TA14}) is customarily connected to the 
spectral energy density of the relic gravitons. It will now be shown
 that the spectral energy density of the relic gravitons can be directly assessed, by numerical means, without resorting to Eq. (\ref{TA14}). 
\subsection{Spectral energy density}
\label{sec2b}
Having presented the standard derivation of the amplitude transfer function we will now focus the attention 
on the transfer function of the spectral energy density. The latter approach leads more directly to the estimate 
of the present value of the spectral energy density. Before discussing in some detail the numerics it is 
appropriate to recall  the construction of an energy-momentum pseudo-tensor 
by following the same approach which has been proven to be successful 
in flat space-time \cite{landau}. In a conformally
flat geometry of the type introduced in Eq. (\ref{EQ2}), the energy 
momentum pseudo-tensor can be derived by following two complementary 
strategies.  The first  one is 
to take the energy-momentum tensor associated with the action of Eq. (\ref{T2}). 
Since each polarization of the graviton in a FRW space-time obeys the evolution equation of a minimally coupled scalar field, it is legitimate to establish that the energy-momentum pseudo-tensor  is just given by the energy-momentum tensor of a pair of scalar degrees of freedom minimally coupled to the geometry. By formally taking the functional derivative 
of Eq. (\ref{T2}) with respect to $\overline{g}_{\mu\nu}$, $T_{\mu}^{\nu}$ becomes  
\begin{equation}
T_{\mu}^{\nu} = \frac{1}{4 \ell_{\mathrm{P}}^2} \biggl[ \partial_{\mu} h_{ij} \partial^{\nu} h^{ij} - \frac{1}{2} \delta_{\mu}^{\nu} \overline{g}^{\alpha\beta} \partial_{\alpha} h_{ij} \partial_{\beta}h^{ij} \biggr] = \frac{1}{2\ell_{\mathrm{P}}^2} \sum_{\lambda} \biggl[ \partial_{\mu} h_{(\lambda)}
 \partial^{\nu} h^{(\lambda)} - \frac{1}{2} \overline{g}^{\alpha\beta} \partial_{\alpha} h_{(\lambda)}
 \partial_{\beta} h_{(\lambda)} \delta_{\mu}^{\nu}\biggr],
\label{TA15}
\end{equation}
where the second equality follows from the first by using that $h_{ij} = \sum_{\lambda} h_{(\lambda)} \epsilon_{ij}^{\lambda}$ 
and that $\epsilon_{ij}^{(\lambda)} \epsilon_{ij}^{(\lambda')} = 2 \delta_{\lambda\lambda'}$. This perspective was adopted and developed, for the first time, in \cite{ford1,ford2} by Ford and Parker. A complementary approach is to use the energy-momentum pseudo-tensor defined from the second-order fluctuations of the Einstein tensor:
\begin{equation}
{\mathcal T}_{\mu}^{\nu} = - \frac{1}{\ell_{\mathrm{P}}^2} \delta^{(2)}_{\rm t} {\cal G}_{\mu}^{\nu},\qquad 
{\mathcal G}_{\mu}^{\nu} = R_{\mu}^{\nu} - \frac{1}{2} \delta_{\mu}^{\nu} R.
\label{TA16}
\end{equation}
where $\delta_{\mathrm{t}}^{(2)}$ denotes the second-order tensor
fluctuation of the corresponding quantity. The latter approach is more directly 
related to the well known flat space-time procedure \cite{landau}.
The approach expressed by Eq. (\ref{TA16})  has been described in 
\cite{isacson1,isacson2} and has been reprised, in a 
related context, by the authors of Refs. \cite{abramo1,abramo2} 
mainly in connection with conventional inflationary models 
where the Universe is always expanding. 

According to Eq. (\ref{TA15}) the 
energy density is given by $\rho^{(1)}_{\mathrm{GW}} = \langle 0| T_{0}^{0} |0\rangle$ where $|0\rangle$ is the state 
annihilated by the creation and destruction operators introduced in Eq. (\ref{T8}):
\begin{equation}
\rho^{(1)}_{\mathrm{GW}}(\tau) 
= \frac{1}{a^4} \int d\ln{k} \frac{k^3}{2\pi^2} \biggl\{ |g_{k}(\tau)|^2 + ( k^2 + {\mathcal H}^2) |f_{k}(\tau)|^2
- {\mathcal H}[ f_{k}^{*}(\tau)g_{k}(\tau) + f_{k}(\tau) g_{k}^{*}(\tau)] \biggr\},
\label{TA17}
\end{equation}
where  $f_{k}(\tau) = F_{k}(\tau) a(\tau)$   and $g_{k}(\tau)= f_{k}'(\tau)$ have been 
introduced.   The superscript appearing in Eqs. (\ref{TA17}) reminds that the energy density refers to the first choice of the energy-momentum tensor given  in Eq. (\ref{TA15}).
According to Eqs. (\ref{T9a}) and (\ref{T9}) 
the tensor mode functions $f_{k}$ and $g_{k}$ obey
\begin{equation}
f_{k}' = g_{k},\qquad g_{k}' = - [ k^2 - ({\mathcal H}^2 + {\mathcal H}')]f_{k}.
\label{TA18}
\end{equation}
 By adopting the approach expressed by Eq. (\ref{TA16}), 
  the energy density  of the relic gravitons  $\rho^{(2)}_{\mathrm{GW}} =  \langle 0 | {\mathcal T}_{0}^{0} |0 \rangle$ become:
\begin{equation}
\rho^{(2)}_{\mathrm{GW}}(\tau) = \int d\ln{k} \frac{k^3}{2\,\pi^2\,a^4} \biggl\{ |g_{k}(\tau)|^2 + ( k^2 -7 {\mathcal H}^2) |f_{k}(\tau)|^2 + 3{\mathcal H}[ f_{k}^{*}(\tau)g_{k}(\tau) + f_{k}(\tau) g_{k}^{*}(\tau)] \biggr\}.
\label{TA19}
\end{equation}
To pass from Eq. (\ref{TA16}) to Eq. (\ref{TA19}) the simplest procedure is to obtain the second-order fluctuation of the Einstein 
tensor, i.e. $\delta^{(2)}{\mathcal G}_{\mu}^{\nu}$. This calculation can be easily carried on by using the results of Eqs. (\ref{rij1})--(\ref{r}).
Furthermore, it should be appreciated that the two energy-momentum pseudo-tensors lead also to different pressures 
and this observation has an impact on the back-reaction problems \cite{mg0}. 
From Eqs. (\ref{TA17})--( and (\ref{TA19}) the corresponding critical fractions are:
\begin{equation}
\Omega_{\mathrm{GW}}^{(1)}(k,\tau) = \frac{1}{\rho_{\mathrm{crit}}} \frac{d \rho^{(1)}_{\mathrm{GW}}}{d\ln{k}},\qquad
\Omega_{\mathrm{GW}}^{(2)}(k,\tau) = \frac{1}{\rho_{\mathrm{crit}}} \frac{d \rho^{(2)}_{\mathrm{GW}}}{d\ln{k}}.
\label{TA20}
\end{equation}
If $ k /{\mathcal H} > 1$, then $f_{k}(\tau)$ will be, in the 
first approximation, plane waves and $ g_{k}(\tau) \simeq \pm i k f_{k}(\tau)$ and 
the two versions of $\Omega_{\mathrm{GW}}(k,\tau)$ will be given by:
\begin{eqnarray}
&&\Omega_{\mathrm{GW}}^{(1)}(k,\tau) = \frac{k^5 \, \ell_{\mathrm{P}}^2}{ 3 \pi^2  a^2 {\mathcal H}^2} \biggl[1 + 
\frac{{\mathcal H}^2}{2 k^2} \biggr] |f_{k}(\tau)|^2 = \frac{k^2}{ 12  {\mathcal H}^2} {\mathcal P}_{\mathrm{T}}(k,\tau)  \biggl[1 + 
\frac{{\mathcal H}^2}{2 k^2} \biggr],
\label{cr1inside}\\
&&\Omega_{\mathrm{GW}}^{(2)}(k,\tau) = \frac{k^5 \, \ell_{\mathrm{P}}^2}{ 3 \pi^2  a^2 {\mathcal H}^2} \biggl[1 -
\frac{7{\mathcal H}^2}{2 k^2} \biggr]|f_{k}(\tau)|^2=\frac{k^2 }{ 12 {\mathcal H}^2} {\mathcal P}_{\mathrm{T}}(k,\tau)\biggl[1 -
\frac{7{\mathcal H}^2}{2 k^2} \biggr] ,
\label{cr2inside}
\end{eqnarray}
where, the second equality follows from the first by recalling that $
|f_{k}(\tau)|^2 = \pi^2 a^2 {\mathcal P}_{\mathrm{T}}(k,\tau)/(4 \ell_{\mathrm{P}}^2 k^3)$
Equations (\ref{cr1inside}) and (\ref{cr2inside}) 
coincide (up to corrections ${\mathcal O}({\mathcal H}^2/k^2)$). This means, physically, that the energy density of the relic gravitons 
is effectively the same no matter what choice of the energy-momentum pseudo-tensor is adopted but provided 
the wavelengths of the gravitons are all inside (i.e. shorter than) the Hubble radius.  
When the given wavelengths are larger than the Hubble radius (i.e. $k \tau \ll 1$), $g_{k} = {\mathcal H} f_{k}$ and the 
corresponding expressions for the spectral energy densities can be easily obtained.
\begin{figure}[!ht]
\centering
\includegraphics[height=6.2cm]{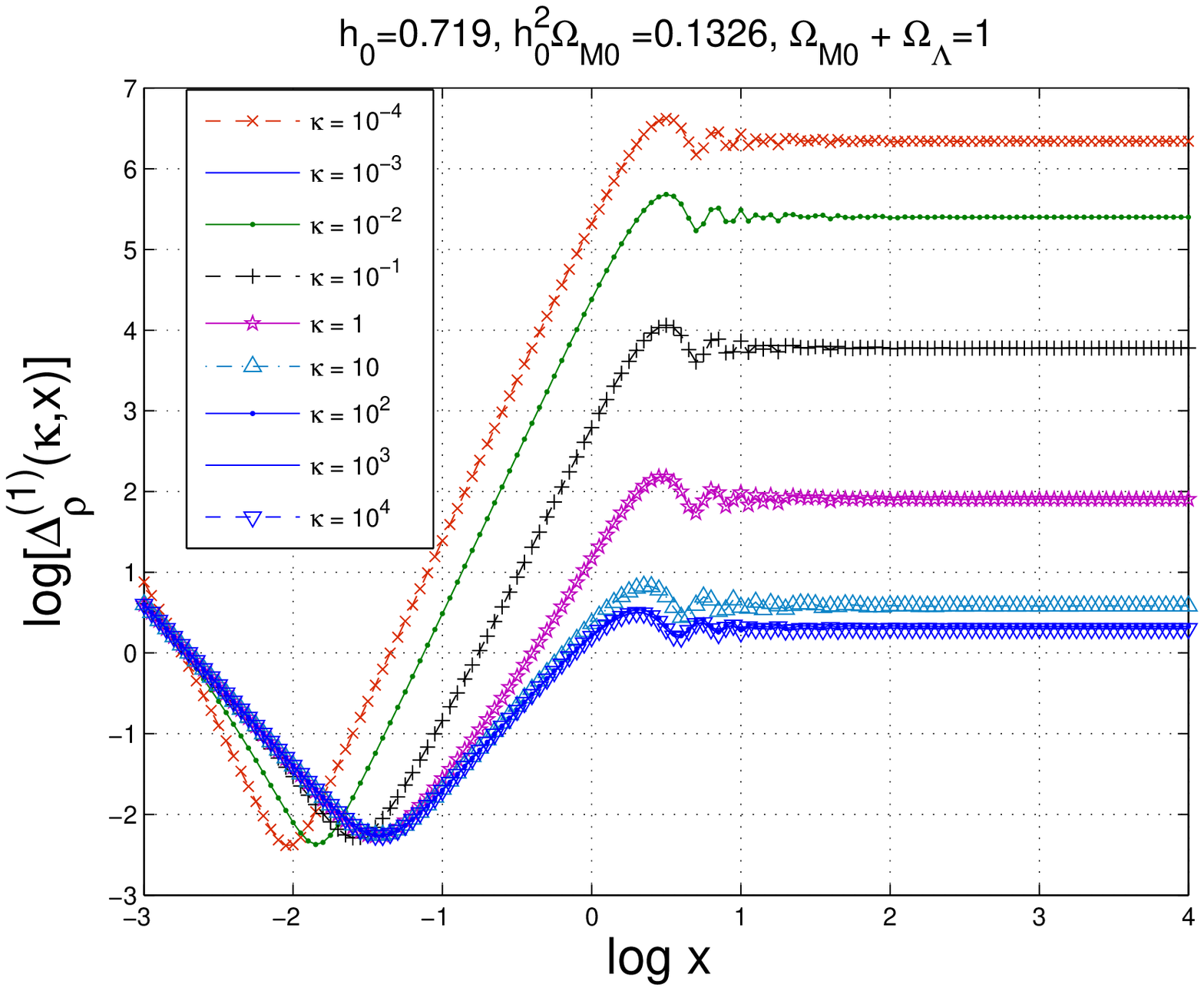}
\includegraphics[height=6.2cm]{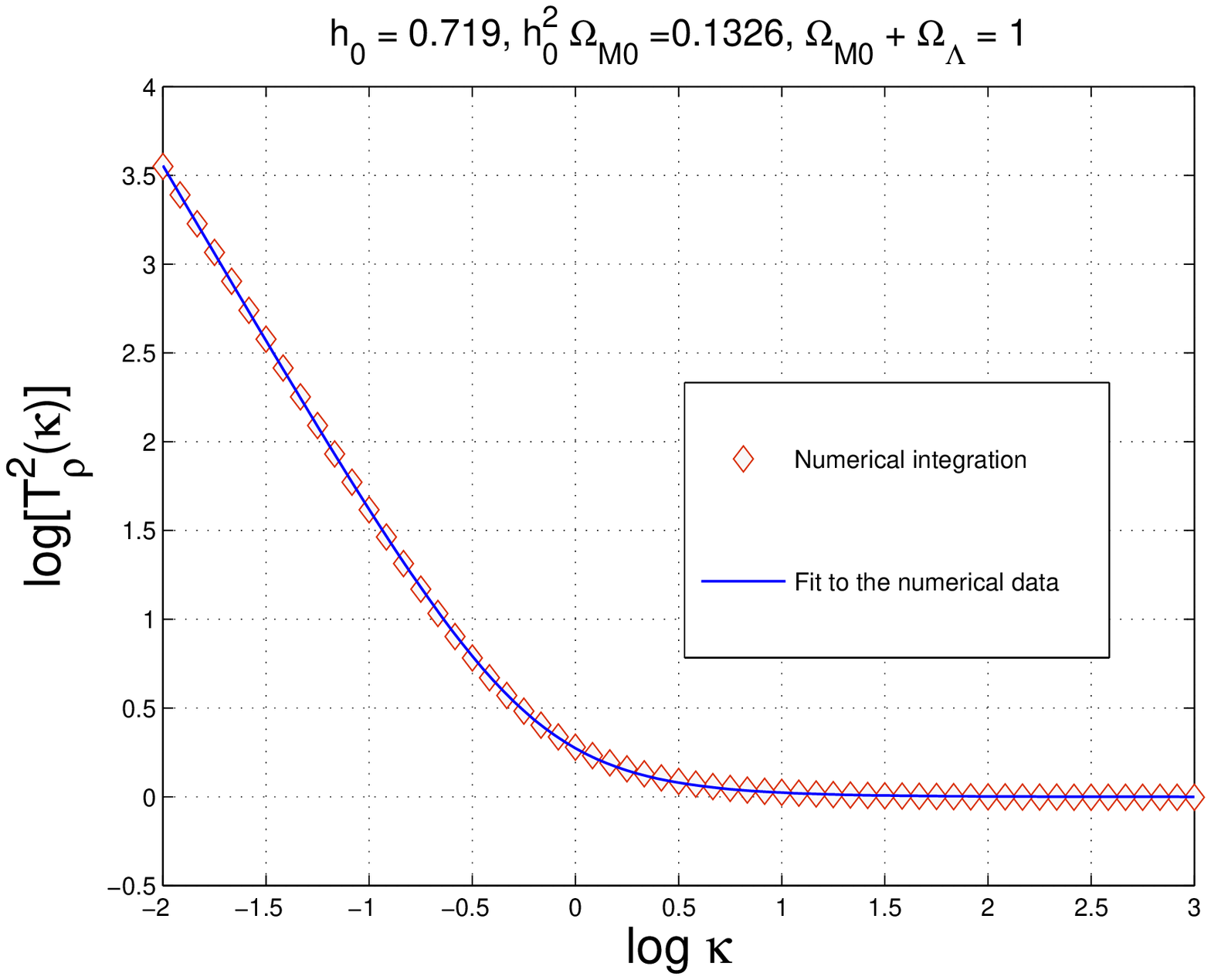}
\caption[a]{The functions given in Eqs. (\ref{DELTA1}) and (\ref{DELTA2}) are
numerically computed (plot at the left) for different 
values of $\kappa$ and in the case of the radiation-matter transition.
 In the plot at the right the transfer function for the energy density is illustrated.}
\label{Figure2}      
\end{figure}
In summary, for modes which are inside the Hubble radius the energy density 
of the relic gravitons can be expressed in terms of the power spectrum as
\begin{equation}
\rho_{\mathrm{GW}}(\tau) = \frac{2}{a^4} \int d\ln{k} \frac{k^2}{2\pi^2} |f_{k}|^2 = \frac{1}{ 4 \ell_{\mathrm{P}}^2 a^2}
\int d\ln{k}\, k^2\, {\mathcal P}_{\mathrm{T}}(k,\tau),
\end{equation}
and the critical fraction of relic gravitons at a given time as:
\begin{equation}
\Omega_{\mathrm{GW}}(k,\tau) = \frac{1}{\rho_{\mathrm{crit}}} \frac{d \rho_{\mathrm{GW}}}{d \ln{k}} = 
\frac{k^2}{12 H^2 a^2} {\mathcal P}_{\mathrm{T}}(k,\tau).
\label{DEF1}
\end{equation}
Specific examples of the numerical calculation of the spectral energy density 
will be given in the following subsection (i.e. subsection \ref{EXCALC}). The first example will be the one of the radiation-matter transition the second
example will be the one of the transition between a stiff phase and the radiation-dominated phase. 

\subsection{Transfer function for the spectral energy density: examples}
\label{EXCALC}
 The idea pursued in the present subsection is to use, as pivot quantity for the numerical integration, 
 not   the  power spectrum ${\mathcal P}(k,\tau)$ but rather the energy density itself.  The evolution equations of the background geometry (i.e. Eqs. (\ref{FL1})--(\ref{FL3})) and of the tensor mode functions (i.e. 
Eq. (\ref{TA18}))  will be solved simultaneously and the energy density computed in one shot.  This program will be illustrated 
in two simple examples, i.e. the radiation-matter transition and the case of a stiff background.
It is useful to point out that Eq. (\ref{TA2aaa}) suggests 
that an appropriate variable for the numerical calculation is exactly $x= k\tau$ whose definition 
we now repeat:
\begin{equation}
x = k \tau = \kappa \biggl(\frac{\tau}{\tau_{\mathrm{eq}}}\biggr), \qquad \kappa = \frac{k}{k_{\mathrm{eq}}}.
\label{defx}
\end{equation}
It is both practical and physically sound to adopt $x$ and $\kappa$ as pivotal variables 
for the numerical integration around the radiation-matter transition. Indeed $x$ is a smooth variable which interpolates between the sub-Hubble regime (where 
Eqs. (\ref{cr1inside})--(\ref{cr2inside}) are valid and the super-Hubble regime where $x > 1$.
The result of the numerical calculation are reported in Fig. \ref{Figure2} in terms of $\Delta^{(1)}_{\rho}(\kappa,x)$ and in terms 
of the transfer function of the energy density (denoted by $T_{\rho}(\kappa)$).  The quantities $\Delta^{(1)}_{\rho}(\kappa, x)$ 
(and, analogously, $\Delta^{(2)}_{\rho}(\kappa, x)$) are nothing but 
\begin{eqnarray}
\Delta^{(1)}_{\rho}(k,\tau) &=& \biggl\{ |g_{k}(\tau)|^2 + ( k^2 + {\mathcal H}^2) |f_{k}(\tau)|^2
- {\mathcal H}[ f_{k}^{*}(\tau)g_{k}(\tau) + f_{k}(\tau) g_{k}^{*}(\tau)] \biggr\},
\label{DELTA1}\\
\Delta^{(2)}_{\rho}(k,\tau) &=&\biggl\{ |g_{k}(\tau)|^2 + ( k^2 -7 {\mathcal H}^2) |f_{k}(\tau)|^2 + 3{\mathcal H}[ f_{k}^{*}(\tau)g_{k}(\tau) + f_{k}(\tau) g_{k}^{*}(\tau)] \biggr\},
\label{DELTA2}
\end{eqnarray}
Equations (\ref{DELTA1}) and (\ref{DELTA2}) are simply related to the spectral energy densities in critical units, i.e.
\begin{equation}
\Omega^{(1)}_{\mathrm{GW}}(k,\tau) = \frac{k^3}{2\pi^2 a^4 \rho_{\mathrm{crit}}} \Delta^{(1)}_{\rho}(k,\tau), \qquad 
\Omega^{(2)}_{\mathrm{GW}}(k,\tau) = \frac{k^3}{2\pi^2 a^4 \rho_{\mathrm{crit}}} \Delta^{(2)}_{\rho}(k,\tau).
\label{DELTA3}
\end{equation}
As a function of $x$ and $\kappa$
$\Delta_{\rho}^{(1,2)}(\kappa,x)$ reaches a constant value when the relevant modes are 
evaluated deep inside the Hubble radius. The energy transfer function which is then defined as: 
\begin{equation}
\lim_{x \gg 1} \Delta_{\rho}^{(1,2)}(\kappa,x) \equiv T^{2}_{\rho}(\kappa)  \Delta_{\rho}^{(1,2)}(\kappa,x_{\mathrm{i}}), \qquad x_{\mathrm{i}} \ll 1.
\label{DELTA4}
\end{equation}
\begin{figure}[!ht]
\centering
\includegraphics[height=6.2cm]{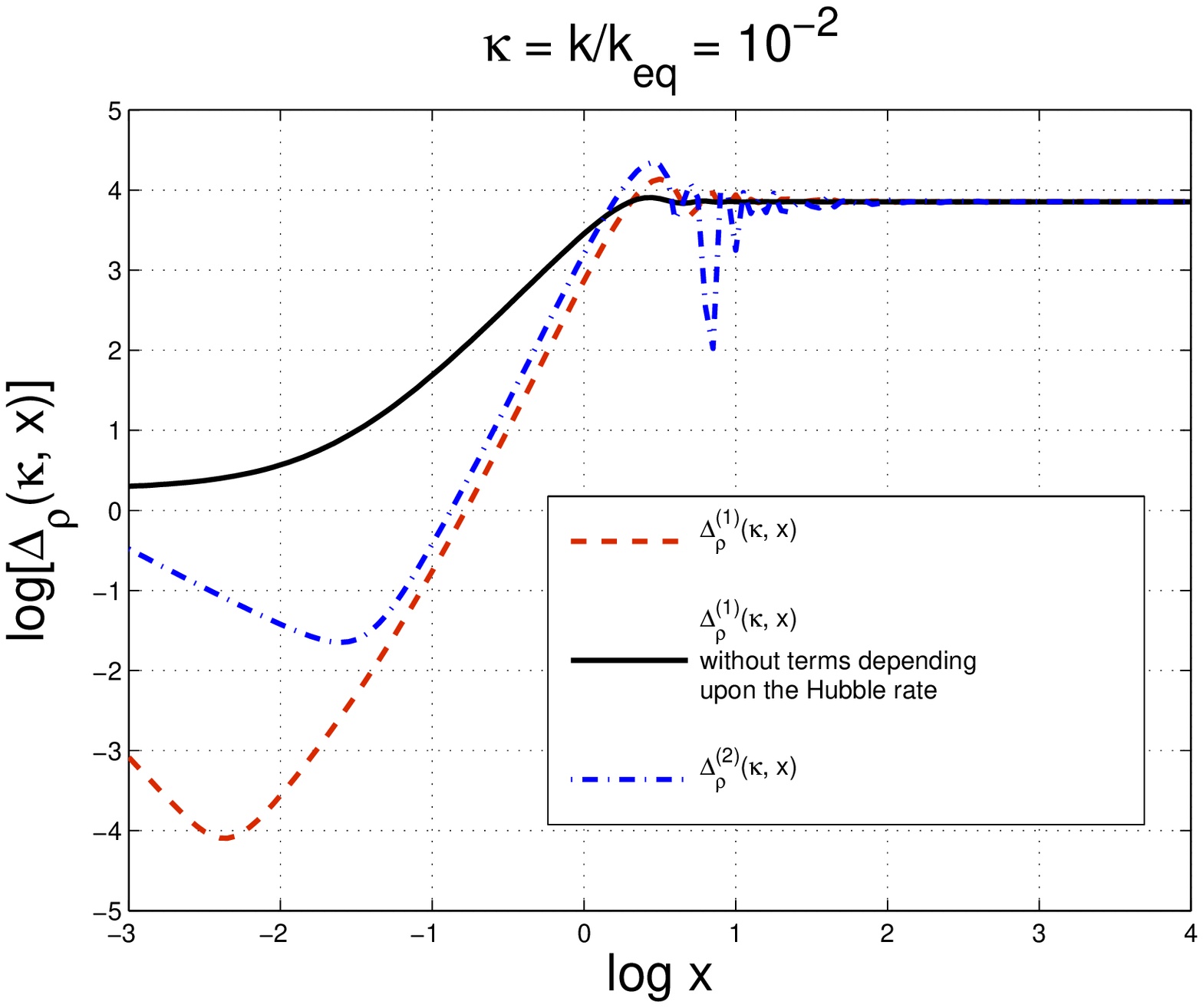}
\includegraphics[height=6.2cm]{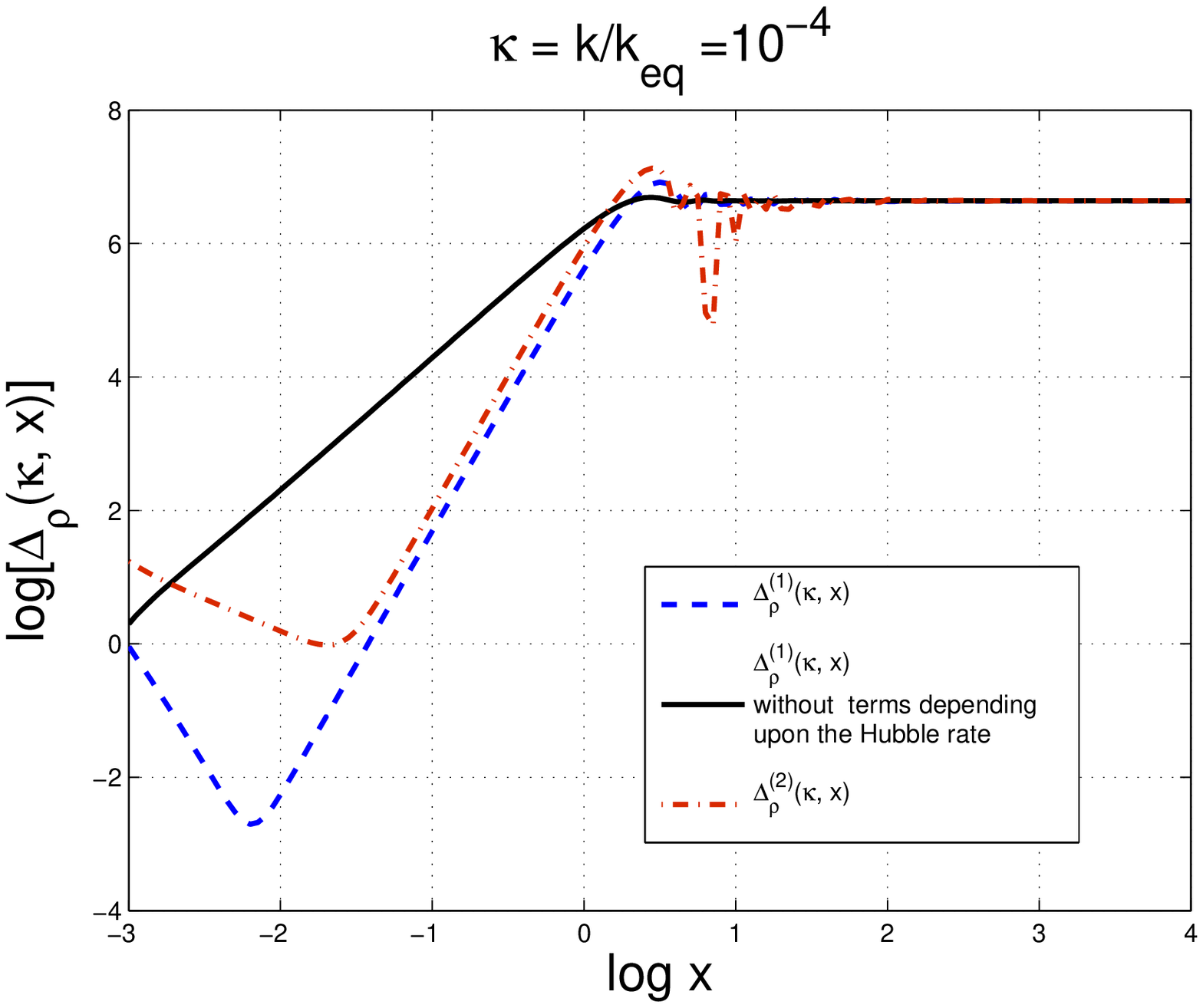}
\caption[a]{The different definitions of energy-momentum pseudo-tensor (i.e. 
Eqs. (\ref{DELTA1}) and (\ref{DELTA2})) are compared in the 
determination of the asymptotic value of the energy transfer function.}
\label{Figure3}      
\end{figure}
The specific form of the energy-momentum tensor is immaterial 
for the determination of $T_{\rho}^2(\kappa)$: different forms of the energy-momentum tensor 
of the relic gravitons will lead to the same result. This occurrence can be appreciated 
from Fig. \ref{Figure3} where $\Delta^{(1,2)}_{\rho}(k,\tau)$ has been 
reported for $\kappa = 10^{-2}$ (plot at the left) and for $\kappa = 10^{-4}$ (plot at the right). 
The dashed and the dot-dashed curves (in both plots) correspond, respectively, to $\Delta^{(1)}(\kappa, x)$ and 
to $\Delta^{(2)}(\kappa, x)$. The full line, in both plots, corresponds 
to the combination
\begin{equation}
 k^2 |f_{k}(\tau)|^2 +  |g_{k}(\tau)|^2= k( |c_{+}(k)|^2 + |c_{-}(k)|^2),
\label{comb1}
\end{equation}
where $c_{\pm}(k)$ are the so-called mixing coefficients which
parametrize, at a given time, the solution for the tensor mode 
functions when the relevant wavelengths are all inside the Hubble radius, i.e. 
\begin{equation}
\overline{f}_{k}(\tau) = \frac{1}{\sqrt{2 k}} \biggl[ c_{+}(k) e^{-i k \tau} + c_{-}(k) e^{i k\tau}\biggr],\qquad 
\overline{g}_{k}(\tau) = - i \sqrt{\frac{k}{2}} \biggl[ c_{+}(k) e^{-i k\tau} - c_{-}(k) e^{i k\tau}\biggr],
\label{BOG1}
\end{equation}
where $\overline{f}_{k}(\tau)$ and $\overline{g}_{k}(\tau)$ are the solutions to leading order in the limit $k\tau\gg1 $.
From Eq. (\ref{BOG1}), $c_{\pm}(k)$ are given by
\begin{equation}
 c_{+}(k) = \frac{e^{i k\tau}}{\sqrt{2 k}}[ k \overline{f}_{k}(\tau) + i \overline{g}_{k}(\tau)],\qquad c_{-}(k) = \frac{e^{-i k\tau}}{\sqrt{2 k}}[ k \overline{f}_{k}(\tau) - i \overline{g}_{k}(\tau)],
\label{BOG2}
\end{equation}
Using Eqs. (\ref{BOG1})--(\ref{BOG2}), 
 Eqs. (\ref{DELTA1})--(\ref{DELTA2}) can be directly assessed in the limit 
  $x = k\tau \gg1$ with the result that 
\begin{equation}
 \Delta^{(1)}_{\rho}(\kappa,x_{\mathrm{f}}) = \Delta^{(2)}_{\rho}(\kappa,x_{\mathrm{f}}) = 
 \kappa ( |c_{+}(\kappa)|^2 + |c_{-}(\kappa)|^2)  + {\mathcal O}\biggl(\frac{1}{x_{\mathrm{f}}}\biggr),
 \label{BOG4}
 \end{equation}
 which proofs that the oscillating contributions are suppressed as $x_{\mathrm{f}}^{-1}$ for $x_{\mathrm{f}} \gg 1$.
 
 To get to the results illustrated in Figs. \ref{Figure2} and \ref{Figure3} the evolution equations of the mode functions have been integrated by setting 
initial conditions deep outside the Hubble radius (i.e. $x = k\tau \ll 1$), by following 
the corresponding quantities through the Hubble crossing (i.e. $ x \simeq 1$) and then, finally, deep inside 
the Hubble radius (i.e. $x \gg 1$).  The initial value of the integration variable $x$ has been chosen to be $x_{\mathrm{i}} = 10^{-5}$.
The integration of the mode functions is most easily performed in terms of appropriately rescaled variables and since these 
rescalings are rather obvious, the relevant details will be omitted.

In the plot at the right of Fig. \ref{Figure2}, the fit to the energy transfer function is reported with the full (thin) 
line on top of the diamonds defining the numerical points. The analytical form of the fit can then be 
written as:
\begin{equation}
T_{\rho}(k/k_{\mathrm{eq}}) = \sqrt{1 + c_{2}\biggl(\frac{k_{\mathrm{eq}}}{k}\biggr) + b_{2}\biggl(\frac{k_{\mathrm{eq}}}{k}\biggr)^2},\qquad c_{2}= 0.5238,\qquad
b_{2}=0.3537.
\label{ENTRANS}
\end{equation}
Equation (\ref{ENTRANS}) permits the accurate evaluation of the spectral energy density of relic gravitons, for instance, in the minimal version of the $\Lambda$CDM paradigm. 

Yet another relevant physical situation for the present considerations is the one where the background geometry, after inflation, 
transits from a stiff epoch to the ordinary radiation-dominated epoch. 
 In the primeval plasma, stiff phases can arise: 
 this idea goes back to the pioneering suggestions 
of Zeldovich \cite{ZEL1} in connection with the entropy problem.  
The approach of Zeldovich was revisited in \cite{mg1,mg2,mg3,mg4}
by supposing that the stiff phase would take place after the inflationary phase 
with the main purpose of identifying a potential source of high-frequency gravitons
which could even be interesting for the LIGO/VIRGO detectors in one of their advanced versions.

\begin{figure}[!ht]
\centering
\includegraphics[height=6.2cm]{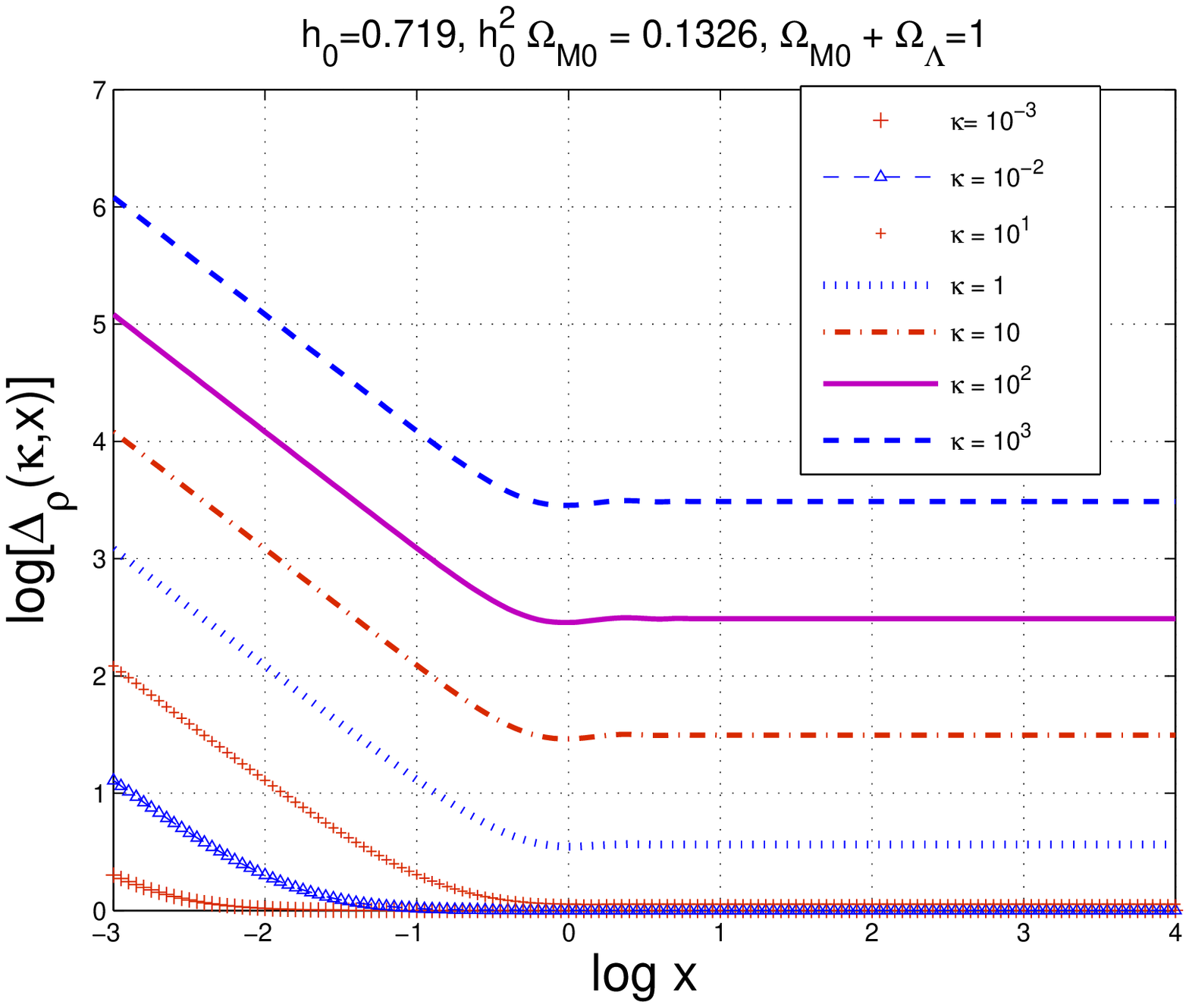}
\includegraphics[height=6.2cm]{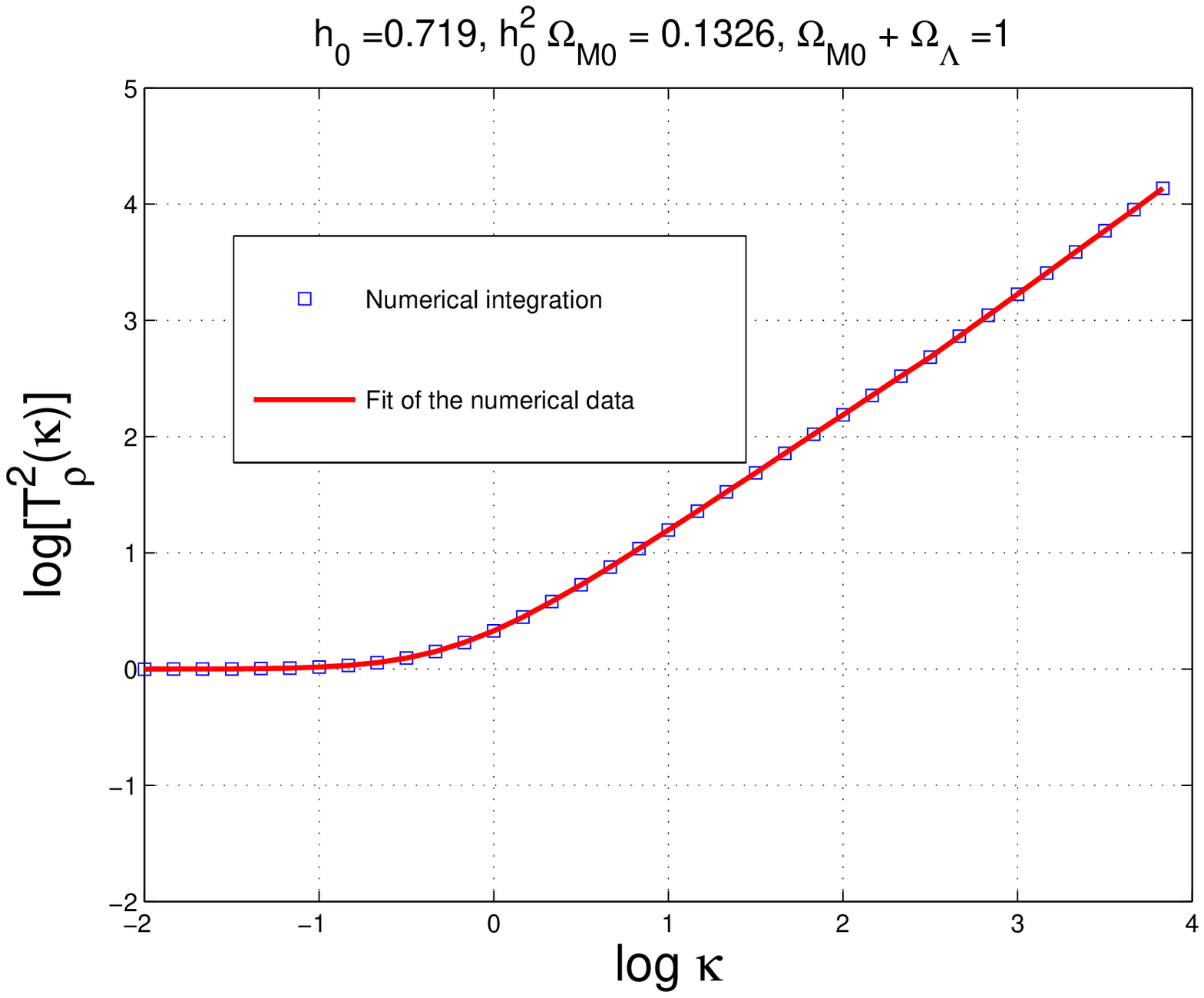}
\caption[a]{The transition between the stiff phase and the radiation phase is illustrated. The energy transfer function increases with the frequency while the opposite is true for the radiation-matter transition (see Fig. \ref{Figure2}). }
\label{Figure4}      
\end{figure}

 At the end of inflation, in a model-independent 
approach, it is plausible to think that the onset of the radiation-dominance could be 
delayed. This may happen, in concrete models, for various reasons. One possibility 
could be that the inflaton field does not decay but rather
changes its dynamical nature by acting as quintessence field \cite{PV} (see also 
\cite{Spok}).  In this kind of situations the geometry passes from a stiff phase where
\begin{eqnarray}
w_{\mathrm{t}}(\tau) &=& \frac{p_{\mathrm{t}}}{\rho_{\mathrm{t}}} > \frac{1}{3}, 
\label{ST1}\\
c^2_{\mathrm{st}}(\tau) &=& \frac{p_{\mathrm{t}}'}{\rho_{\mathrm{t}}'} = w_{\mathrm{t}} - \frac{w_{\mathrm{t}}'}{ 3 {\mathcal H} (w_{\mathrm{t}} + 1)} = w_{\mathrm{t}} - \frac{1}{3} \frac{d \ln{( w_{\mathrm{t}} + 1)}}{d \ln{a}}>  \frac{1}{3},
\label{ST2}
\end{eqnarray}
to a radiation-dominated phase where $c_{\mathrm{st}} = 1/\sqrt{3}$.
According to Eqs. (\ref{ST1}) and (\ref{ST2}), $c_{\mathrm{st}}^2 = w_{\mathrm{t}}$ iff the (total) barotropic 
index is constant in time. In the limiting case  $w_{\mathrm{t}} = 1 = c_{\mathrm{st}}^2$ 
and the speed of sound coincides with the speed of light. As argued 
in \cite{SpS}, barotropic indices $w_{\mathrm{t}} >1$ would not be 
compatible with causality (see, however, \cite{kessence1}).  
As in the case of the matter-radiation transition the transfer function only depends upon $\kappa$ which is defined, this time, 
as $\kappa = k/k_{\mathrm{s}}$, where $k_{\mathrm{s}} = \tau_{\mathrm{s}}^{-1}$ and $\tau_{\mathrm{s}}$ parametrizes 
the transition time. A simple analytical form of the transition regime is given by 
\begin{equation}
a(y) = a_{\mathrm{s}} \sqrt{y^2 + 2 y}, \qquad y = \frac{\tau}{\tau_{\mathrm{s}}},\qquad 
\tau_{\mathrm{s}} = \frac{1}{a_{\mathrm{i}} H_{\mathrm{i}}} \sqrt{\frac{\rho_{\mathrm{Si}}}{ \rho_\mathrm{Ri}}},
\label{ST3}
\end{equation}
where, by definition, $\rho_{\mathrm{si}} = \rho_{\mathrm{s}}(\tau_{\mathrm{i}})$ and $\rho_{\mathrm{Ri}} = 
\rho_{\mathrm{R}}(\tau_{\mathrm{i}})$. 
Equation (\ref{ST3}) is a solution of Eqs. (\ref{FL1})--(\ref{FL3}) when the radiation is present together 
with a stiff component which has, in the case of Eq. (\ref{ST3}) a sound speed which equals the speed of light. In the limit 
$y \to 0$ the scale factor expands as $a(y) = \sqrt{2 y}$ while,  in the opposite limit, 
$a(y) \simeq y$. 
In Fig. \ref{Figure4} (plot at the left) $\Delta_{\rho}(\kappa, x)$ is illustrated for different values of $\kappa$. We shall not dwell 
here (again) about the possible different forms of the energy momentum 
pseudo-tensor.  The bottom line will always be that, provided the energy density is evaluated deep inside the Hubble radius the different approaches 
to the energy density of the relic gravitons give the same result. From the numerical points reported in Fig. \ref{Figure4} (plot 
at the right) the semi-analytical form of the transfer function becomes, this time, 
\begin{equation}
T_{\rho}^2(k/k_{\mathrm{s}}) = 1.0  + 0.204\,\biggl(\frac{k}{k_{\mathrm{s}}}\biggr)^{1/4} - 0.980 \,\biggl(\frac{k}{k_{\mathrm{s}}}\biggr)^{1/2}  + 3.389 \biggl(\frac{k}{k_{\mathrm{s}}}\biggr) -0.067\,\biggl(\frac{k}{k_{\mathrm{s}}}\biggr)\ln^2{(k/k_{\mathrm{s}})},
\label{ST10}
\end{equation}
where $k_{\mathrm{s}} = \tau_{\mathrm{s}}^{-1}$. The value of $k_{\mathrm{s}}$ can either be 
computed in an explicit model\footnote{ In the context of quintessential 
inflation \cite{PV} (see also \cite{mg2,mg3}) $\rho_{\mathrm{Ri}} \simeq H_{\mathrm{i}}^4$ \cite{ford3}.} or it can be left as a free parameter. In section \ref{sec3}
both strategies will be explored by privileging, however, a model-independent approach.
Taking into account that the energy density of the inflaton will be exactly $\rho_{\mathrm{si}} \simeq H_{\mathrm{i}}^2 \overline{M}_{\mathrm{P}}^2$,  
the value of $k_{\mathrm{s}}$ (as well as the duration of the stiff phase) will be determined, grossly speaking, by 
$H_{\mathrm{i}}/\overline{M}_{\mathrm{P}}$.  
\subsection{Analytical estimates of the mixing coefficients}
\label{sec2c}
To obtain a fit of the transfer function for the spectral energy density 
it is useful to be aware of the analytical results which should always 
be reproduced by the numerical analysis when $\kappa$ is sufficiently 
larger than $1$.  This is the purpose of the present subsection where 
it will be shown that the semi-analytical results are consistent with the numerical 
evaluations which are, however, intrinsically more accurate.

Consider the transition from a generic accelerated phase  to a decelerated stage of expansion. In this 
situation, by naming the transition point $-\tau_{1}$, the continuous 
and differentiable form of the scale factors can be written as:
\begin{eqnarray}
&& a_{\mathrm{i}}(\tau) = \biggl( - \frac{\tau}{\tau_{1}}\biggl)^{-\beta}, 
\qquad \tau < - \tau_{1},
\label{infsc}\\
&& a_{\mathrm{s}}(\tau) = \biggl[ \frac{\beta}{\alpha}\biggl(\frac{\tau}{\tau_{1}} + 1\biggr) + 1\biggr]^{\alpha}, \qquad 
\tau \geq -\tau_{1},
\label{stsc}
\end{eqnarray}
where the scale factors are continuous and differentiable at the transition 
point which has been generically indicated as $\tau_{1}$. 
The pump fields of the tensor mode functions turn out to be:
\begin{equation}
\frac{a_{\mathrm{i}}''}{a_{\mathrm{i}}} = \frac{\beta(\beta + 1)}{\tau^2},\qquad 
\frac{a_{\mathrm{s}}''}{a_{\mathrm{s}}} = \frac{\alpha (\alpha - 1)}{\biggl[
\tau + \biggl(\frac{\alpha}{\beta} + 1\biggr) \tau_{1}\biggr]^2}.
\end{equation}
The solution of Eq. (\ref{TA18}) can then be written as:
\begin{eqnarray}
f_{\mathrm{i}}(\tau) &=& \frac{{\mathcal N}}{\sqrt{ 2 k}} \sqrt{- x}
 H_{\nu}^{(1)}( - x),\qquad \tau <  - \tau_{1},\qquad x = k \tau,
 \nonumber\\
\tilde{f}_{\mathrm{s}}(\tau) &=&  \frac{\sqrt{y}}{\sqrt{ 2 k}}[ {\mathcal M}
c_{+}(k)H_{\lambda}^{(2)}(y)  + 
{\mathcal M}^{*} c_{-}(k)H_{\lambda}^{(1)}(y)], \qquad \tau \geq -\tau_{1},
\end{eqnarray}
where  $y = k \tau + k\tau_{1} \biggl(1 + \frac{\alpha}{\beta}\biggr)$ and 
where
\begin{equation}
{\mathcal N} = \sqrt{\frac{\pi}{2}} e^{i(\nu + 1/2)\pi/2}, \qquad 
{\mathcal M} = \sqrt{\frac{\pi}{2}} e^{- i(\lambda + 1/2)\pi/2}.
 \end{equation}
The continuity of the tensor mode functions at the transition point  [i.e. 
$ f_{\mathrm{i}}(-\tau_{1}) = \tilde{f}_{\mathrm{s}}(-\tau_{1})$ and 
 $g_{\mathrm{i}}(-\tau_{1}) = \tilde{g}_{\mathrm{s}}(-\tau_{1})$] implies that the mixing coefficients are given by: 
 \begin{eqnarray}
 c_{+}(k) &=& \frac{i \pi}{8 \sqrt{\alpha \beta}} e^{i \pi (\nu +\lambda)/2} 
 \{[\beta ( 2 \lambda + 1) + \alpha ( 2 \nu + 1)] H_{\nu}^{(1)}(x_1) H_{\lambda}^{(1)}(y_{1}) 
 \nonumber\\
 &-& 2 \alpha x_{1} [H_{\lambda}^{(1)}(y_{1}) H_{\nu+1}^{(1)}(x_1) + H_{\nu}^{(1)}(x_{1}) H_{\lambda + 1}^{(1)}(y_{1})]\}, 
 \nonumber\\
 c_{-}(k) &=&  \frac{i\pi}{8 \sqrt{\alpha \beta}} e^{i \pi (\nu - \lambda)/2} 
 \{[\beta ( 2 \lambda + 1) + \alpha ( 2 \nu + 1)] H_{\nu}^{(1)}(x_1) H_{\lambda}^{(2)}(y_{1}) 
 \nonumber\\
&-& 2 \alpha x_{1} [H_{\lambda}^{(2)}(y_{1}) H_{\nu+1}^{(1)}(x_1) + H_{\nu}^{(1)}(x_{1}) H_{\lambda + 1}^{(2)}(y_{1})]\},
\end{eqnarray} 
where, according to the notations previously established, $y_{1}
= y(-\tau_{1}) = (\alpha/\beta) x_{1}$.  The case $\alpha = \beta = 1$ corresponds 
to a transition from the inflationary phase 
to a radiation-dominated phase. In this case we do know which are the mixing 
coefficients. The previous expressions give us:
\begin{equation}
c_{-}(k) = \frac{e^{ 2 i x_{1}}}{2 x_{1}^2},\qquad 
c_{+}(k) = \biggl( 1 - \frac{1}{2 x_{1}^2} + \frac{i}{x_{1}}\biggr),
\label{pureDS}
\end{equation}
which clearly agree with previous results  \cite{rub,inflsp}. 
In the case of Eq. (\ref{pureDS}) $|c_{+}(k)|^2 - |c_{-}(k)|^2 =1$ and 
$ k^4 |c_{-}(k)|^2$ is exactly scale-invariant.
Another interesting situation is the one of the transition from inflation to stiff, 
 i.e. $\beta = 1$, $\alpha = 1/2$, $y_{1} = x_{1}/2$ which leads 
 to a logarithmic enhancement at small wavenumbers \cite{mg1,mg2}.
 In this situation the mixing coefficients can be written as:
 \begin{eqnarray}
 c_{-}(k) &=& \sqrt{\frac{\pi}{2}} \frac{( i - 1 )}{4 x_1^{3/2}} e^{i x_1}\biggl\{ \sqrt{2} e^{- i x_{1}/2} [ x_{1}^2 + 6 i x_{1} - 12] H_{0}^{(2)}(x_{1}/2) 
 \nonumber\\
 &+&  (i + x_{1}) [ i x_{1} H_{1}^{(2)}(x_{1}/2) - 3 i H_{0}^{(2)}(x_1/2)]\biggr\},
 \label{mix1}\\
 c_{+}(k) &=&\sqrt{\frac{\pi}{2}} \frac{( i + 1 )}{4 \sqrt{x_{1}}} e^{i x_1}\biggl\{x_{1} H_{0}^{(1)}(x_{1}/2) + i ( i + x_{1})  H_{1}^{(1)}(x_{1}/2)\biggr\}.
 \label{mix2}
 \end{eqnarray}
The above result can be expanded in for $x_{1}\ll 1$ and the result is:
\begin{eqnarray}
c_{+}(k) &=& \frac{-0.398( 1 -\, i)}{{x_1}^{\frac{3}{2}}}  + \sqrt{x_1}\,
   [ \left( 0.131 + 0.338\, i \right)  - 0.149\left( 1 - i  \right) \ln{x_{1}}]  + {\mathcal O}(x_{1}^{3/2}),
 \label{mix1a}\\
 c_{-}(k) &=& \frac{\left( 7.031 - 1.723\,i \right)  -  16.68 \left( 1+ \,i   \right) \,\ln{x_{1}}}{x_{1}^{3/2}} 
\nonumber\\
 &+& \sqrt{x_1}\,\left[ \left( -0.621 + 0.265 \,i  \right)  + 0.282\left(1 + i  \right) \, \ln{x_{1}} \right] + {\mathcal O}({x_{1}}^{3/2}).
\label{mix2a}
\end{eqnarray}
The logarithms arising in Eqs. (\ref{mix1a}) and (\ref{mix2a}) explain why, in Eq. (\ref{ST10}), the transfer 
function of the spectral energy density contains logarithms.
In spite of the fact that semi-analytical estimates can pin down the slope of the transfer functions in different intervals, they are insufficient for a faithful account of more realistic situations where the slow-roll corrections are relevant and when 
other dissipative effects (such as neutrino fee streaming) are taken into account.

\subsection{Exponential damping of the mixing coefficients}  
\label{sec2d}
In a model-independent perspective, it can be argued that the relic gravitons are also characterized by a maximal 
frequency which is related to the modes which are maximally amplified. 
Let us consider, for instance, the case 
of the $\Lambda$CDM paradigm where the inflationary phase is almost suddenly followed by the radiation-dominated 
phase. By denoting the transition time as $\tau_{\mathrm{i}}$, it is plausible to think that all the modes of the field such that 
$k > a_{\mathrm{i}} H_{\mathrm{i}} \simeq \tau_{\mathrm{i}}^{-1}$ are exponentially suppressed \cite{bir,gar}.  For the modes $k\tau_{\mathrm{i}} > 1$, 
the pumping action of the gravitational field is practically 
absent.  The wavenumber $k_{\mathrm{max}}$ 
(which is related to the maximal frequency introduced in Eq. (\ref{EQ5})) 
is the maximally amplified wavenumber which can be determined 
by requiring $k \simeq \tau_{\mathrm{i}}^{-1}$:
\begin{equation}
k_{\mathrm{max}} = 3.5661\times 10^{22} \,\biggl(\frac{\epsilon}{0.01}\biggr)^{1/4} 
\biggl(\frac{{\mathcal A}_{\mathcal R}}{2.41 \times 10^{-9}}\biggr)^{1/4}
\biggl(\frac{h_{0}^2 \Omega_{\mathrm{R}0}}{4.15 \times 10^{-5}}\biggr)^{1/4} \,\, \mathrm{Mpc}^{-1},
\label{W1}
\end{equation}
where the typical values of the slow-roll parameter have been derived by taking into account 
that, in the absence of running of the tensor spectral index, $r_{\mathrm{T}} = 16 \epsilon$; since, 
according to the WMAP 5-yr data alone, $r_{\mathrm{T}} < 0.43$,  $\epsilon \leq 0.01$.
Note that $\nu_{\mathrm{max}}= 2\pi k_{\mathrm{max}} =117.45 \times (\pi \epsilon {\mathcal A}_{\mathcal R})^{1/4}\, \mathrm{GHz}$ 
for the same typical values of the $\Lambda$CDM parameters.
\begin{figure}[!ht]
\centering
\includegraphics[height=6.2cm]{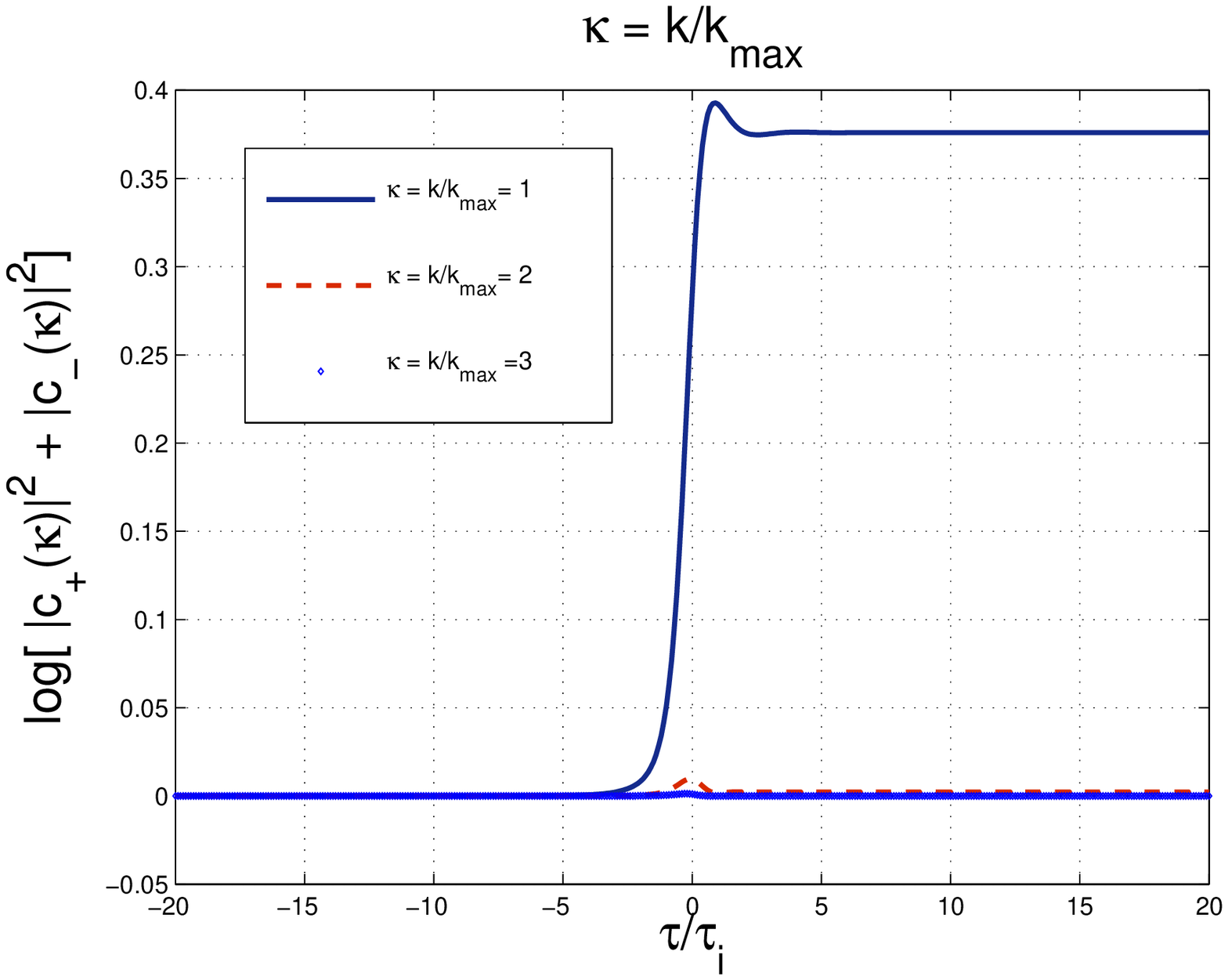}
\includegraphics[height=6.2cm]{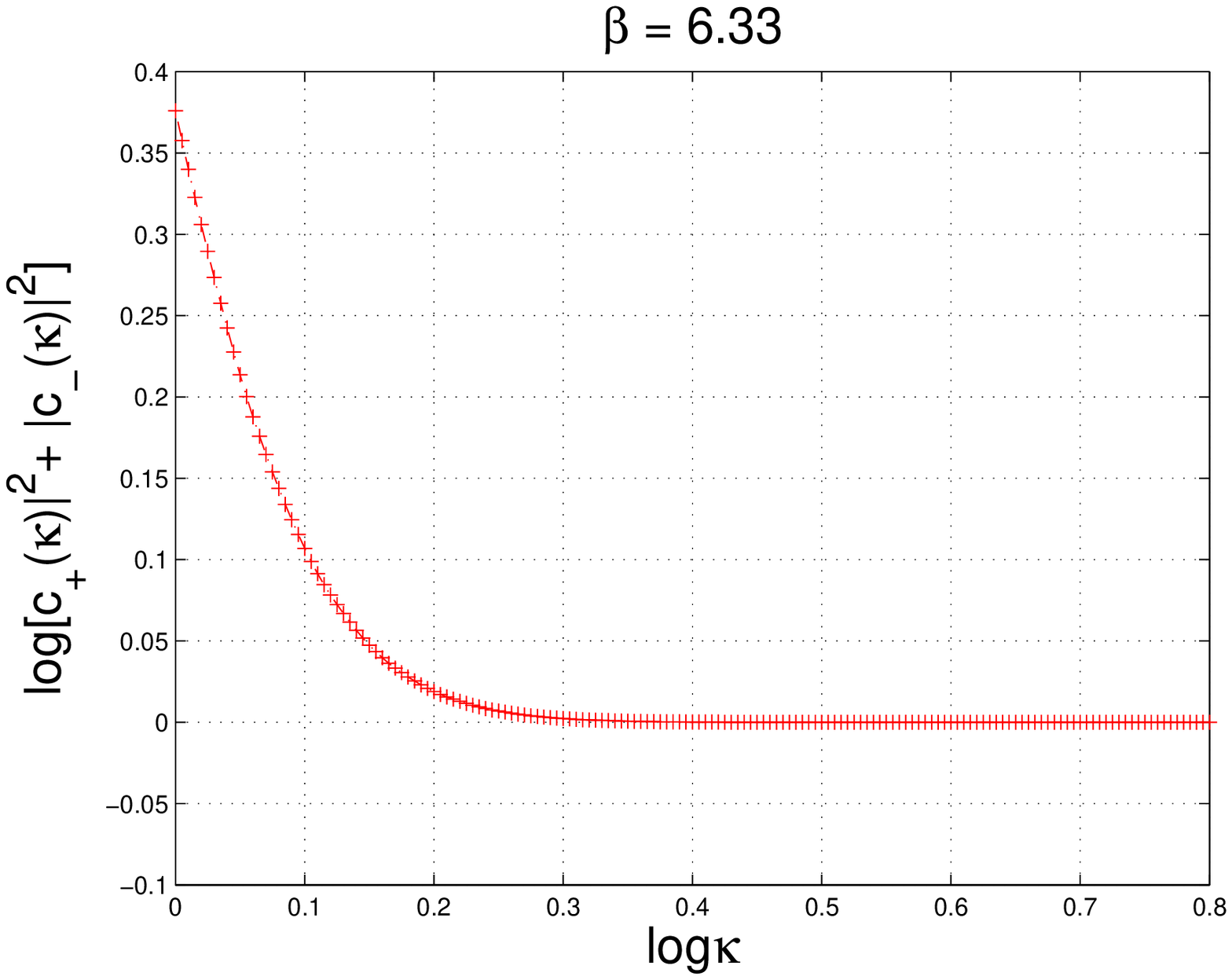}
\caption[a]{The time evolution of the mixing coefficients is reported at the left(on the horizontal axis the scale is linear). 
The exponential decay of the mixing coefficients is illustrated in the plot at the right. }
\label{Figure5}      
\end{figure}
For phenomenological purposes it can be also interesting to know what kind of exponential suppression 
we can expect. From the analysis of various transitions it emerges that the mixing coefficients for $k > k_{\mathrm{max}}$ (or $\nu > \nu_{\mathrm{max}}$) will satisfy 
\begin{equation}
|c_{+}(k)|^2 - |c_{-}|^2 =1 , \qquad |c_{+}(k)|^2 + |c_{-}(k)|^2 =  e^{- 2 \beta \frac{k}{k_{\mathrm{max}}}} + 1.
\label{W5}
\end{equation}
From Eq. (\ref{W5}) we can easily argue that, for $k > k_{\mathrm{max}}$, $|c_{+}(k)| \to 1$ and 
$|c_{-}(k)| \simeq 2^{-1/2} \exp{[- \beta k/k_{\mathrm{max}}]}$.  The point is then to estimate 
the value of $\beta$ which depends on the nature of the transition regime. Typically, however, $\beta > 2$ for sufficiently 
smooth transitions. To justify this statement it is interesting 
to consider the following toy model where the scale factor interpolates between a quasi-de Sitter phase and a radiation-dominated phase:
\begin{equation}
a(\tau) = a_{\mathrm{i}} [ \tau +  \sqrt{\tau^2 + \tau_{\mathrm{i}}^2}].
\label{W6}
\end{equation}
For  $\tau \to - \infty$ (i.e. $\tau \ll - \tau_{\mathrm{i}}$) , $a(\tau) \simeq - a_{\mathrm{i}}/\tau$ and the 
quasi de-Sitter dynamics is recovered. In the opposite 
limit (i. e. $\tau \gg + \tau_{\mathrm{i}}$), $a(\tau) \simeq a_{\mathrm{i}}\, \tau$ and the radiation dominance 
is recovered. 
In Fig. \ref{Figure5} (plot at the left) the exponential damping of the mixing coefficients is numerically illustrated. The 
curve at the top (full line) illustrates the case $\kappa = 1$. The cases $\kappa= 2$ and $\kappa = 3$ are barely 
distinguishable at the bottom of the plot. Notice, always in the right plot, the rather narrow 
range of times which are reported in a linear scale. In the plot at the right the asymptotic values of the mixing 
coefficients are reported for different values of $\kappa = k/k_{\mathrm{max}}$. By fitting the numerical data with 
with an equation of the form given in Eq. (\ref{W5}), the value of $\beta= 6.33$. 
Different examples can be presented on the same line of the one discussed in Fig. \ref{Figure5}. 
While it is clear, from the numerical data, that the decay is indeed exponential, the value of $\beta$ may well vary for different
models of the transition. The latter observation is effectively equivalent to a rescaling of
$k_{\mathrm{max}}$ for different models of inflation-radiation transition. By positing, for instance 
that $k_{\mathrm{max}} \to  \tilde{k}_{\mathrm{max}}/\beta$ we will have a new $\tilde{k}_{\mathrm{max}}$ which 
differs slightly from $k_{\mathrm{max}}$. It is clear that this indetermination on the maximal frequency 
of the relic graviton spectrum can only be solved by endorsing a given model (i.e. by theoretical prejudice) 
or by having direct measurements at those frequencies (which seems to be unlikely in the near future).  
\begin{figure}[!ht]
\centering
\includegraphics[height=6.2cm]{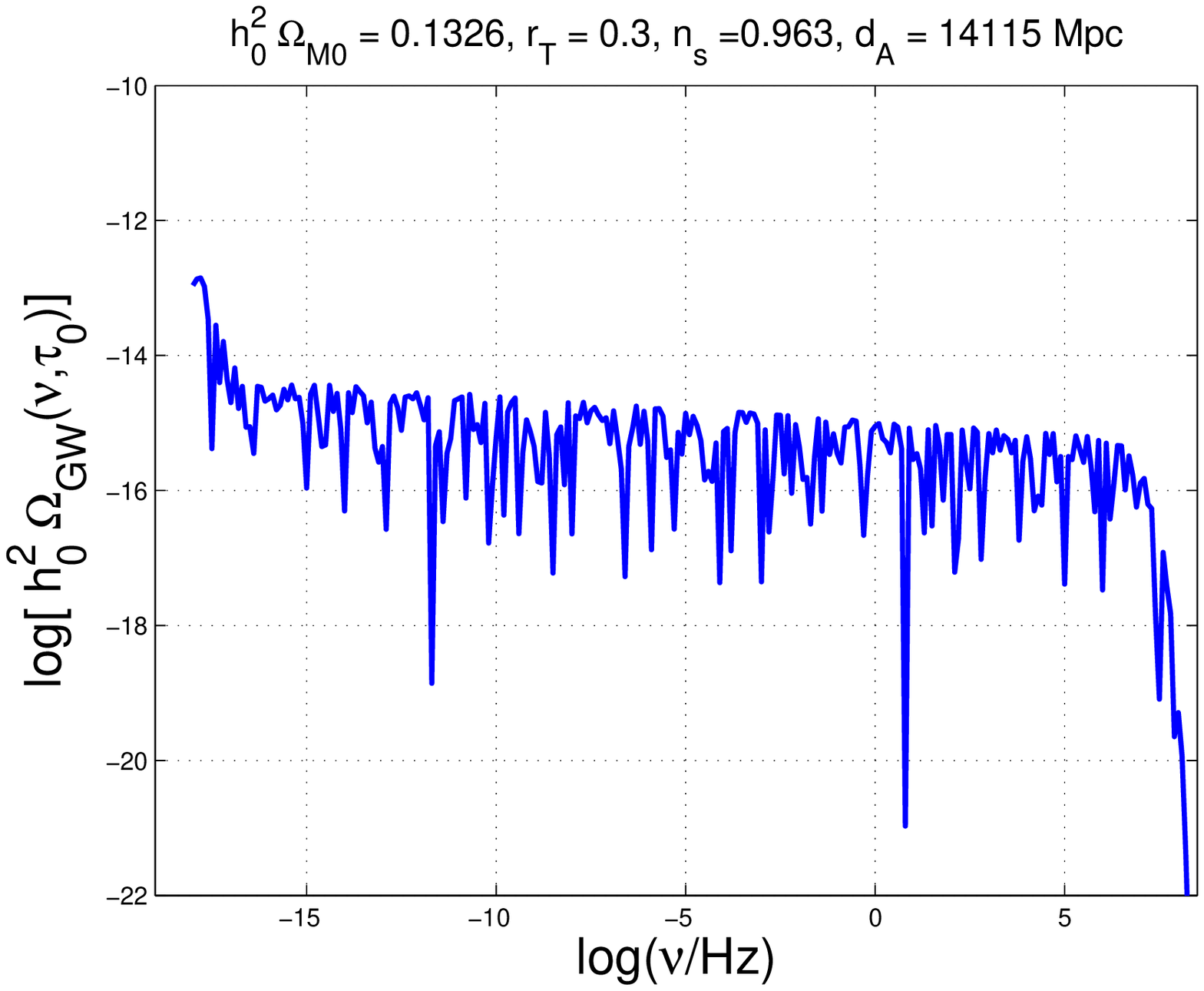}
\includegraphics[height=6.2cm]{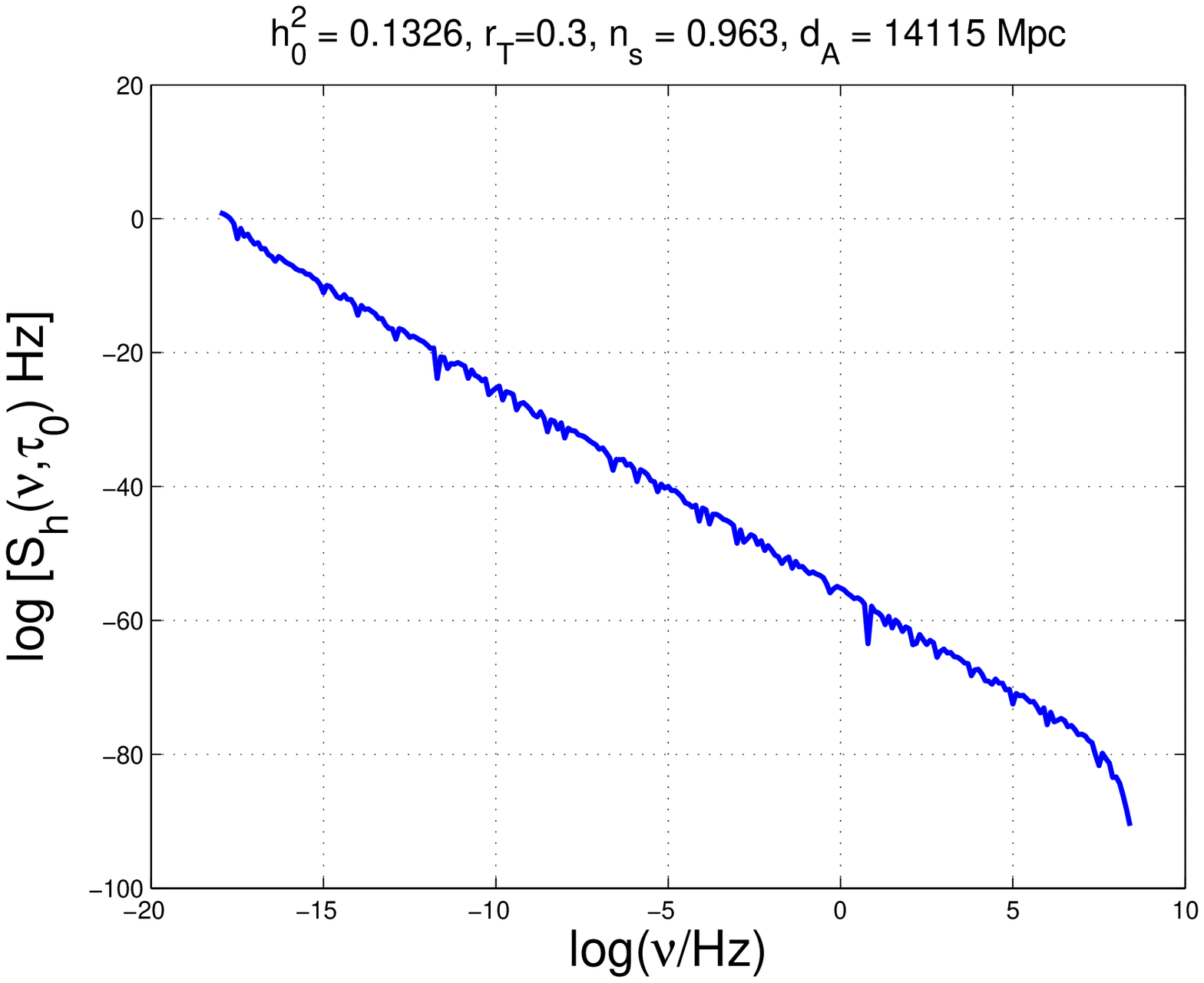}
\caption[a]{The spectral energy density of relic gravitons in critical units (plot at the left). The strain amplitude is instead reported in the plot at the right. Note that 
while $h_{0}^2 \Omega_{\mathrm{GW}}(\nu,\tau_{0})$ is dimensionless, $S_{h}(\nu,\tau_{0})$ has dimensions of $\mathrm{Hz}^{-1}$. The fiducial 
set of parameters used corresponds to the best fit to the WMAP 5-yr data. }
\label{Figure6}      
\end{figure}
\renewcommand{\theequation}{3.\arabic{equation}}
\setcounter{equation}{0}
\section{Nearly scale-invariant spectra}
\label{sec3}
The transfer function of the spectral energy density has been numerically computed in the previous section and the numerical results have been corroborated by appropriate semi-analytical estimates. We are then ready for an explicit 
calculation of the spectral energy density  in the $\Lambda$CDM 
scenario.  In  subsection \ref{sec31} the spectral energy density will be computed in terms of the amplitude transfer function and also directly in terms of the transfer function for the spectral energy density. Explicit calculations will show that the latter method is more accurate. Subsection \ref{sec32} is devoted to various late time effects 
(e.g. neutrino anisotropic stress, late dominance of dark-energy, progressive diminishment of the number of relativistic species)  which are certainly present and which affect the amplitude of the spectral energy density.  Finally, in subsection 
\ref{sec33} current (and foreseen) sensitivities of wide-band 
interferometers will be briefly compared to the (unfortunately minute) $\Lambda$CDM signal.
\subsection{Spectral energy density in the $\Lambda$CDM paradigm}
\label{sec31} 
In Fig. \ref{Figure6} $h_{0}^2 \Omega_{\mathrm{GW}}(\nu,\tau_{0})$ and the strain amplitude 
$S_{h}(\nu,\tau_{0})$ are computed using the transfer function for the amplitude discussed (and rederived) in Eqs. (\ref{TA8}), (\ref{TA9}) and (\ref{TA10}). 
The strain amplitude appearing in the plot at the right of Fig. \ref{Figure6}
is related to the spectral energy density as in Eq. (\ref{T11}) 
which also implies that $\Omega_{\mathrm{GW}}(\nu,\tau_{0})$ 
 can be expressed in terms of $S_{h}(\nu,\tau_{0})$. Indeed, according to Eq. (\ref{T11}), 
${\mathcal P}_{\mathrm{T}}(k,\tau) = 4 \nu S_{h}(\nu,\tau)$ and the 
spectral energy density becomes:
\begin{equation}
\Omega_{\mathrm{GW}}(\nu,\tau) = \frac{4 \pi^2}{3 {\mathcal H}^2} \nu^3  S_{h}(\nu,\tau),
\label{DEF2}
\end{equation}
where, in natural units, $k = 2 \pi \nu$.  By making more explicit the numerical factors and by inverting Eq. (\ref{DEF2}) in terms of  $S_{h}(\nu,\tau_{0})$ we obtain:
\begin{equation}
S_{h}(\nu,\tau_{0}) = 7.981\times 10^{-43} \,\,\biggl(\frac{100\,\mathrm{Hz}}{\nu}\biggr)^3 \,\, h_{0}^2 \Omega_{\mathrm{GW}}(\nu,\tau_{0})\,\, \mathrm{Hz}^{-1},
\label{DEF3}
\end{equation}
where $H_{0} = 3.24078\times 10^{-18}\,\,h_{0}\,\,\mathrm{Hz}$.
The oscillations of Fig. \ref{Figure6}  are related to the way the transfer 
function for the amplitude is derived. There are complementary 
forms of the strategy leading to the results of Fig. \ref{Figure6} (see, for instance, \cite{page}). The plots of Fig. \ref{Figure6} have been obtained by using directly Eq. (\ref{TA8})--(\ref{TA10}) inside Eq. (\ref{DEF1}).  This is the procedure 
used, originally, in \cite{tur1} (see also \cite{EF1}). One could also define the transfer function as in Eq. (\ref{TA9}) and then, at the level of the 
spectral energy density, compute $g_{k}(\tau) = f_{k}'(\tau)$ (see Appendix E of \cite{page}).  By working with the transfer function of the 
tensor amplitude the spectral energy density for frequencies $\nu \gg \nu_{\mathrm{eq}}$ is given by:
\begin{eqnarray}
h_{0}^2 \Omega_{\mathrm{GW}}(\nu,\tau_{0}) &=& {\mathcal N}_{h} \,\, r_{\mathrm{T}} \,\, \biggl(\frac{\nu}{\nu_{\mathrm{p}}}\biggr)^{n_{\mathrm{T}}} e^{- 2\beta \frac{\nu}{\nu_{\mathrm{max}}}},
\label{scaleinv1}\\
{\mathcal N}_{h} &=& 7.992 \times 10^{-15} \biggl(\frac{h_{0}^2 \Omega_{\mathrm{M}0}}{0.1326}\biggr)^{-2} 
\biggl(\frac{h_{0}^2 \Omega_{\mathrm{R}0}}{4.15\times 10^{-5}}\biggr) \biggl(\frac{d_{\mathrm{A}}}{1.4115 \times 10^{4}\, \mathrm{Mpc}}\biggr)^{-4},
\label{scaleinv2}
\end{eqnarray}
where $d_{\mathrm{A}}(z_{*})$ is the (comoving) angular diameter distance to decoupling. The dependence upon $d_{A}(z_{*})$ arises 
because we have to estimate $\tau_{0}$. In Eqs. (\ref{scaleinv1})--(\ref{scaleinv2}) (as well as in the program used for the 
numerical calculations) there are two complementary options: the first one is to give the angular diameter distance 
to decoupling (which is directly inferred from the CMB data). In the case of the 5-yr WMAP data alone, 
$d_{A}(z_{*}) = 14115 \,\,\mathrm{Mpc}_{-191}^{+188}$: this  approach 
 has been followed, for instance, in \cite{EF1}. In a complementary 
 perspective it is
also possible to take the best fit value of the total matter fraction (i.e. $\Omega_{\mathrm{M}0}= 0.258$ for the case of the WMAP
5-yr data alone) and compute the comoving angular diameter distance according to the well know expression for spatially flat Universes: 
\begin{equation}
d_{\mathrm{A}}(z_{*}) =  \frac{1}{H_{0}} \int_{0}^{z_{*}} \frac{d z}{\sqrt{\Omega_{\mathrm{M}0} ( 1 + z)^3 + \Omega_{\Lambda} + \Omega_{\mathrm{R}0} (1 + z)^4}} = \frac{3.375}{H_{0}}= 14072\,\, \mathrm{Mpc},\qquad z_{*} = 1090.
\label{scaleinv3}
\end{equation}
The latter strategy has been used, for instance, in \cite{page}. The two strategies are
 compatible and, moreover, this 
explains why, in Eq. (\ref{scaleinv2}) the dependence upon $\Omega_{\mathrm{M}0}$ does not cancel. 
In Eq. (\ref{scaleinv1}) $n_{\mathrm{T}}$ denotes, as usual, the tensor spectral index which can be also written as 
\begin{equation}
n_{\mathrm{T}} =  - 2 \epsilon + \frac{\alpha_{\mathrm{T}}}{2} \ln{(k/k_{\mathrm{p}})}, \qquad \alpha_{\mathrm{T}} = \frac{r_{\mathrm{T}}}{8}\biggl[(n_{\mathrm{s}} -1) + 
\frac{r_{\mathrm{T}}}{8}\biggr],
\label{int3}
\end{equation}
If $\alpha_{\mathrm{T}} =0$ Eq. (\ref{TA4}) is recovered and the spectral index is independent on the frequency. In the case when $\alpha_{\mathrm{T}}\neq 0$ 
and it is given by Eq. (\ref{int3}) the spectral index does depend upon the frequency: in the jargon this is often dubbed by saying the the spectral 
index runs.
  The frequency-dependent correction (i.e. $\alpha_{\mathrm{T}}$) contains the scalar spectral index $n_{\mathrm{s}}$ 
and this is why the value of $n_{\mathrm{s}}$ is mentioned in the parameters of Fig. \ref{Figure6}.  
The last remark concerning the result of Eq. (\ref{scaleinv2}) is that, in the limit, 
 $\nu \gg \nu_{\mathrm{eq}}$, the oscillating terms have been appropriately averaged: this is
 done by setting the terms going as $\cos^2{(2\pi \nu\tau_{0})}$ to $1/2$. The latter procedure has been employed, for instance, in the analyses of \cite{EF1,wnu3}.
This procedure is justified in semi-analytical terms but rather odd in a fully numerical 
context. In what follows it will be argued that there is no need of this type 
of tricks if the spectral energy density is computed directly without passing through 
the transfer function of the tensor amplitude.

\begin{figure}[!ht]
\centering
\includegraphics[height=6.2cm]{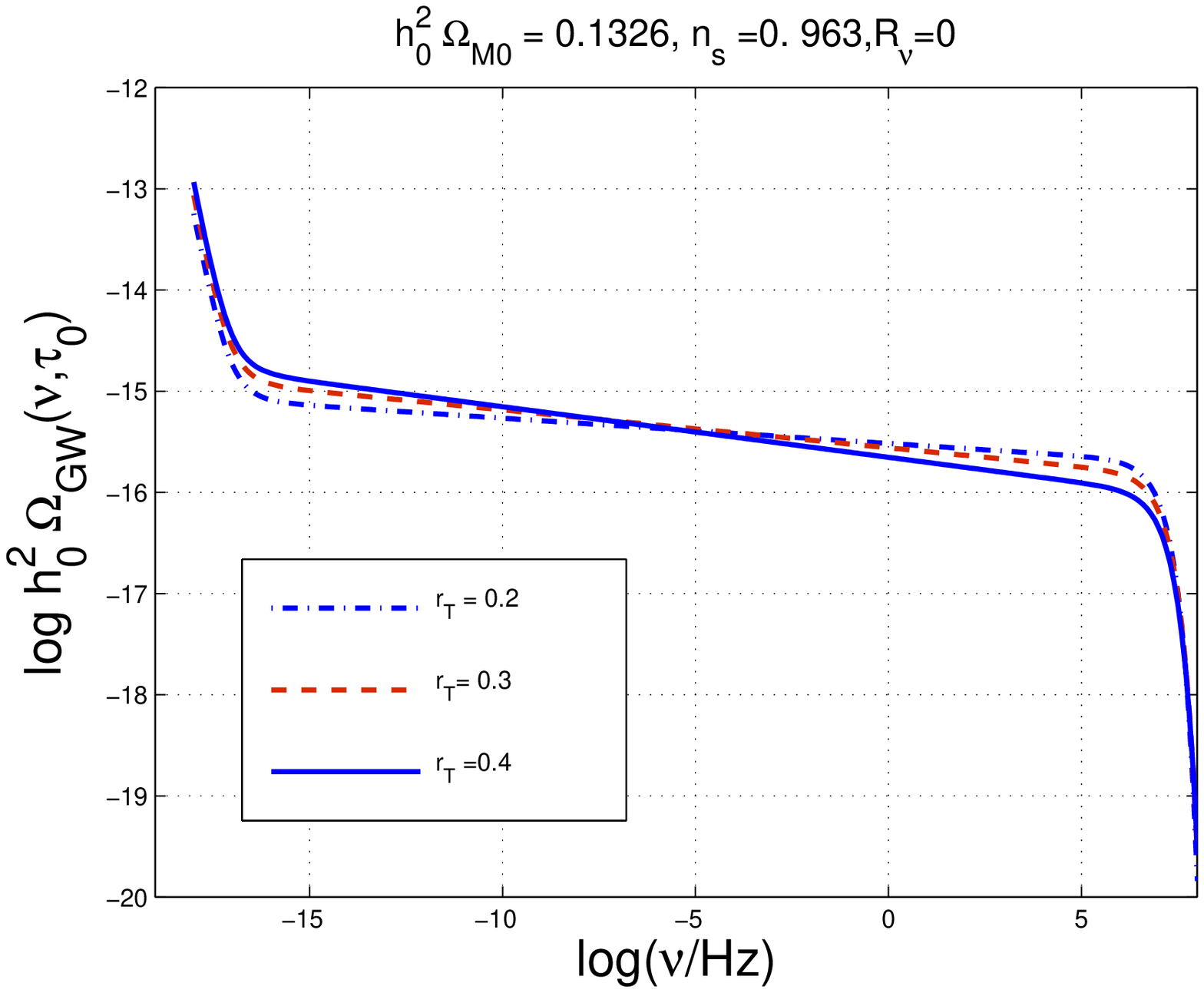}
\includegraphics[height=6.2cm]{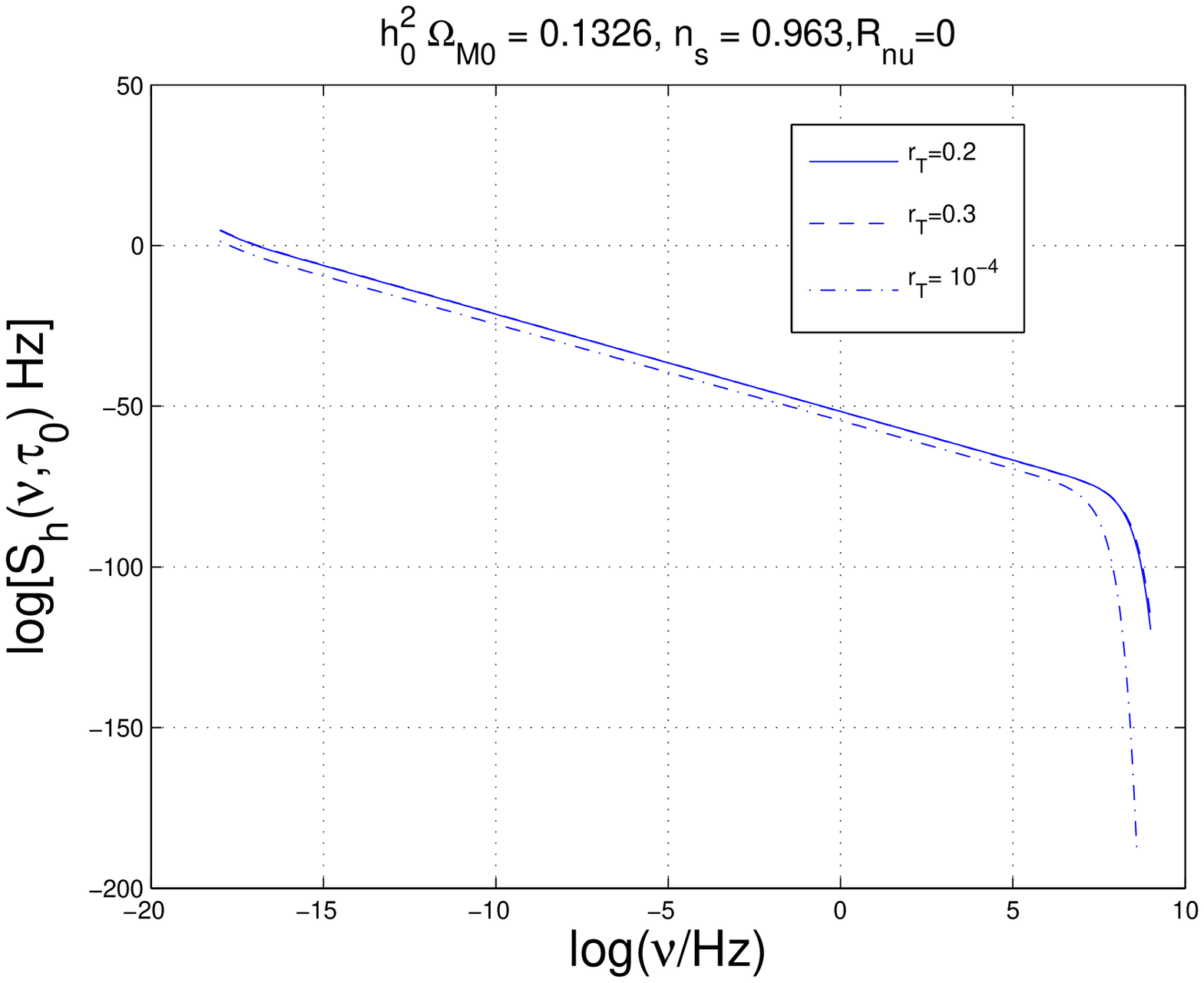}
\caption[a]{The spectral energy density of the relic gravitons (plot at the left) and the related $S_{h}(\nu,\tau_{0})$ (plot at the left) 
for different values of $r_{\mathrm{T}}$ and for the same set of fiducial parameters 
illustrated in Fig. \ref{Figure6}.}
\label{Figure7}      
\end{figure}

Along this perspective, the results of Fig. \ref{Figure6} should then be compared with Fig. \ref{Figure7} where the transfer function for the spectral energy density has been 
consistently employed. 
In Fig. \ref{Figure7} the spectral energy density of the relic gravitons as well as 
$S_{h}(\nu,\tau_{0})$ are reported for different values 
of $r_{\mathrm{T}}$ and for the same fiducial set of parameters used in Fig. \ref{Figure6}. The first salient feature emerging from the comparison 
of Figs. \ref{Figure6} and \ref{Figure7} is that the oscillatory behaviour disappear. 
The spectra of Fig. \ref{Figure7} have been obtained from the direct integration 
of the mode functions but can be parametrized, according to Eq. (\ref{ENTRANS}) as 
\begin{eqnarray}
h_{0}^2 \Omega_{\mathrm{GW}}(\nu,\tau_{0}) &=& {\mathcal N}_{\rho}  T^2_{\rho}(\nu/\nu_{\mathrm{eq}}) r_{\mathrm{T}} \biggl(\frac{\nu}{\nu_{\mathrm{p}}}\biggr)^{n_{\mathrm{T}}} e^{- 2 \beta \frac{\nu}{\nu_{\mathrm{max}}}}
\label{scaleinv4}\\
{\mathcal N}_{\rho} &=& 4.165 \times 10^{-15} \biggl(\frac{h_{0}^2 \Omega_{\mathrm{R}0}}{4.15\times 10^{-5}}\biggr).
\label{scaleinv5}
\end{eqnarray}
By comparing Eqs. (\ref{scaleinv1})--(\ref{scaleinv2}) to Eqs. (\ref{scaleinv4})--(\ref{scaleinv5}), the amplitude for $\nu\gg \nu_{\mathrm{eq}}$ differs 
by a factor which is roughly a factor $2$. This occurrence is not surprising since  Eqs. (\ref{scaleinv1})--(\ref{scaleinv2}) have been obtained by averaging over the 
oscillations (i.e. by replacing cosine squared with $1/2$) and by imposing that $|g_{k}| = k |f_{k}|$. In Fig. \ref{Figure7} the impact of the variation of $n_{\mathrm{T}}$ is also illustrated. Recalling that the WMAP 5-yr data alone sugggest $r_{\mathrm{T}}< 0.4$, 
the variation of the spectral energy density is more pronounced than the change of the strain power spectrum. This is because of the steepness of $S_{h}(\nu,\tau_{0})$ in frequency. 
\subsection{Anisotropic stress and dark-energy contribution}
\label{sec32}
The considerations of the previous subsection suggest that the results 
obtainable with the transfer function of the spectral energy density 
seem to be intrinsically more accurate. The obvious question is of course 
if we need this precision. There are two answers to this kind of questions. 
The first one is that, of course, the accuracy in the estimate of the 
$\Lambda$CDM plateau is necessary for comparing the theoretical 
predictions with the data. Therefore it would be strange to treat very accurately 
the tensor contribution to CMB anisotropies but not to wide-band detectors.
The second issue is more theoretical. In the recent past the community 
investigated various late time effects which can modify the $\Lambda$CDM plateau 
for $\nu \gg \nu_{\mathrm{eq}}$. All these effects compete with the accuracy 
which is inherent in the estimate of the transfer function. This will be the subject 
of the present subsection. 

Let us therefore start by noticing that, so far, the evolution of the tensor modes has been treated as if the anisotropic stress of the fluid was absent. 
After neutrino decoupling, the neutrinos free stream and the effective energy-momentum tensor acquires, to first-order in the amplitude of the plasma fluctuations, an anisotropic stress, i.e. 
\begin{equation}
\delta T_{i}^{j} = - \delta p  \delta_{i}^{j} + \Pi_{i}^{j},\qquad \partial_{i} \Pi_{j}^{i}= \Pi_{i}^{i} =0.
\label{ANIS1}
\end{equation}
The presence of the anisotropic stress clearly affects the evolution  the tensor 
modes whose evolution is then dictated by 
\begin{equation}
{h_{i}^{j}}'' + 2 {\mathcal H} {h_{i}^{j}}' - \nabla^2 h_{i}^{j} = - 16 \pi G a^2 \Pi_{i}^{j}.
\label{ANIS2}
\end{equation}
Equation (\ref{ANIS2}) reduces to an integro-differential equation which has been analyzed in \cite{wnu1} (see also \cite{wnu2,wnu3,wnu4}). 
The overall effect of collisionless particles is a reduction 
of the spectral energy density of the relic gravitons. Assuming that the only collisionless 
species in the thermal history of the Universe are the neutrinos, the amount 
of suppression can be parametrized by the function
\begin{equation}
{\mathcal F}(R_{\nu}) = 1 -0.539 R_{\nu} + 0.134 R_{\nu}^2 
\label{ANIS3}
\end{equation}
where $R_{\nu}$ is the fraction of neutrinos in the radiation plasma, i.e. 
\begin{equation}
R_{\nu} = \frac{r}{r + 1}, \qquad r = 0.681 \biggl(\frac{N_{\nu}}{3}\biggr),\qquad R_{\gamma} + R_{\nu} = 1. 
\label{ANIS4}
\end{equation}
In Eq. (\ref{ANIS4}) $N_{\nu}$  represents the number of massless neutrino families. In the standard $\Lambda$CDM scenario the 
neutrinos are taken to be massless. 
\begin{figure}[!ht]
\centering
\includegraphics[height=6.2cm]{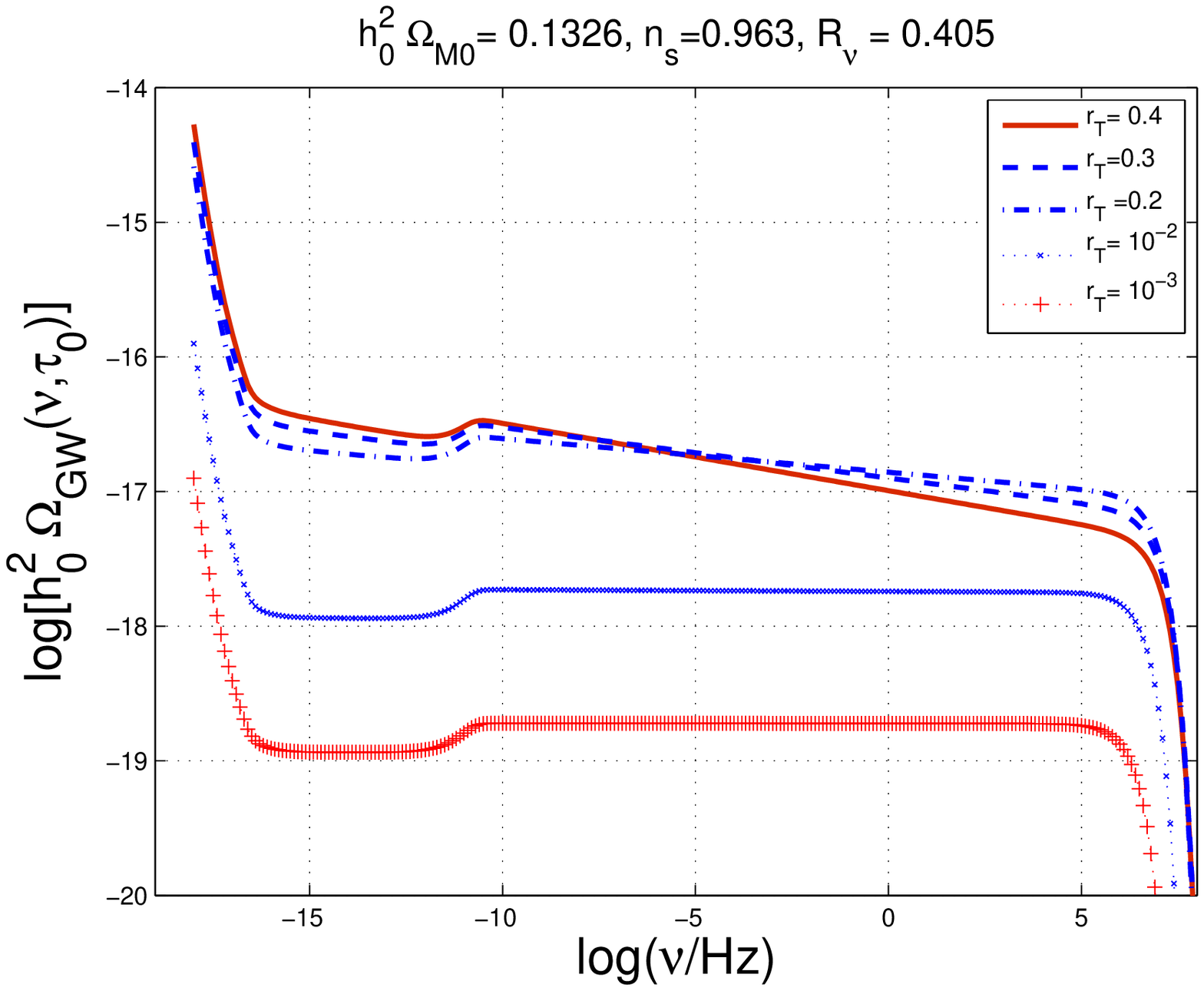}
\includegraphics[height=6.2cm]{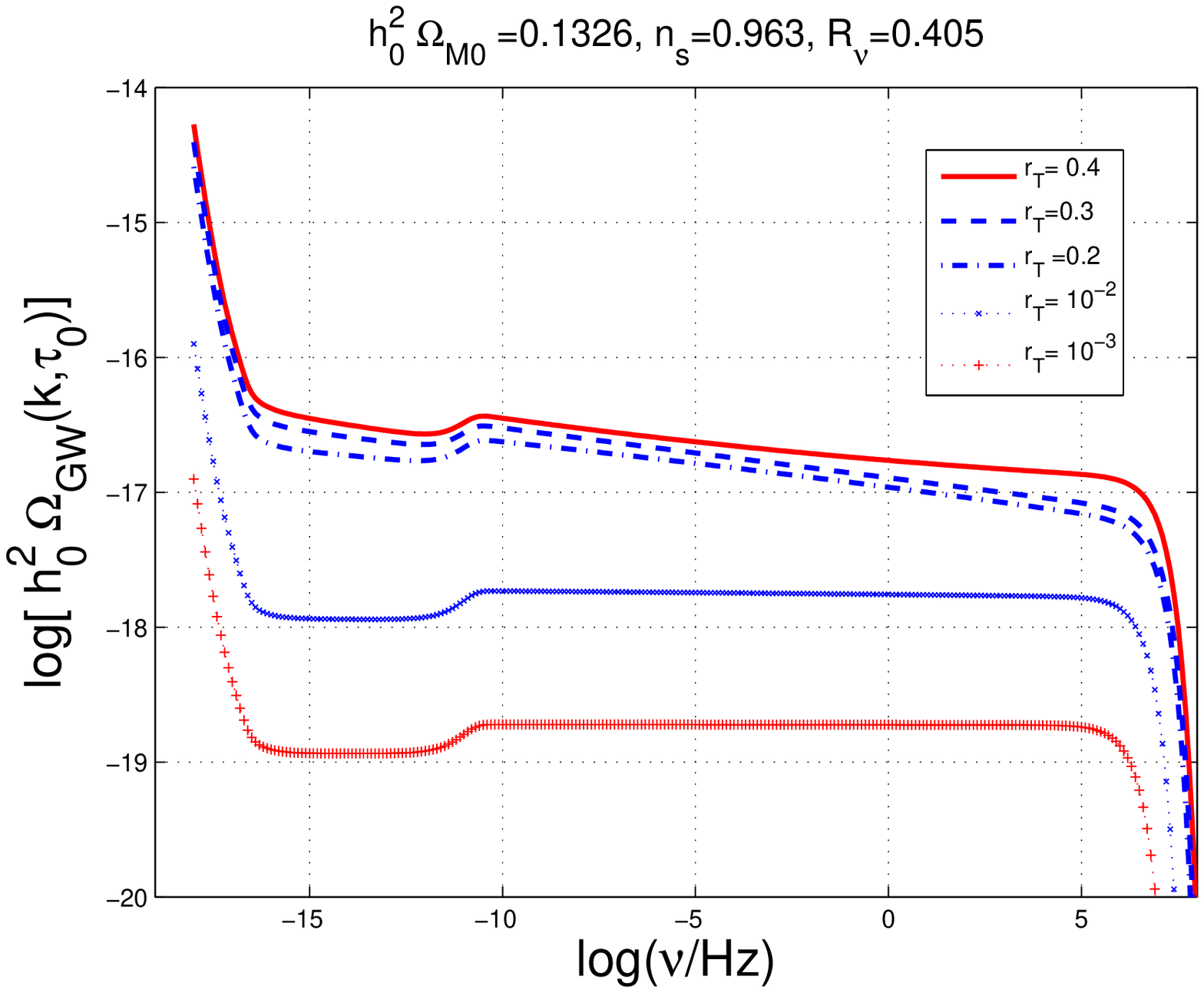}
\caption[a]{In both plots the contribution of the neutrino free streaming and of the 
variation in the number of degrees of freedom has been taken into account. In the plot at the right 
the spectral index has been allowed to depend upon the frequency.}
\label{Figure8}      
\end{figure}
In the case $R_{\nu}=0$ (i.e. in the absence of collisionless patrticles) there is no suppression. If, on the contrary, 
$R_{\nu} \neq 0$ the suppression can even reach one order of magnitude. In the case $N_{\nu} = 3$, 
$R_{\nu} = 0.405$ and the suppression of the spectral energy density is proportional 
to ${\mathcal F}^2(0.405)= 0. 645$. This suppression will be effective for relatively 
small frequencies which are larger than $\nu_{\mathrm{eq}}$ and smaller than the
frequency corresponding to the Hubble radius at the time of big-bang nucleosynthesis, i.e.  $\nu_{\mathrm{bbn}}$ of Eq. (\ref{EQ4}).

The effect of neutrino free streaming has been included in Fig. \ref{Figure8} together with the damping effect associated with the (present) dominance of the dark energy component. The redshift of $\Lambda$-dominance is defined as 
\begin{equation}
1 + z_{\Lambda} = \biggl(\frac{a_{0}}{a_{\Lambda}}\biggr) = 
 \biggl(\frac{\Omega_{\Lambda}}{\Omega_{\mathrm{M}0}}\biggr)^{1/3}.
\label{LAM1}
\end{equation}
Consider now the mode which will be denoted as $k_{\Lambda}$, i.e. the mode 
reentering the Hubble radius at $\tau_{\Lambda}$. By definition $k_{\Lambda} = H_{\Lambda} a_{\Lambda}$ must hold. But for $\tau > \tau_{\Lambda}$ 
is constant, i.e. $H_{\Lambda} \equiv H_{0}$ where $H_{0}$ is the present value of the Hubble rate.
Using now Eq. (\ref{LAM1}), it can be easily shown that $k_{\Lambda} = (\Omega_{M0}/\Omega_{\Lambda})^{1/3}k_{\mathrm{H}}$ 
where $k_{\mathrm{H}} = a_{0} H_{0}$. The frequency interval between $\nu_{\mathrm{H}}$ and $\nu_{\Lambda}$ is rather  tiny. 
Indeed, it turns out that $\nu_{\Lambda} = k_{\Lambda}/(2\pi)$ is given by 
\begin{equation}
 \nu_{\Lambda} = 2.607 \times 10^{-19}  \biggl(\frac{h_{0}}{0.719}\biggr) \biggl(\frac{\Omega_{\mathrm{M}0}}{0.258}\biggr)^{1/3} \biggl(\frac{\Omega_{\Lambda}}{0.742}\biggr)^{1/3} \,\, \mathrm{Hz}.
\label{LAM3}
\end{equation}
For the same choice of parameters of Eq. (\ref{LAM3}), $\nu_{\mathrm{H}} = H_{0}/(2\pi) = 3.708 \times 10^{-19}$ Hz 
which is not so different than $\nu_{\Lambda} =  2.607 \times 10^{-19}$ Hz.  
The adiabatic damping of the mode function across $\tau_{\Lambda}$ reduces the amplitude of the spectral energy density by a factor 
$(\Omega_{\mathrm{M}0}/\Omega_{\Lambda})^2$. For the typical choice of parameters of Eq. (\ref{LAM3}) we have that 
the suppression is of the order of $0.12$. This class of effects has been repeatedly in a number of recent papers 
\cite{zh1,zh2}.  The essence of the effect is captured by the following observation. Consider a mode $k$ which reenters before  $\tau_{\Lambda}$.
The present value of the amplitude $F_{k}(\tau) = f_{k}(\tau)/a(\tau)$ will be adiabatically suppressed since, as repeatedly stressed, in this 
regime $f_{k}(\tau)$ will simply be plane waves. Consequently, defining as $\tilde{F}_{k_{*}}$ the amplitude   at $k_{*} = H_{*} a_{*}$ when the given 
mode crosses the Hubble radius, we will also have that
 \begin{equation}
F_{k}(\tau_{0}) = \biggl(\frac{a_{k_{*}}}{a_{\Lambda}}\biggr)_{\mathrm{mat}} \biggl(\frac{a_{\Lambda}}{a_{0}}\biggr)_{\Lambda} \tilde{F}_{k_{*}} \equiv 
\biggl(\frac{k}{k_{\mathrm{H}}}\biggr)^{-2} \biggl(\frac{\Omega_{\mathrm{M}0}}{\Omega_{\Lambda}}\biggr)  \tilde{F}_{k_{*}},
\label{LAM4}
\end{equation}
where the subscripts (in the first equality) denote the time range over which the corresponding redshift is computed, i.e. either matter-dominated 
or $\Lambda$-dominated stages. The second equality follows from the first one by appreciating that $a(k_{*}) \simeq \tau_{*}^{2} \simeq k^{-2}$ and by using 
Eq. (\ref{LAM1}). Equation (\ref{LAM4}) implies, immediately, that the spectral energy density of relic gravitons is corrected in two different fashions. 
For $\nu < \nu_{\mathrm{H}}$ the frequency dependence will be different and will be proportional to 
$\Omega_{\mathrm{GW}}(\nu,\tau_{0}) \propto (\nu/\nu_{\mathrm{H}})^{n_{\mathrm{T}} -2} (\Omega_{\mathrm{M}0}/\Omega_{\Lambda})^{2}$.
Vice versa, in the range $\nu > \nu_{\mathrm{H}}$ the frequency dependence will be exactly the one already computed but, overall, the 
amplitude will be smaller by a factor $(\Omega_{\mathrm{M}0}/\Omega_{\Lambda})^2$. Two comments are in order. 
The modification of the frequency dependence is only effective between\footnote{We are here enforcing the 
usual terminology stemming from the powers of $10$: aHz (for atto Hz i.e. $10^{-18}$ Hz), fHz (for femto Hz, i.e.   $10^{-15}$ Hz) 
and so on.} $0.36$ aHz and $0.26$ aHz: this effect is therefore unimportant and customarily ignored (see, for instance, \cite{EF1,zh1}) 
for phenomenological purposes. On the other hand, 
the overall suppression going as $(\Omega_{\mathrm{M}0}/\Omega_{\Lambda})^{2}$ must be taken properly into account on the same footing of other sources of suppression of the spectral energy density.

There is, in principle, a third effect which may arise and it has to do with the variation of the effective number of relativistic species.  
The total energy density and the total entropy density of the plasma can be written as 
\begin{equation}
\rho_{\mathrm{t}} = g_{\rho}(T) \frac{\pi^2}{30} T^4,\qquad s_{\mathrm{t}} = g_{\mathrm{s}}(T) \frac{2 \pi^2}{45} T^3.
\label{EFF1}
\end{equation}
For temperatures much larger than the top quark mass, all the known species of the minimal standard model of particle interactions are in local thermal 
equilibrium, then $g_{\rho} = g_{\mathrm{s}} = 106.75$. Below, $T \simeq 175$ GeV the various species 
start decoupling, the notion of thermal equilibrium is replaced by the notion of kinetic equilibrium and the 
time evolution of the number of relativistic degrees of freedom effectively changes the evolution of the Hubble rate. 
In principle if a given mode $k$ reenters the Hubble radius at a temperature $T_{k}$ the spectral energy density 
of the relic gravitons is (kinematically) suppressed by a factor which can be written as (see, for instance, \cite{zh1})
 \begin{equation}
 \biggl(\frac{g_{\rho}(T_{k})}{g_{\rho0}}\biggr)\biggl(\frac{g_{\mathrm{s}}(T_{k})}{g_{\mathrm{s}0}}\biggr)^{-4/3}.
 \label{EFF2}
 \end{equation}
At the present time  $g_{\rho0}= 3.36$ and $g_{\mathrm{s}0}= 3.90$. In general terms the effect parametrized by 
Eq. (\ref{EFF2}) will cause a frequency-dependent suppression, i.e. a further modulation of the spectral 
energy density $\Omega_{\mathrm{GW}}(\nu,\tau_{0})$.  The maximal suppression one can expect 
can be obtained by inserting into Eq. (\ref{EFF2}) the highest possible number of degrees of freedom. 
So, in the case of the minimal standard model this would imply that the suppression (on $\Omega_{\mathrm{GW}}(\nu,\tau_{0})$)
 will be of the order of $0.38$. In popular supersymmetric extensions of the minimal standard models $g_{\rho}$ and $g_{s}$  can be as high as, approximately, $230$. This will bring down the figure given above to $0.29$.
 
All the effects estimated in the last 
part of the present section (i.e. free streaming, dark energy, evolution of relativistic degrees of freedom) have common features.
Both in the case of the neutrinos and in the case of the evolution of the relativistic degrees of freedom the potential impact of the effect
could be larger. For instance, suppose that, in the early Universe, the particle model has many more degrees of freedom and many more 
particles which can free stream, at some epoch.
 At the same time 
we can say that all the aforementioned effects {\em decrease} rather than {\em increasing} the spectral energy density.
Taken singularly, each of the effects will decrease $\Omega_{\mathrm{GW}}$ by less than  one 
order of magnitude. The net result of the combined effects will then be, roughly,  a suppression of $\Omega_{\mathrm{GW}}(\nu,\tau_{0})$
which is of the order of $3 \times 10^{-2}$ (for $10^{-16}\,\,\mathrm{Hz} < \nu <10^{-11} \, \mathrm{Hz}$) and of the order of $4\times 10^{-2}$ for $\nu > 10^{-11}$Hz.
The impact of the various damping effects is self-evident by looking at Fig. \ref{Figure8}.  The effects of the neutrinos is visible in the intermediate region where the spectrum exhibits a shallow depression. 
\subsection{Sensitivity of wide-band interferometers to $\Lambda$CDM signal}
\label{sec33}
\begin{figure}[!ht]
\centering
\includegraphics[height=6.2cm]{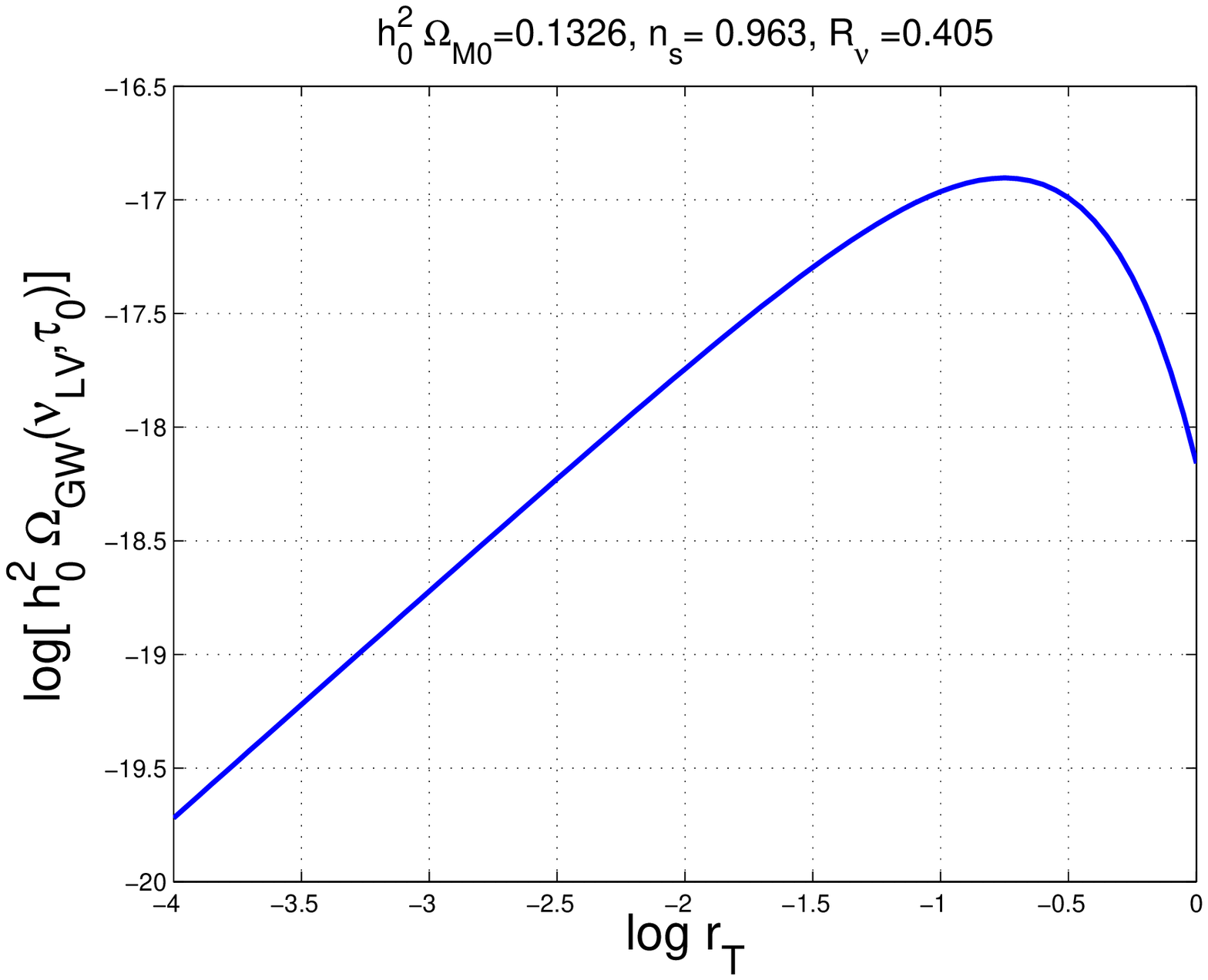}
\includegraphics[height=6.2cm]{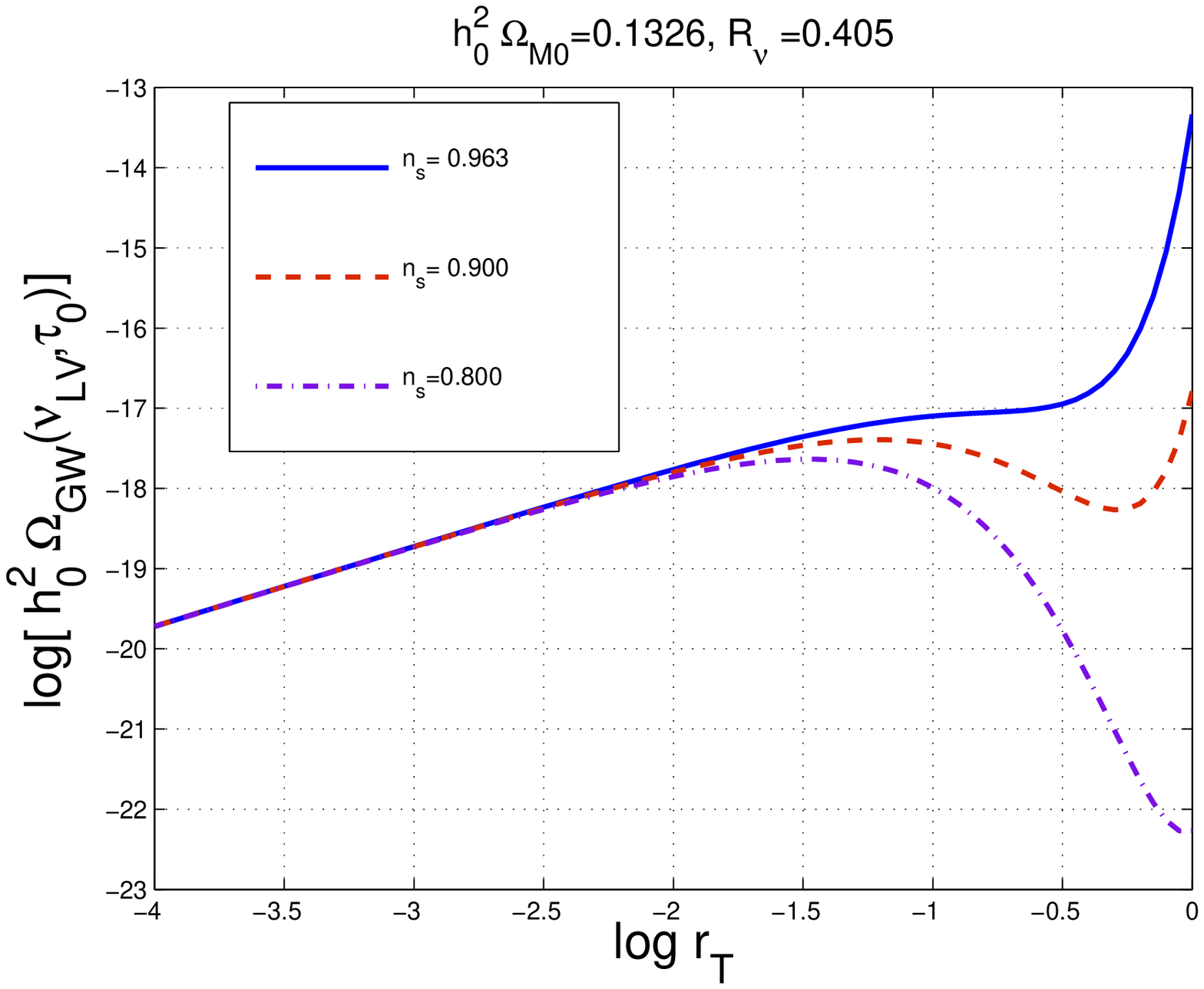}
\caption[a]{The spectral energy density of the relic gravitons in the context of the $\Lambda$CDM model evaluated at 
the Ligo/Virgo frequency as a function of the tensor-to-scalar ratio. In the plot at the left $\alpha_{\mathrm{T}} =0$ (i.e. the spectral index 
does not ``run" and it is independent upon the frequency. In the plot at the right $\alpha_{\mathrm{T}}$ is given as in Eq. (\ref{int3}) and the spectral index does 
depend upon the frequency. }
\label{Figure9}      
\end{figure}

It is now interesting to compare the ideas discussed in the present section with the sensitivity of wide-band interferometers. In Fig. \ref{Figure9}  the spectral density of relic gravitons is reported, at a specific frequency,
as a function of $r_{\mathrm{T}}$.
The specific frequency at which $\Omega_{\mathrm{GW}}(\nu,\tau_{0})$ is computed is given, as indicated by $\nu_{\mathrm{LV}}= 100$Hz.
The subscript LV is a shorthand notation for Ligo/Virgo.  If Fig. \ref{Figure9} (plot at the left) the tensor 
spectral index is frequency-independent (i.e. $\alpha_{\mathrm{T}} =0$ in Eq. (\ref{TA4})). In the same Fig. \ref{Figure9} (plot at the right) $n_{\mathrm{T}}$ is allowed to 
run and $\alpha_{\mathrm{T}}$ is given, in terms of the scalar spectral index 
$n_{\mathrm{T}}$ as in Eq. (\ref{int3}).

It is the moment of comparing the theoretical signal with the current sensitivity of wide-band interferometers. This figure can be assessed, for instance, from 
Ref. \cite{LIGOS2} (see also \cite{LIGOS1,LIGOS3}) where the current limits on the presence of an isotropic 
background of relic gravitons have been illustrated. According to the Ligo collaboration 
(see Eq. (19) of Ref. \cite{LIGOS2}) the spectral energy density of a putative 
(isotropic) background of relic gravitons can be parametrized as\footnote{To be 
completely faithful with the Ligo parametrization the variable $\beta$ will not be 
changed. It should be borne in mind, however, that $\beta$ is used, in the present paper, 
to quantify the theoretical error on the maximal frequency of the relic graviton spectrum
(see e.g. Eq. (\ref{W5}) and discussion therein).}:
\begin{equation}
\Omega_{\mathrm{GW}}(\nu,\tau_{0}) = \Omega_{\mathrm{GW},\beta} \biggl(\frac{\nu}{100\,\mathrm{Hz}}\biggr)^{\beta + 3}.
\label{LIGOpar}
\end{equation}
It is worth mentioning that the parametrization of Eq. (\ref{LIGOpar}) fits very well 
with Fig.  \ref{Figure9} where the pivot frequency $\nu_{\mathrm{LV}}=100 
\mathrm{Hz}$ coincides 
with the pivot frequency appearing in the parametrization (\ref{LIGOpar}). For 
the scale-invariant case (i.e. $\beta= -3$ in eq. (\ref{LIGOpar}))
the Ligo collaboration sets a $90 \%$ upper limit of $1.20\times 10^{-4}$ on 
the amplitude appearing in Eq. (\ref{LIGOpar}), i.e. $\Omega_{\mathrm{GW},-3}$.
Using different sets of data (see \cite{LIGOS1,LIGOS3}) the Ligo collaboration 
manages to improve the bound even by a factor $2$ getting down to 
$6.5\times 10^{-5}$. Keeping an eye on Fig. \ref{Figure9} shows that the current
LIGO sensitivity is still too small. 

As far as the $\Lambda$CDM model is concerned, direct detection looks equally hopeless also for the advanced interferometers. 
In the case of an exactly scale invariant spectrum
the correlation of the two (coaligned) LIGO detectors with 
central corner stations in Livingston (Lousiana) and in Hanford 
(Washington) might reach a sensitivity to a flat spectrum 
which is \cite{mg4,mg5,mg6}
 \begin{equation}
h_0^2\,\, \Omega_{\rm GW}(\nu_{\mathrm{LV}},\tau_{0}) \simeq 6.5 \times 10^{-11} \,\, 
\biggl(\frac{1\,\,\mathrm{yr}}{T} \biggr)^{1/2}\,\,\mathrm{SNR}^2, \qquad \nu_{\mathrm{LV}} =0.1 \,\, \mathrm{kHz}
\label{SENS}
\end{equation} 
where $T$ denotes the observation time and $\mathrm{SNR}$ is the signal to noise ratio.  Equation (\ref{SENS}) is in close agreement with the 
sensitivity of the advanced Ligo apparatus \cite{LIGO} to an exactly scale-invariant spectral energy density \cite{int1,int2,int3}.  Equation (\ref{SENS}) together with the 
plots of Fig. \ref{Figure9} suggest that the relic graviton 
background predicted by the $\Lambda$CDM paradigm is not directly 
observable by wide-band interferometers in their advanced version.
The minuteness of $h_{0}^2 \Omega_{\mathrm{GW}}(\nu_{\mathrm{LV}},\tau_{0})$ stems directly from the 
assumption that the inflationary phase is  suddenly followed by the radiation-dominated phase. 
\renewcommand{\theequation}{4.\arabic{equation}}
\setcounter{equation}{0}
\section{Scaling violations at high frequencies}
\label{sec4}
According to the results of the previous section, even in the future, 
the sensitivity of wide-band interferometers will be insufficient to reach 
into the parameter space of the $\Lambda$CDM scenario. The accurate techniques 
introduced in the present paper seem therefore a bit pleonastic. In this section the opposite will be argued insofar as the spectral energy density of relic gravitons 
may well be increasing (rather than decreasing) as a function of the frequency $\nu$. 

The late and early time effects conspire, in the $\Lambda$CDM paradigm 
to make the spectral energy density slightly decreasing at high frequencies 
(see e.g. Fig. \ref{Figure8}).  Different thermal histories allow for scaling violations which may also go in the opposite direction and make the spectral energy density increasing (rather than decreasing as in the $\Lambda$CDM case) for typical
 frequencies larger than a pivotal frequency $\nu_{\mathrm{s}}$ which is related to the total duration of the stiff phase.
If the stiff phase takes place before  BBN, then $\nu_{\mathrm{s}} > 10^{-2}$ nHz. If the 
stiff phase takes place for equivalent temperatures larger than $100$ GeV, then $\nu_{\mathrm{s}} \geq \mu\mathrm{Hz}$. 
Finally, if the stiff phase takes place for $T \geq 100$ TeV,  then $\nu_{\mathrm{s}} > \mathrm{mHz}$.  

In the early Universe, 
the dominant energy condition might be violated and this observation will also 
produce scaling violations in the spectral energy density  \cite{DOC1}. 
If we assume the validity of the $\Lambda$CDM paradigm, a violation 
of the dominant energy condition implies that, during an early stage of the life of the Universe, the effective enthalpy density of the sources driving the geometry was negative and this may happen in the presence 
of bulk viscous stresses \cite{DOC1} (see also \cite{DOC2,DOC3} for interesting reprises of this idea). In what follows the focus will be on the more mundane possibility that the thermal history of the plasma includes a phase where the  speed of sound was close to the speed of light.  In  Eqs. (\ref{ST1}) and (\ref{ST2}) the stiff evolution has been 
parametrized in terms of an effective (i.e. fluid) description which can be 
realized in diverse models not necessarily related to a fluid behaviour. If the energy-momentum tensor of the sources of the geometry is provided
by a scalar degree of freedom (be it for instance $\varphi$) The effective energy density, pressure 
and anisotropic stress of $\varphi$ will then be, respectively, 
\begin{eqnarray}
&& \rho_{\varphi} =\biggl(\frac{{\dot{\varphi}}^2}{2} + V \biggr) + \frac{1}{2a^2} (\partial_{k} \varphi)^2, \qquad p_{\varphi} = \biggl(\frac{\dot{\varphi}^2}{2} - V\biggr) - \frac{1}{6 a^2} (\partial_{k} \varphi)^2,
\label{PPHI}\\
&& \Pi_{i}^{j}(\varphi) = - \frac{1}{a^2} \biggl[\partial_{i}\varphi \partial^{j}\varphi - \frac{1}{3} (\partial_{k}\varphi)^2 \delta_{i}^{j} \biggr].
\label{PijPHI}
\end{eqnarray}
Equation (\ref{PPHI}) imply that the effective barotropic index for the scalar system under discussion is simply given by 
\begin{equation}
w_{\varphi} = \frac{p_{\varphi}}{\rho_{\varphi}} = 
\frac{\biggl(\frac{\dot{\varphi}^2}{2} - V\biggr) - \frac{1}{6 a^2} (\partial_{k} \varphi)^2}{\biggl(\frac{{\dot{\varphi}}^2}{2} + V \biggr) + \frac{1}{2a^2} (\partial_{k} \varphi)^2}.
\label{wPHI}
\end{equation}
If $\dot{\varphi}^2 \gg V$ and $\dot{\varphi}^2 \gg (\partial_{k}\varphi)^2/a^2$, then $p_{\varphi} \simeq \rho_{\varphi}$: in this 
regime the scalar field behaves as a stiff fluid. If $V \gg \dot{\varphi}^2 \gg (\partial_{k}\varphi)^2/a^2$, then $w_{\varphi} \simeq -1$: in this regime 
the scalar field is an inflaton candidate. Finally if $ (\partial_{k}\varphi)^2/a^2 \gg  \dot{\varphi}^2 $ and $ (\partial_{k}\varphi)^2/a^2 \gg  V$, then $w_{\varphi} \simeq -1/3$:
in this regime the system is gradient-dominated. Of course also intermediate situations are possible (or plausible).

From the purely phenomenological point of view it is not forbidden 
(by any phenomenological consideration) to have a sufficiently long stiff 
phase. This was the point of view invoked in \cite{mg1} (see also \cite{mg2,mg3}) 
where it was also suggested that the spectral energy density of relic gravitons 
may increase with frequency. The presence of a phase dominated by the 
kinetic energy of a scalar degree of freedom (typical of quintessence models) 
became more compelling also in the light of the formulation of the so-called 
quintessential inflationary models \cite{PV} where the inflaton field practically does 
not decay and it is identified with the quintessence field. 
If there is some delay between the end of inflation and the onset of radiation 
the maximal wavenumber of the spectrum will be given by: 
\begin{equation}
k_{\mathrm{max}} = M_{\mathrm{P}} \biggl(\frac{H}{M_{\mathrm{P}}}\biggr)^{1 - \alpha}
 \biggr(\frac{H_{\mathrm{r}}}{M_{\mathrm{P}}}\biggl)^{\alpha - 1/2} 
 \biggl(\frac{H_{\mathrm{eq}}}{M_{\mathrm{P}}}\biggr)^{1/2} \biggl(\frac{a_{\mathrm{eq}}}{a_{0}}\biggr)
 \label{STFR1}
 \end{equation}
 where $\alpha = 2/[3(w_{\mathrm{t}} +1)]$ is related to the specific kind of stiff dynamics (indeed, $w_{\mathrm{t}} >1/3$). Equation 
 (\ref{STFR1}) can also be written as 
 \begin{equation}
k_{\mathrm{max}} = M_{\mathrm{P}} \Sigma^{-1}
 \biggl(\frac{H_{\mathrm{eq}}}{M_{\mathrm{P}}}\biggr)^{1/2} \biggl(\frac{a_{\mathrm{eq}}}{a_{0}}\biggr), \qquad 
 \nu_{\mathrm{M}} = k_{\mathrm{M}}/(2\pi).
 \label{STFR2}
 \end{equation}
 where 
 \begin{equation}
 \Sigma = \biggl(\frac{H}{M_{\mathrm{P}}}\biggr)^{\alpha-1 }
 \biggr(\frac{H_{\mathrm{r}}}{M_{\mathrm{P}}}\biggl)^{1/2-\alpha }.
 \label{STFR3}
\end{equation}
In the case $\Sigma = {\mathcal O}(1)$ (as it happens in the case $\alpha = 1/3$ if the initial radiation is in the 
form of quantum fluctuations) $\nu_{\mathrm{M}} \simeq 100\,\,\mathrm{GHz}$, more precisely:
\begin{equation}
\nu_{\mathrm{max}} = 1.177 \times 10^{11} \Sigma^{-1} 
\biggl(\frac{h_{0}^2 \Omega_{\mathrm{R}0}}{4.15 \times 10^{-5}}\biggr)^{1/4}\,\,\mathrm{Hz}.
\label{STFR4}
\end{equation}
 The strategy will now be to parametrize the violation of scale invariance in terms 
of the least possible number of parameters, i.e the frequency $\nu_{\mathrm{s}}$ (defining the region of the spectrum at which the scaling violations take place) and 
 the slope of the spectrum arising during the stiff phase. Of course 
 the frequency $\nu_{\mathrm{s}}$ can be dynamically related to the frequency of the maximum and, consequently, 
the first parameter can be trated for $\Sigma$. The slope of the spectrum during the stiff phase depends 
upon the total barotropic index and can therefore be traded for $w_{\mathrm{t}}$.
Assuming the presence of a single stiff (post-inflationary) phase we will have that 
\begin{equation}
k_{\mathrm{s}} =  M_{\mathrm{P}} \biggl(\frac{H_{\mathrm{eq}}}{M_{\mathrm{P}}}\biggr)^{1/2} \biggl(\frac{a_{\mathrm{eq}}}{a_{0}}\biggr) \sqrt{\frac{H_{\mathrm{r}}}{M_{\mathrm{P}}}}= M_{\mathrm{P}} \biggl(\frac{H_{\mathrm{eq}}}{M_{\mathrm{P}}}\biggr)^{1/2} 
\biggl(\frac{a_{\mathrm{eq}}}{a_{0}}\biggr)  \Sigma^{ 1/(1 - 2\alpha)}  \biggl(\frac{H}{M_{\mathrm{P}}}\biggr)^{(\alpha -1)/(2 \alpha -1)} ,
\label{STFR5}
\end{equation}
where the second equality follows from the first by using  the relation of $H_{\mathrm{r}}$ to $\Sigma$ dictated by Eq. (\ref{STFR5}).  From Eq. (\ref{STFR5}) 
the frequency turns out to be:
\begin{equation}
\nu_{\mathrm{s}} =  1.173 \times 10^{11} \Sigma^{ 1/(1-2\alpha)} \,\,(\pi \epsilon {\mathcal A}_{{\mathcal R}})^{\frac{\alpha -1}{2(2\alpha -1)}}\,\,\biggl(\frac{h_{0}^2 \Omega_{\mathrm{R}0}}{4.15 \times 10^{-5}}\biggr)^{1/4}\,\,\mathrm{Hz}.
\label{STFR6}
\end{equation}
The quantity $\Sigma$ is always smaller than $1$ or, at most, of order $1$.  
This is what happens within specific models. For instance, 
if the radiation present at the end of inflation comes from amplified quantum fluctuations (i.e. Gibbons-Hawking radiation), 
quite generically, at the end of inflation $\rho_{\mathrm{r}} \simeq H^4$. More specifically 
\begin{equation}
\rho_{\mathrm{r}}  = \frac{\pi^2}{30} N_{\mathrm{eff}} T_{H}^4 = \frac{N_{\mathrm{eff}} H^4}{480 \pi^2}.
\label{STFR13}
\end{equation}
In Eq. (\ref{STFR13}) $N_{\mathrm{eff}}$ is the number of species contributing 
to the quantum fluctuations during the quasi-de Sitter stage of expansion. 
In \cite{ford3} (see also \cite{mg2,mg3,PV}) it has been argued that this quantity could be evaluated using a perturbative expansion valid in the limit of quasi-conformal 
coupling. It should be clear that $N_{\mathrm{eff}}$ is conceptually 
different from the number of relativistic degrees of freedom $g_{\rho}$. 
Given $H$ and $N_{\mathrm{eff}}$ the length of the stiff phase is fixed, in this case,  by \cite{PV}
\begin{equation}
\lambda H^4 \biggl(\frac{a_{\mathrm{i}}}{a_{\mathrm{r}}}\biggr)^{4} =  H^2 M_{\mathrm{P}}^2  \biggl(\frac{a_{\mathrm{i}}}{a_{\mathrm{r}}}\biggr)^{3(w +1)} 
= H^2 M_{\mathrm{P}}^2  \biggl(\frac{a_{\mathrm{i}}}{a_{\mathrm{r}}}\biggr)^{2/\alpha}, 
\label{STFR14}
\end{equation}
where we used the fact that $\alpha = 2/[3(w +1)]$ and where we defined  $\lambda= N_{\mathrm{eff}}/(480\pi^2)$.
Equation (\ref{STFR14}) implies that
\begin{equation}
\biggl(\frac{a_{\mathrm{i}}}{a_{\mathrm{r}}}\biggr) = \lambda^{\frac{\alpha}{2 - 4\alpha}} \biggl(\frac{H}{M_{\mathrm{P}}}\biggr)^{\frac{\alpha}{1 - 2 \alpha}}, \qquad 
\biggl(\frac{H_{\mathrm{r}}}{M_{\mathrm{P}}}\biggr) = \lambda^{\frac{1}{2 ( 1 - 2 \alpha)}} \biggl(\frac{H}{M_{\mathrm{P}}}\biggr)^{\frac{2 ( 1 - \alpha)}{(1 -2 \alpha)}}.
\label{STFR15}
\end{equation}
\begin{figure}[!ht]
\centering
\includegraphics[height=6.2cm]{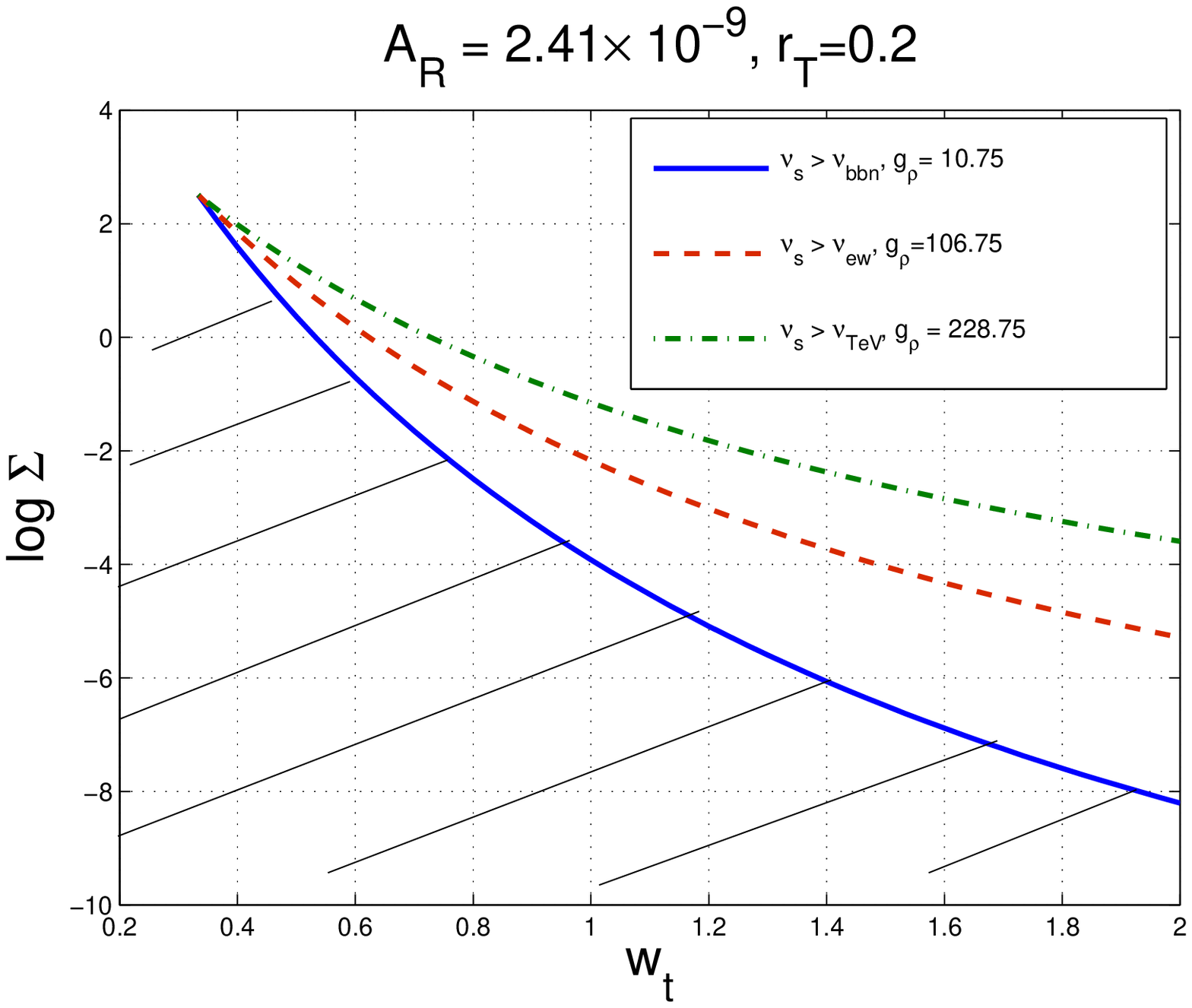}
\includegraphics[height=6.2cm]{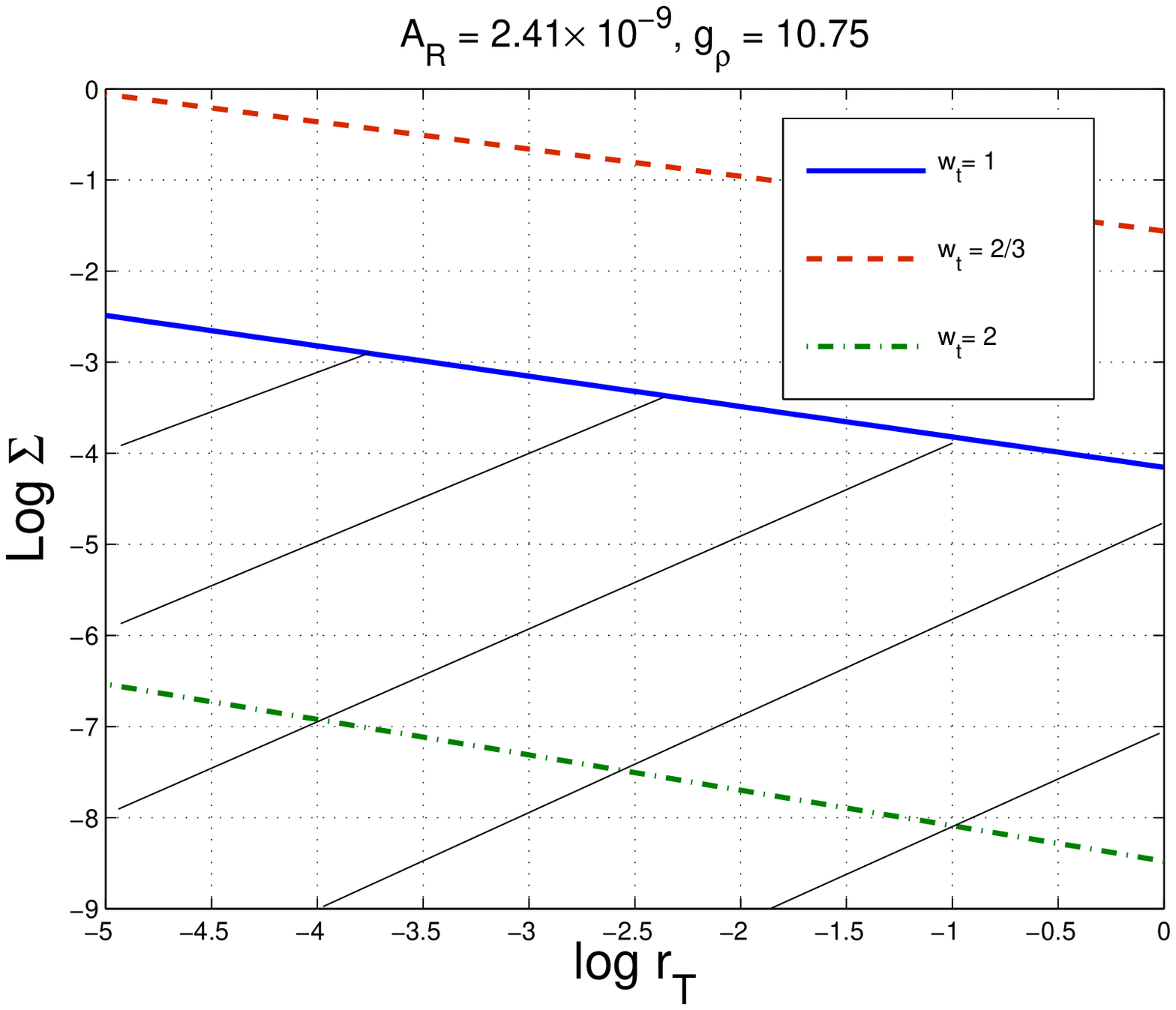}
\caption[a]{In the plot at the left the bounds on $\nu_{\mathrm{s}}$ are illustrated in terms of $w_{\mathrm{t}}$ (i.e. the barotropic index of the stiff phase). The quantity 
$\Sigma$ (defined in Eq. (\ref{STFR3})) depends, both, on the Hubble rate 
at the end of inflation and on the Hubble rate at the onset of radiation. 
At the right the exclusion region is phrased in terms of $r_{\mathrm{T}}$ (i.e. the 
tensor to scalar ratio) for different values of $w_{\mathrm{t}}$. The shaded areas represent the excluded regions.}
\label{Figure10}      
\end{figure}
Using the second relation in Eq. (\ref{STFR15})  and  Eq. (\ref{STFR3}),
 it turns out that  $\Sigma = \lambda^{1/4}$, which is always smaller than $1$ and, at most, ${\mathcal O}(1)$.  Instead of endorsing an explicit model by pretending to know 
the whole thermal history of the Universe in reasonable detail, it is more 
productive to keep $\Sigma$ as a free parameter  and to require that the  scaling violations in the spectral energy density will take place before BBN.  
The variation of $\Sigma$, $w$ and $r_{\mathrm{T}}$
can be simultaneously bounded.
The essential constraint which must be enforced in any model of scaling violations implies that the frequency $\nu_{\mathrm{s}}$ must necessarily exceed $\nu_{\mathrm{bbn}}$ (see Eq. (\ref{EQ4})).
This requirement guarantees that the stiff dynamics will be over by the time light nuclei start being formed.
In a complementary approach one might also require that $\nu_{\mathrm{s}} > \nu_{\mathrm{ew}}$ where 
$\nu_{\mathrm{ew}}$ corresponds to the value of the Hubble rate at the electroweak 
epoch, i.e. 
\begin{equation}
\nu_{\mathrm{ew}} = 3.998\times 10^{-6} \biggl(\frac{g_{\rho}}{106.75}\biggr)^{1/4} \biggl(\frac{T_{*}}{100\,\,\mathrm{GeV}}\biggr) 
\biggl(\frac{h_{0}^2 \Omega_{\mathrm{R}0}}{4.15 \times 10^{-5}}\biggr)^{1/4}\,\,\mathrm{Hz}.
\label{EW2}
\end{equation}
Finally, yet a different requirement could be to impose that $\nu > \nu_{\mathrm{Tev}}$ where $\nu_{\mathrm{TeV}}$ is defined as
\begin{equation}
\nu_{\mathrm{TeV}} = 4.819\times 10^{-3} \biggl(\frac{g_{\rho}}{228.75}\biggr)^{1/4} \biggl(\frac{T_{*}}{100\,\,\mathrm{TeV}}\biggr) 
\biggl(\frac{h_{0}^2 \Omega_{\mathrm{R}0}}{4.15 \times 10^{-5}}\biggr)^{1/4}\,\,\mathrm{Hz}.
\label{EW3}
\end{equation}
\begin{figure}[!ht]
\centering
\includegraphics[height=6.2cm]{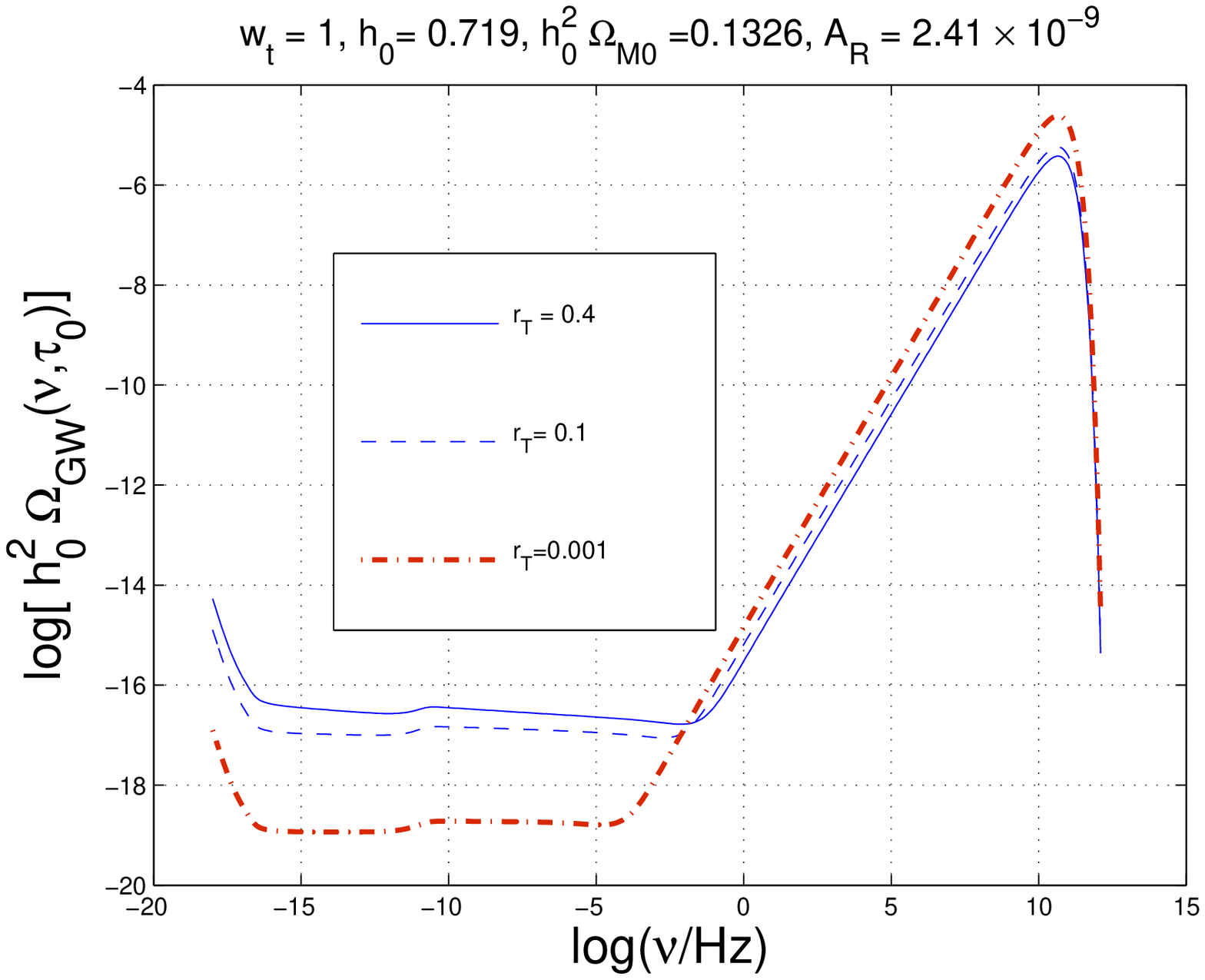}
\includegraphics[height=6.2cm]{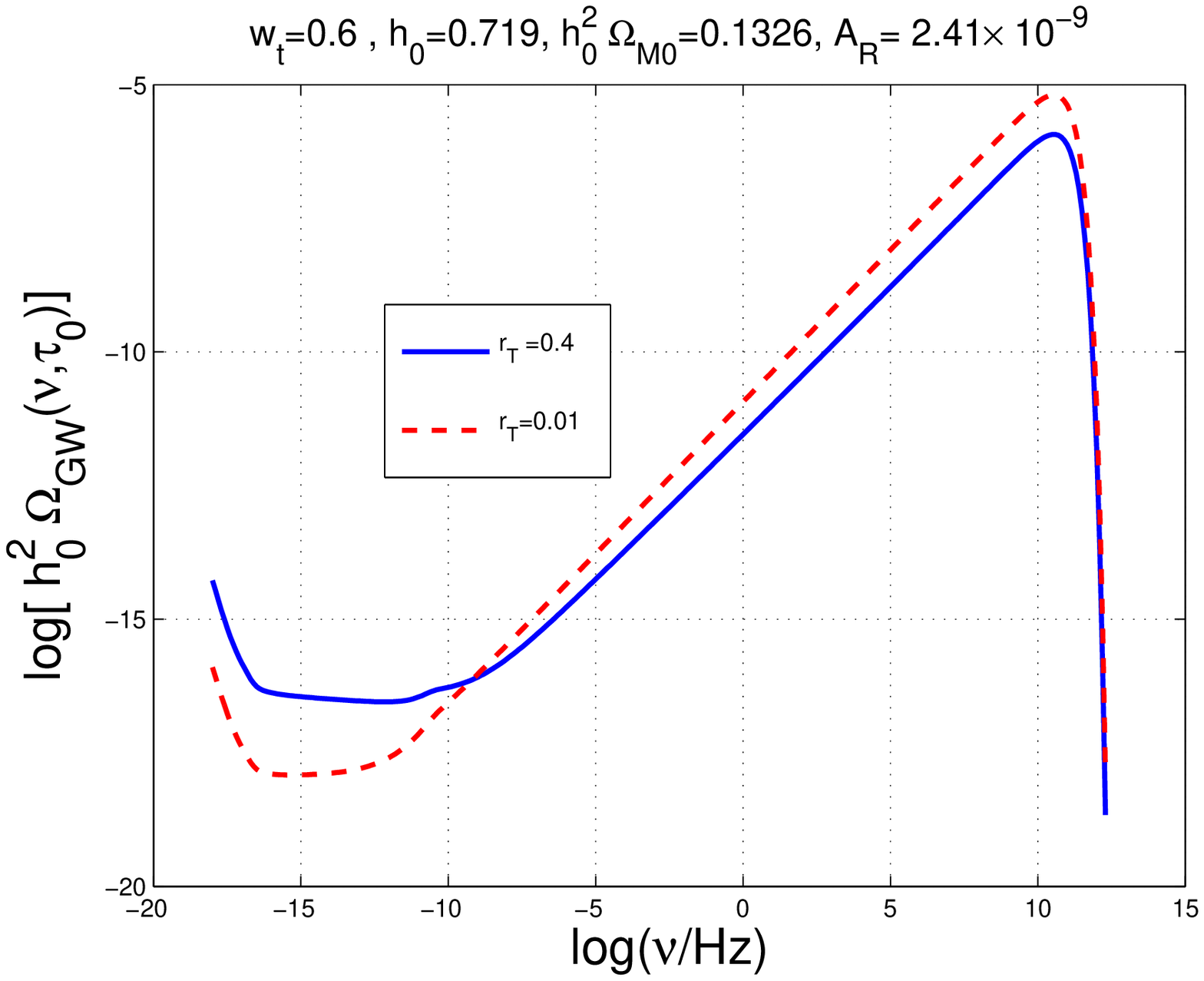}
\caption[a]{The spectral energy density of the relic gravitons coming from the stiff ages. 
In the plot at the left $w_{\mathrm{t}} =1$ while in the plot at the 
right $w_{\mathrm{t}} = 0.6$. In both plots the value of  $\Sigma$ has been 
fixed to $0.15$.}
\label{Figure11}      
\end{figure}
The condition $\nu > \nu_{\mathrm{Tev}}$ (as opposed to $\nu > \nu_{\mathrm{ew}}$) would imply that the stiff age did already finish by the time 
the Universe had a temperature of the order of $100$ TeV when, presumably, the number of relativistic degrees of freedom 
was much larger than in the minimal standard model \footnote{In Eq. (\ref{EW3}) the typical value of $g_{\rho}$ is 
the one arising in the minimal supersymmetric extension of the standard model.}. 

The constraints on $\Sigma$, $w_{\mathrm{t}}$ and $r_{\mathrm{T}}$ are summarized in Fig. \ref{Figure10}
The value of $\Sigma$ controls the position of the frequency at which 
the nearly scale-invariant slope of the spectrum will be violated. The barotropic index $w_{\mathrm{t}}$ 
is taken to be always larger than $1/3$ (by definition of stiff fluid) and with a maximal value of $1$. In Fig. \ref{Figure10} values 
of $w_{\mathrm{t}}$ as large as $2$ have been allowed just for completeness since some authors 
like to speculate that models with $w_{\mathrm{t}} >1$ do not violate causality 
constraints.  It is amusing to notice that the cases $w_{\mathrm{t}} >1$ 
are, anyway, totally irrelevant from the phenomenological point of view. In 
these cases, in fact, the detectability prospects are forlorn (see section \ref{sec5}).

In Fig. \ref{Figure10} (plot at the right) the different curves denote, respectively, the cases $\nu_{\mathrm{s}}= \nu_{\mathrm{bbn}}$ 
(full line), $\nu_{\mathrm{s}}= \nu_{\mathrm{ew}}$ (dashed line) and $\nu_{\mathrm{s}}= \nu_{\mathrm{TeV}}$ (dot-dashed line).
To be compatible with the corresponding constraint we have to be above each curve and the shaded region are excluded. Of course since different curves are present, we decided to shade the region which might correspond, according to theoretical prejudice to the most typical choice of parameters. 
\renewcommand{\theequation}{5.\arabic{equation}}
\setcounter{equation}{0}
\section{Relic gravitons and from the stiff age}
\label{sec5}
\subsection{Spectral energy density in the minimal T$\Lambda$CDM  scenario}
The conclusion of the previous section has been that it is indeed plausible 
to parametrize the scaling violations (at high frequency) in terms of two  parameters, i.e. 
the typical frequency $\nu_{\mathrm{s}}$ at which scaling violations occur and the typical slope of the spectral 
energy density for $\nu > \nu_{\mathrm{s}}$.  The latter framework has been dubbed T$\Lambda$CDM for 
tensor-$\Lambda$CDM \cite{mg7}. 
The two supplementary parameters physically depend upon 
 the sound speed during the stiff phase (i.e. $c_{\mathrm{st}}$) and
the threshold frequency (i.e. $\nu_{\mathrm{s}}$). Besides $c_{\mathrm{st}}$ 
and $\nu_{\mathrm{s}}$, there will also be $r_{\mathrm{T}}$ 
which controls, at once, the normalization and the slope of the 
low-frequency  branch of the spectral energy density.  The 
remaining six parameters of the underlying $\Lambda$CDM model 
will be fixed, just for illustration, to the best fit of the 5-yr WMAP data alone 
\cite{WMAP51,WMAP52,WMAP53,WMAP54,WMAP55}. In the numerical 
program used to compute the spectral energy density of the relic gravitons 
the putative values of the cosmological parameters can be changed at wish.

In spite of their present sensitivities \cite{LIGOS2} (see also Eq. (\ref{LIGOpar}) and discussion therein)
terrestrial interferometers might be able, one day,  to provide a 
prima facie evidence of relic gravitons. The present 
numerical approach will then be instrumental not only in setting upper limits 
but also in providing global fits of the cosmological observables
within the T$\Lambda$CDM model. According to this perspective, in the future
the three (now available) cosmological data sets 
will be complemented by the observations 
of the wide-band interferometers. Therefore, different 
choices of cosmological parameters (like, for instance, the various 
critical fractions of matter and dark energy) could slightly change 
the typical frequencies of the relic graviton spectrum as well as other 
features in the low-frequency region of the spectral energy density.

In the absence of any tensor contribution (i.e. $r_{\mathrm{T}}=0$) 
 the 5-yr WMAP data alone imply: 
\begin{equation}
(\Omega_{\mathrm{b}0},\, \Omega_{\mathrm{c}0}, \,\Omega_{\mathrm{\Lambda}}, \,h_{0}, \,n_{\mathrm{s}},\,\tau)= (0.0441,\, 0.214,\, 0.742,\, 0.719,\, 0.963,\,0.087).
\label{best1}
\end{equation}
If the tensors are included (i.e. $r_{\mathrm{T}} \neq 0$) but without any running 
of the scalar spectral index the parameters (inferred from the 
5-yr best fit to the WMAP data alone) slightly change and become: 
\begin{equation}
(\Omega_{\mathrm{b}0},\, \Omega_{\mathrm{c}0}, \,\Omega_{\mathrm{\Lambda}}, \,h_{0}, \,n_{\mathrm{s}},\,\tau,r_{\mathrm{T}})= (0.0417,\, 0.188,\, 0.770,\, 0.751,\, 0.986,\,0.090, < 0.43).
\label{best2}
\end{equation}
In Eq. (\ref{best2}) the last entry of the array contains $r_{\mathrm{T}}$ and it is 
actually an upper limit ($95\%$ CL) corresponding to the first row appearing in Tab.
\ref{MOD1}.  
Various other examples could be provided by considering, for instance, 
the combinations listed in Tab. \ref{MOD1}. 
In the numerical examples reported here, 
 the $\Lambda$CDM parameters will be fixed to their 
best fit values as they are reported in Eq. (\ref{best1}).  In this situation the 
tensor contribution will be parametrized not only by  
$r_{\mathrm{T}}$, but also by $c_{\mathrm{st}}$ and $\nu_{\mathrm{s}}$. 
The bounds on $r_{\mathrm{T}}$ are spelled out in Tab. \ref{MOD1}. 

In both plots of Fig. \ref{Figure11} the parameters are fixed to the 
values reported in Eq. (\ref{best1}).  
In Fig. \ref{Figure11} (plot at the left)  the $\Lambda$CDM scenario is 
complemented by a stiff phase with $w_{\mathrm{t}}=1$ and for different values of $r_{\mathrm{T}}$.  Always in Fig. \ref{Figure11} the value of the barotropic index is slightly reduced from 1 to  $w_{\mathrm{t}} =0.6$. In 
both plots of Fig. \ref{Figure11}, $\alpha_{\mathrm{T}}\neq 0$ and its value\footnote{If not otherwise specified, the value of the scalar spectral index used to compute $\alpha_{\mathrm{T}}$ is consistent with the 5-yr best fit to the WMAP data alone.} is given by Eq. (\ref{int3}).  The effect associated with a slight frequency variation  of the tensor spectral 
index is rather modest so that it can be hardly distinguished from 
 $\alpha_{\mathrm{T}} =0$ except when  $r_{\mathrm{T}}$ is sufficiently large. 
 A similar occurrence can be observed in the two plots reported in Fig. \ref{Figure9}. 
 
The infrared branch of the spectrum in both plots of Fig. \ref{Figure11} reproduces the results of Fig. \ref{Figure8}. 

As soon as the frequency increases from the aHz up to the nHz (and even larger) the spectral energy density 
increases sharply in comparison with the nearly scale-invariant case (see, e.g. Figs. 
\ref{Figure8} and \ref{Figure9}) where the spectral energy density was, 
for $\nu > \mathrm{nHz}$, at most ${\mathcal O}(10^{-16})$.
In the case of Fig. \ref{Figure11} the spectral energy density is clearly much larger. 
The accuracy in the determination of the infra-red branch of the spectrum is a condition 
for the correctness of the estimate of the spectral energy density of the high-frequency branch. The plots of 
Fig. \ref{Figure11} demonstrate that the low-frequency bounds on $r_{\mathrm{T}}$ do not forbid a larger signal at higher frequencies. 

A decrease of  $r_{\mathrm{T}}$ implies a suppression of the nearly scale-invariant plateau  in the region 
$\nu_{\mathrm{eq}} < \nu <\nu_{\mathrm{s}}$. At the same time the amplitude of the spectral energy density still 
increases for frequencies larger than the frequency of the elbow (i.e. $\nu_{\mathrm{s}}$). 
The latter trend can be simply understood since, at high frequency,  
the transfer function for the spectral energy density grows faster than the power spectrum of inflationary origin. 
For instance, in the case $w_{\mathrm{t}} =1$ and neglecting logarithmic corrections,
$\Omega_{\mathrm{GW}}(\nu,\tau_{0}) \propto \nu^{n_{\mathrm{T}}+1}$  for $\nu\gg \nu_{\mathrm{s}}$. Now, recall that 
$n_{\mathrm{T}}$ is given by Eq. (\ref{int3}). If $r_{\mathrm{T}}\to 0$, the combination 
 $(n_{\mathrm{T}} +1)$ will be much closer to $1$ than in the case when, say, $r_{\mathrm{T}} \simeq 0.3$.
 This aspect can be observed in both plots of Fig. \ref{Figure11} 
 where different values of $r_{\mathrm{T}}$ have been reported.  By decreasing the $w_{\mathrm{t}}$ from $1$ to, say, $0.6$ the extension of the nearly flat plateau gets narrower. 
 This is also a general effect which is particularly evident by comparing the two plots of Fig. \ref{Figure11}.  
 \begin{figure}[!ht]
\centering
\includegraphics[height=6.2cm]{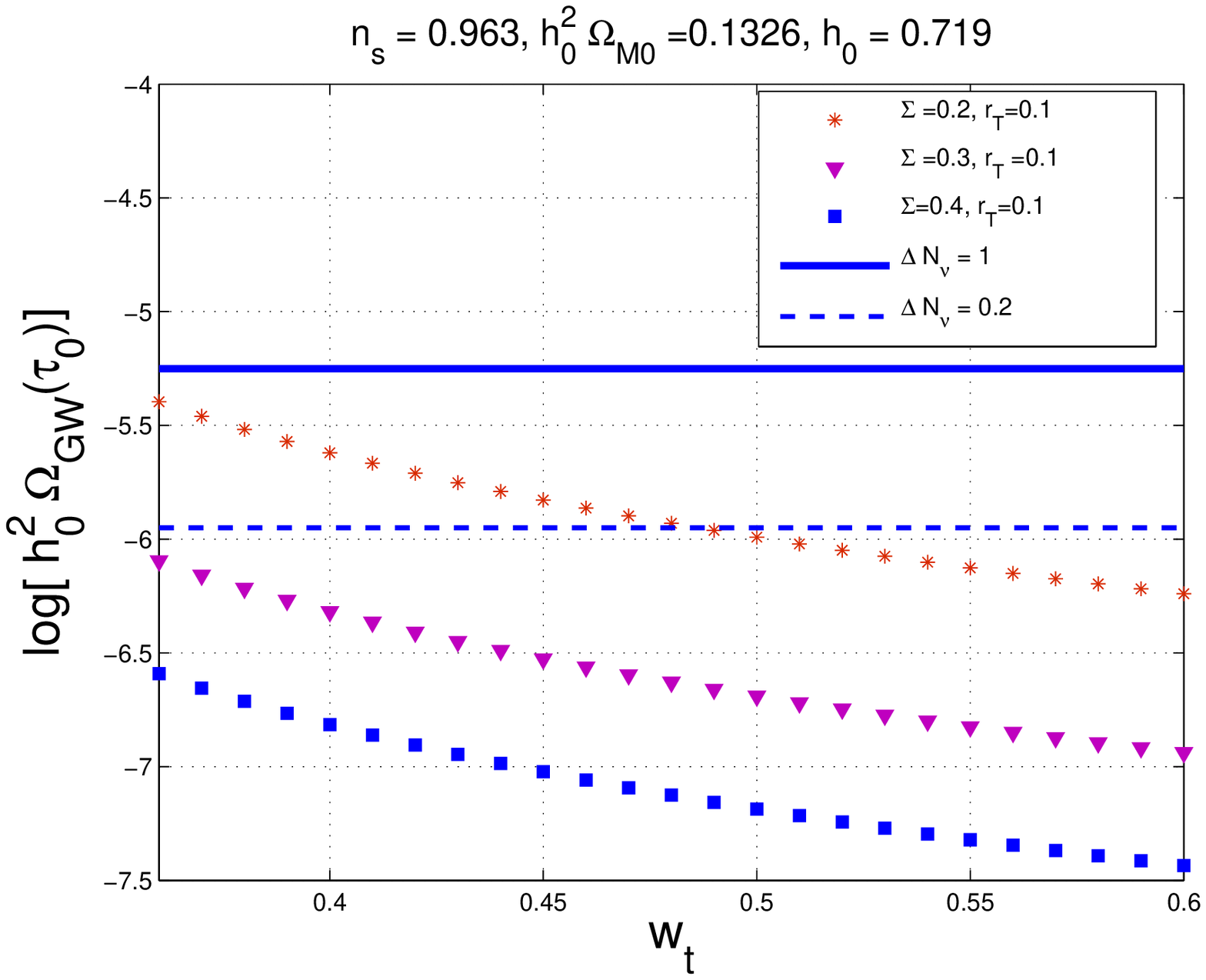}
\includegraphics[height=6.2cm]{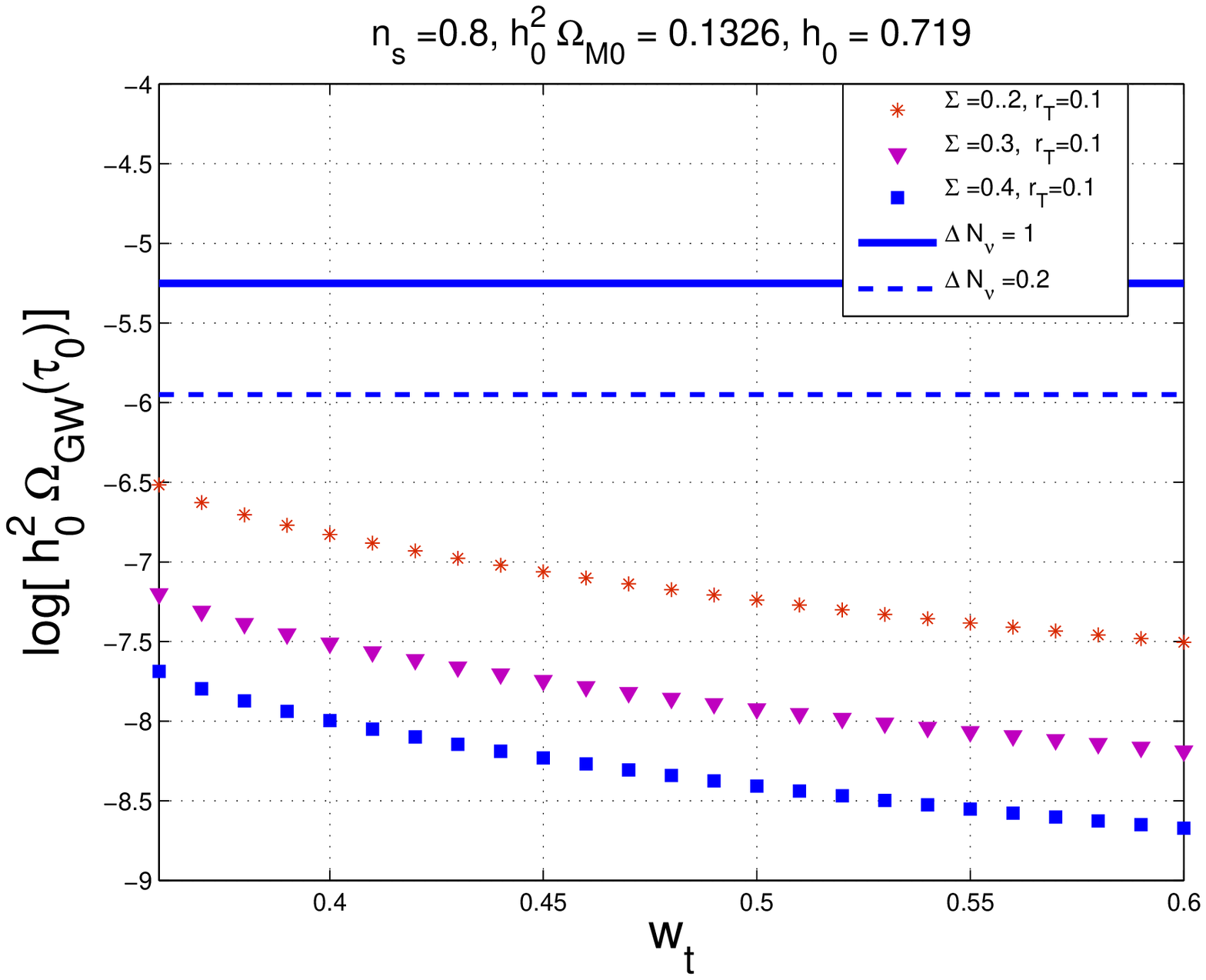}
\caption[a]{The bounds stemming from the amount of extra-relativistic species at the epoch 
of the synthesis of light nuclei are applied to the relic graviton spectra from the stiff epoch. As indicated the parameters of the underlying $\Lambda$CDM model 
are fixed to the best fit derived from the 5-yr WMAP data alone (see also Eq. (\ref{best1})).}
\label{Figure12}      
\end{figure}
 The slope of the high-frequency branch of the graviton energy spectrum can be easily deduced with 
 analytic methods and it turns out to to be 
\begin{equation} 
\frac{d\ln{\Omega_{\mathrm{GW}}}}{d\ln{\nu}}= 
 \frac{6 w_{\mathrm{t}} -2}{3 w_{\mathrm{t}} +1},\qquad \nu> \nu_{\mathrm{s}},
 \label{estimate}
 \end{equation}
  up to logarithmic corrections. 
 The  result of Eq. (\ref{estimate}) stems  from the simultaneous integration of the  background evolution equations and of the tensor mode functions according to the techniques described in section \ref{sec3}. 
 The semi-analytic estimate of the 
 slope (see \cite{mg1}) agrees with the results obtained by means 
 of the transfer function of the spectral energy density. 
 In Fig. \ref{Figure4} (plot at the left), for $\kappa = k/k_{\mathrm{s}} >1$ 
 $T^2_{\rho}(\kappa) \simeq \kappa$ which is consistent with Eq.  (\ref{estimate}) in the case $w_{\mathrm{t}}=1$. The logarithmic corrections arising in the case 
 $w_{\mathrm{t}} =1$ (see, for instance, Eq. (\ref{ST10})) have a simple analytic 
 interpretation which is evident from the results reported in Eqs. (\ref{mix1})--(\ref{mix2}) 
 and (\ref{mix1a})--(\ref{mix2a})  for the mixing coefficients in the case $w_{\mathrm{t}} = 1$. 
 \subsection{Phenomenological constraints}
 The spectra illustrated in Fig. \ref{Figure11} 
 (as all the spectra stemming from the stiff ages) 
 must be compatible not only with the CMB constraints (bounding, from 
 above, the value of $r_{\mathrm{T}}$) but also with other two classes 
 of constraints, i.e. the pulsar timing constraints \cite{pulsar1,pulsar2} and the big-bang nucleosynthesis constraints \cite{bbn1,bbn2,bbn3}. 
The pulsar timing constraint demands
\begin{equation}
\Omega(\nu_{\mathrm{pulsar}},\tau_{0}) < 1.9\times10^{-8},\qquad 
\nu_{\mathrm{pulsar}} \simeq \,10\,\mathrm{nHz},
\label{PUL}
\end{equation}
where $\nu_{\mathrm{pulsar}}$ roughly corresponds to the inverse 
of the observation time along which the pulsars timing has been monitored.  Assuming the maximal growth of the spectral energy density and the minimal 
 value of $\nu_{\mathrm{s}}$, i.e. $\nu_{\mathrm{bbn}}$ we will have 
\begin{equation}
h_{0}^2 \Omega_{\mathrm{GW}}(\nu,\tau_{0}) \propto \nu, \qquad \nu 
\geq \nu_{\mathrm{s}} \simeq \nu_{\mathrm{bbn}}.
\label{PUL2}
\end{equation}
Since $\nu_{\mathrm{pulsar}} \simeq 10^{3} \nu_{\mathrm{bbn}}$,
Eq. (\ref{PUL2}) implies  that 
$h_{0}^2\Omega_{\mathrm{GW}}(\nu_{\mathrm{pulsar}},\tau_{0}) \simeq 10^{-13}$  or even $10^{-14}$ depending upon 
$r_{\mathrm{T}}$.  But this value is always much smaller than the constraint stemming from pulsar timing measurements. If either $\nu_{\mathrm{s}} \gg \nu_{\mathrm{bbn}}$ 
or $c_{\mathrm{st}} < 1$ the value of 
$h_{0}^2\Omega_{\mathrm{GW}}(\nu_{\mathrm{pulsar}},\tau_{0})$ 
will be even smaller\footnote{This conclusion follows immediately from the hierarchy between $\nu_{\mathrm{pulsar}}$ and 
$\nu_{\mathrm{bbn}}$. If either $c_{\mathrm{st}}< 1$ or $\nu_{\mathrm{s}} \gg \nu_{\mathrm{bbn}}$, $h_{0}^2 \Omega_{\mathrm{GW}}$ can only grow very little and certainly much less than 
required to violate the bound of Eq. (\ref{PUL}).}. Consequently, even in the extreme cases when the 
frequency of the elbow  is 
close to $\nu_{\mathrm{bbn}}$, the spectral energy density is always much smaller than the requirement of Eq. (\ref{PUL}). 
The conclusion is that the pulsar timing bound is not constraining  for the T$\Lambda$CDM model. 

It is well known that the most significant constraint on the stiff spectra stems from BBN
\cite{mg0,mg1}. Being massless,  gravitons  can increase the expansion rate at the 
BBN epoch. To avoid the overproduction 
of  $^{4}\mathrm{He}$, the number of relativistic species 
must be bounded from above.
The BBN bound is customarily expressed in terms of (equivalent) extra fermionic 
species.
According to Eq. (\ref{EFF1}), during the radiation-dominated era, the energy density of the plasma 
can be written as $\rho_{\mathrm{t}} = g_{\rho} (\pi^2/30) T^4$ where $T$ denotes here the common (thermodynamic) temperature. 
An (ultra)relativistic fermion species with
two internal degrees of freedom
and in thermal equilibrium contributes $2\cdot7/8 = 7/4 = 1.75$ to $g_{\rho}$.  Before
neutrino decoupling the contributing relativistic particles are photons,
electrons, positrons, and $N_\nu = 3$ species of neutrinos, giving
$g_{\rho} = 10.75$.

The neutrinos have decoupled before electron-positron annihilation so that they
do not contribute to the entropy released in the annihilation.
While they are relativistic, the neutrinos still retain an equilibrium energy
distribution, but after the annihilation
their (kinetic) temperature is lower, $T_\nu = (4/11)^{1/3}T$.  Thus
$g_{\rho} = 3.36$ {\em after} electron-positron annihilation.
By now assuming that there are some additional relativistic degrees of
freedom, which also have decoupled by the time of electron-positron
annihilation,  or just some additional component $\rho_x$ to the energy
density with a radiation-like equation of state (i.e. $p_{x} = \rho_{x}/3$), the effect on the 
expansion rate will be the same as that of having
some (perhaps a fractional number of)
additional neutrino species.  Thus its contribution can be represented by
replacing $N_\nu$ with $ N_\nu + \Delta N_{\nu}$ in the above.  Before
electron-positron annihilation we have $\rho_x = (7/8)\Delta N_{\nu} \rho_\gamma$
and after electron-positron annihilation we have
$\rho_x = (7/8) (4/11)^{4/3} \,\Delta N_{\nu} \,\rho_{\gamma} \simeq 0.227\,\Delta N_{\nu} \,\rho_\gamma$.

The critical fraction of CMB photons can be directly computed from the 
value of the CMB temperature and it is notoriously given by
$h_{0}^2 \Omega_\gamma \equiv \rho_\gamma/\rho_{\mathrm{crit}} = 2.47\times10^{-5}$.
If the extra energy density component has stayed radiation-like until today,
its ratio to the critical density, $\Omega_x$, is given by
\begin{equation}
h_{0}^2   \Omega_x \equiv h^2\frac{\rho_x}{\rho_{\mathrm{c}}} = 5.61\times10^{-6}\Delta N_{\nu} 
\biggl(\frac{h_{0}^2 \Omega_{\gamma0}}{2.47 \times 10^{-5}}\biggr).
\end{equation}
If the additional species are relic gravitons, then  \cite{bbn1,bbn2,bbn3}: 
\begin{equation}
h_{0}^2  \int_{\nu_{\mathrm{bbn}}}^{\nu_{\mathrm{max}}}
  \Omega_{{\rm GW}}(\nu,\tau_{0}) d\ln{\nu} = 5.61 \times 10^{-6} \Delta N_{\nu} 
  \biggl(\frac{h_{0}^2 \Omega_{\gamma0}}{2.47 \times 10^{-5}}\biggr),
\label{BBN1}
\end{equation}
where $\nu_{\mathrm{bbn}}$ and $\nu_{\mathrm{max}}$ are given, respectively, by Eqs. (\ref{EQ4}) and (\ref{STFR4}).
Thus the constraint of Eq. (\ref{BBN1}) arises from the simple consideration 
that new massless particles could eventually increase the expansion rate 
at the epoch of BBN. 
The extra-relativistic species do not have to be, however, fermionic \cite{bbn2} 
and therefore the bounds on $\Delta N_{\nu}$ can be translated  into bounds 
on the energy density of the relic gravitons. 

A review of the constraints on  $\Delta N_{\nu}$ can be found 
in \cite{bbn2} . Depending on the combined data sets (i.e. various light elements abundances and different combinations of CMB observations), the standard BBN scenario implies that the bounds on $\Delta N_{\nu}$ range from $\Delta N_{\nu} \leq 0.2$ 
to $\Delta N_{\nu} \leq 1$. Similar figures, 
depending on the priors of the analysis, have been obtained in a more recent analysis \cite{bbn3}.  All the relativistic species present inside the 
Hubble radius at the BBN contribute to the potential increase in the expansion rate and this  explains why the integral in Eq. (\ref{BBN1}) must be performed from $\nu_{\mathrm{bbn}}$ to $\nu_{\mathrm{max}}$ (see also 
\cite{mg2} where this point was stressed in the framework of a specific model).  

The existence of the exponential suppression for $\nu>\nu_{\mathrm{max}}$  (see Fig. \ref{Figure11})
guarantees the convergence of the integral also in the case when the integration 
is performed up to $\nu \to \infty$. The constraint of Eq. (\ref{BBN1})  can be relaxed in some 
non-standard nucleosynthesis scenarios \cite{bbn2}, but, in what follows, the 
validity of Eq. (\ref{BBN1}) will be enforced by adopting 
$\Delta N_{\nu} \simeq 1$  which implies, effectively 
\begin{equation}
h_{0}^2  \int_{\nu_{\mathrm{bbn}}}^{\nu_{\mathrm{max}}}
  \Omega_{{\rm GW}}(\nu,\tau_{0}) d\ln{\nu} < 5.61\times 10^{-6} \biggl(\frac{h_{0}^2 \Omega_{\gamma0}}{2.47 \times 10^{-5}}\biggr). 
\label{BBN2}
\end{equation}

The models illustrated in Fig. \ref{Figure11} are on the verge 
of saturating the bounds of Eqs. (\ref{BBN1})--(\ref{BBN2}).  This conclusion 
stems directly from the form of spectral energy density: the broad spike  dominates
the (total) energy density of relic gravitons which are inside the Hubble radius 
at the time of big bang nucleosynthesis. A practical way 
of enforcing the bounds of Eqs. (\ref{BBN1}) and (\ref{BBN2})  is to 
integrate around the maximum of the curves depicted in Fig. \ref{Figure11}. 
\begin{figure}[!ht]
\centering
\includegraphics[height=6.2cm]{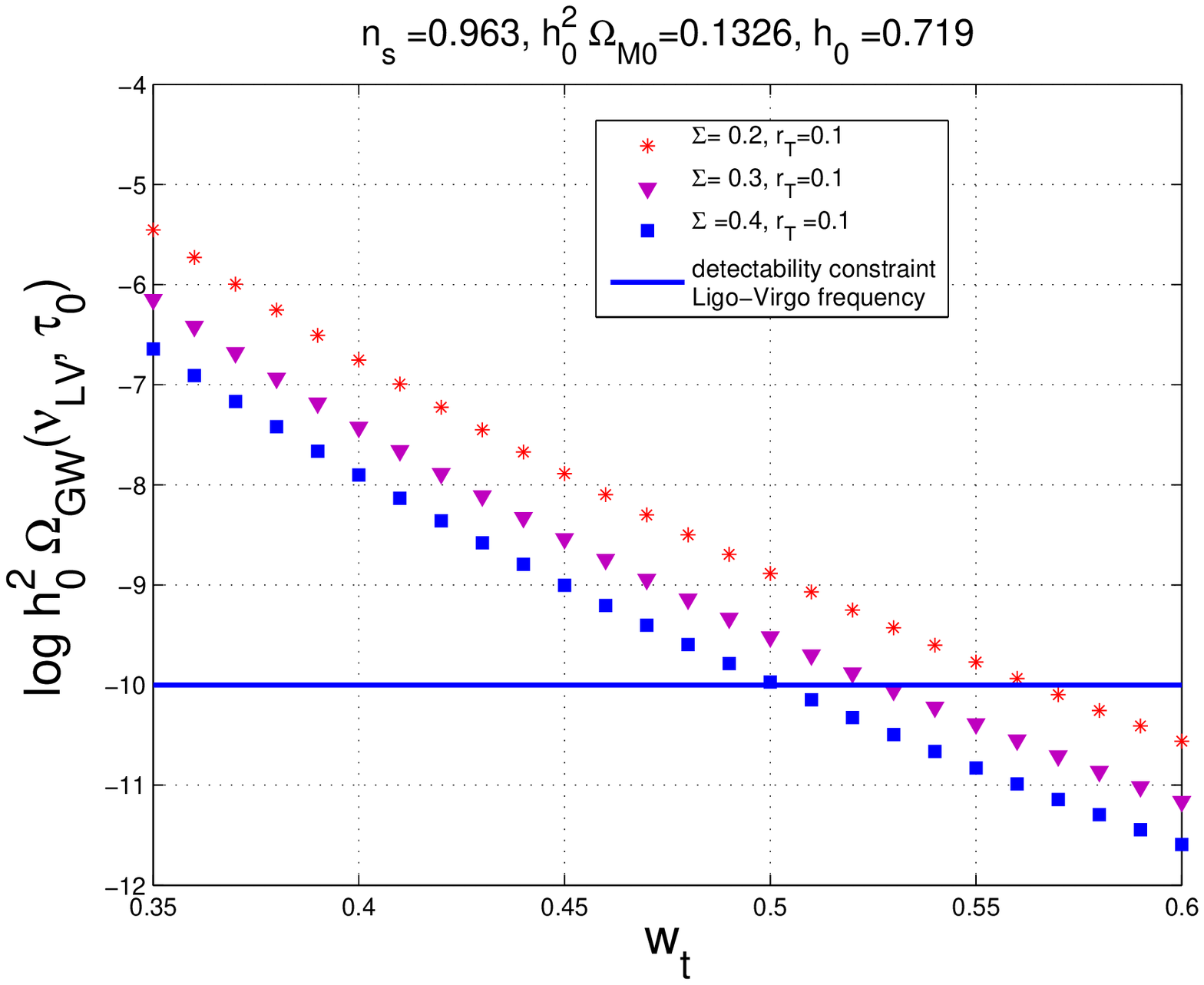}
\includegraphics[height=6.2cm]{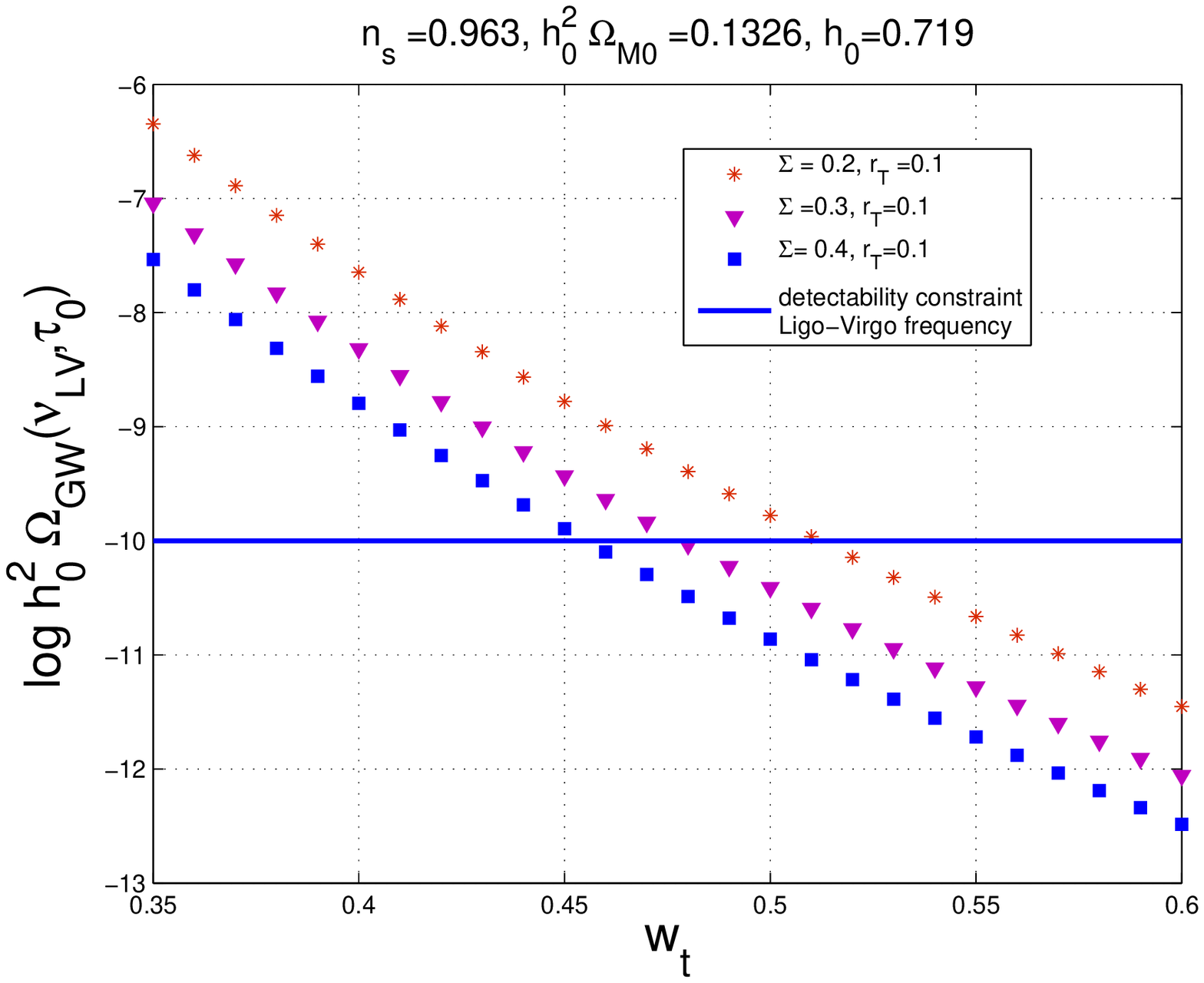}
\caption[a]{The detectability constraints (full lines in both plots)
 stemming from the putative sensitivities of wide-band interferometers in their 
advanced version. The points corresponding to the spectral energy density should lie above the full lines to be potentially interesting for those instruments.}
\label{Figure13}      
\end{figure}
In Fig. \ref{Figure12}  the energy density of the relic gravitons inside the Hubble 
radius at the nucleosynthesis epoch is reported in the case $r_{\mathrm{T}} =0.1$ and for different 
values of $\Sigma$. In the plot at the left $n_{\mathrm{s}} =0.963$ as implied by the WMAP 5-yr data 
alone.  In the plot at the right $n_{\mathrm{s}} =0.8$. The two horizontal lines illustrate
the bounds of Eqs. (\ref{BBN1})--(\ref{BBN2}) in the cases $\Delta N_{\nu} =1$ (full line) and $\Delta N_{\nu} = 0.2$
(dashed line) which are, respectively, the least constraining and the most constraining situations contemplated by 
current analyses.  In both cases the allowed region of the parameter space stays below the horizontal lines.
As the scalar spectral index diminishes, the constraints are better satisfied since 
$n_{\mathrm{s}}$ controls $\alpha_{\mathrm{T}}$ and, consequently,  the frequency dependence of the tensor spectral index $n_{\mathrm{T}}$ (see Eq. (\ref{int3})) in the case $\alpha_{\mathrm{T}} \neq 0$.
\begin{figure}[!ht]
\centering
\includegraphics[height=6.2cm]{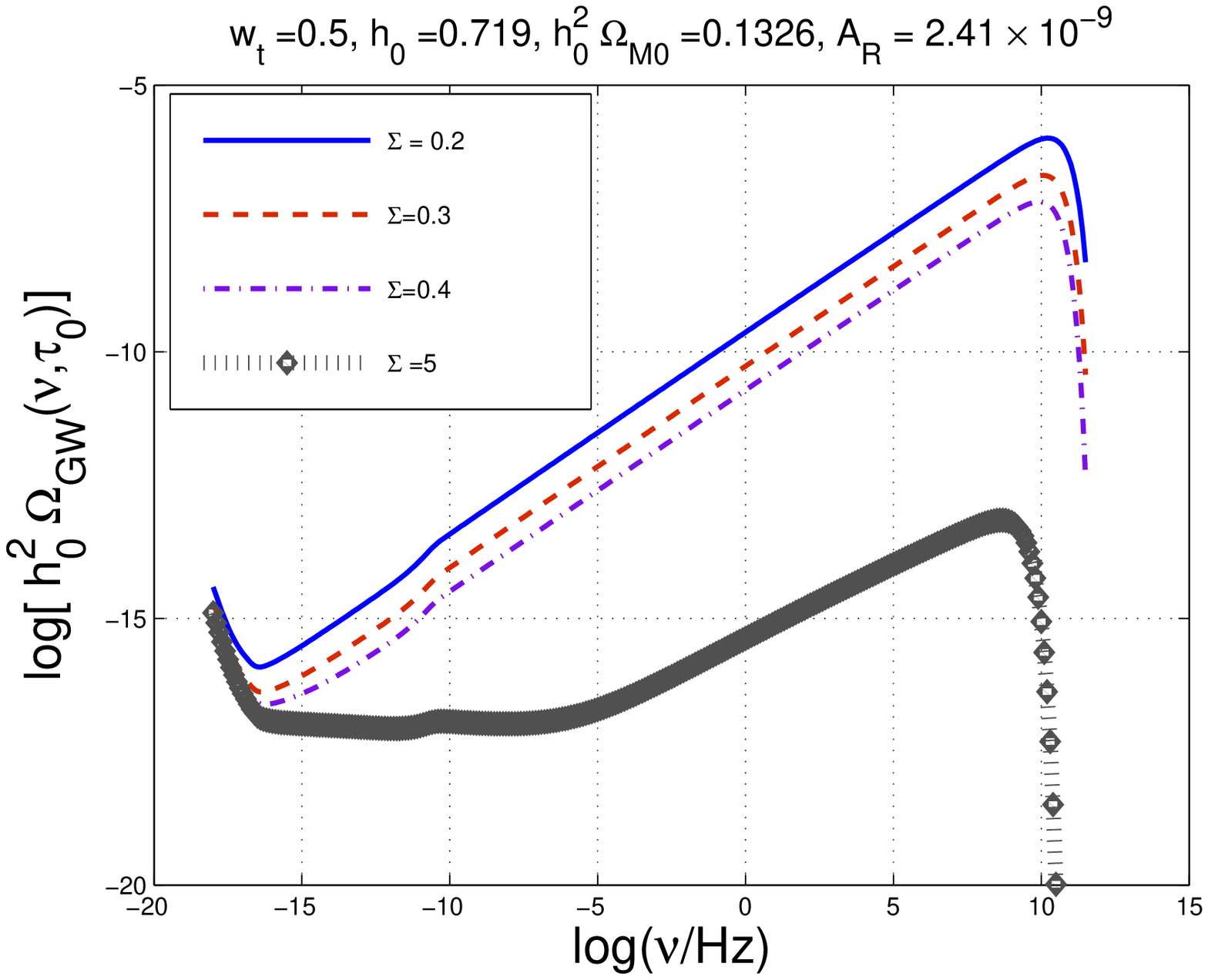}
\includegraphics[height=6.2cm]{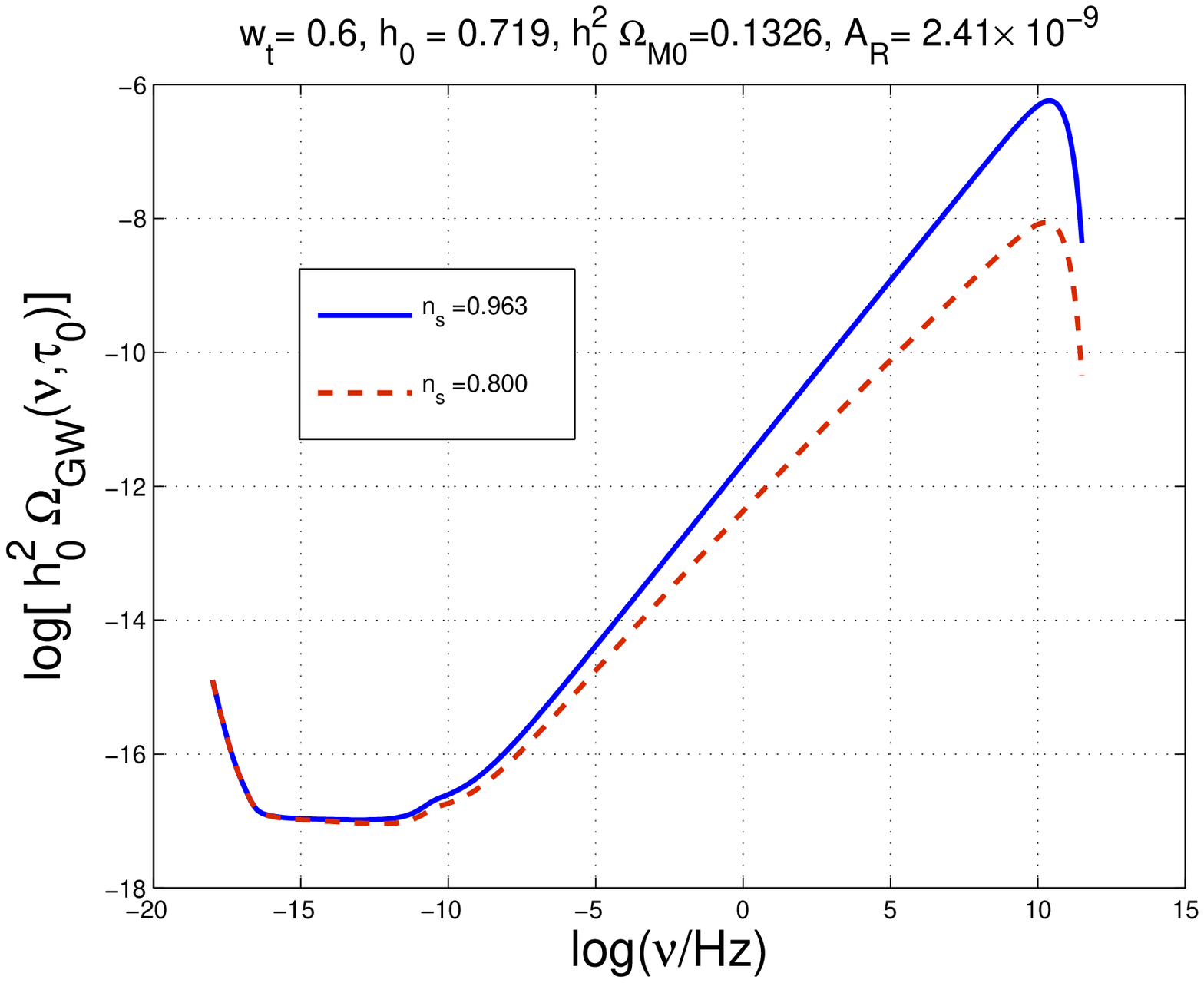}
\caption[a]{The spectral energy density is illustrated for small values of $w_{\mathrm{t}}$
and different values of $\Sigma$ (plot at the left).
In the plot at the right $\Sigma =0.2$ and $w_{\mathrm{t}} =0.6$.}
\label{Figure14}      
\end{figure}

\subsection{Detectability prospects}
The results presented in the previous subsection suggest that if 
$r_{\mathrm{T}}$  is bounded 
 from above by the cosmological data sets (see e.g. Tab. \ref{MOD1}),
a detectable  signal is expected for $\nu_{\mathrm{LV}} \simeq 100$ Hz for $0.35 < w_{\mathrm{t}} < 0.61$. In this case, following the parametrization of the Ligo 
collaboration we could say that the expected signal can be parametrized as 
\begin{equation}
\Omega_{\mathrm{GW}}(\nu,\tau_{0}) = \overline{\Omega}_{\mathrm{GW}} \biggl(\frac{\nu}{100\, \mathrm{Hz}}\biggr)^{n_{\mathrm{T}} + \frac{6 w_{\mathrm{t}} -2}{3 w_{\mathrm{t}} +1}},
\label{LIGOpar2}
\end{equation}
which mirrors Eq. (19) of Ref. \cite{LIGOS2} where the upper limits on the 
amplitude $\overline{\Omega}_{\mathrm{GW}}$ have been set.
This range turns out to be compatible with the bounds 
of Eqs. (\ref{BBN1})--(\ref{BBN2}). The relation (\ref{LIGOpar2}) could be used by the experimenters 
to set bounds on $\overline{\Omega}_{\mathrm{GW}}$ in the same way as upper bounds are obtained in the case of 
nearly scale-invariant spectra (see Eq. (\ref{LIGOpar}) and discussion therein).
Clearly, from the theoretical point of view, 
$\overline{\Omega}_{\mathrm{GW}}$ changes by varying the various 
T$\Lambda$CDM parameters.

For instance, by lowering $w_{\mathrm{t}}$, $h_{0}^2 \Omega_{\mathrm{GW}}(\nu,\tau_{0})$ 
increases for $\nu = \nu_{\mathrm{LV}}\simeq 0.1\, \mathrm{kHz}$.
This  trend can be inferred from Fig. \ref{Figure13} where the spectral energy density is evaluated exactly
for $\nu = \nu_{\mathrm{LV}}$.  To be detectable by wide band interferometers 
the parameters of the T$\Lambda$CDM must lie above the full lines.
The region of low barotropic indices emerging neatly from Fig. \ref{Figure13},
 leads to spectral energy densities which are progressively flattening as $w_{\mathrm{t}}$ diminishes towards $1/3$.  
Low values of $w_{\mathrm{t}}$ bring the frequency of the elbow, i.e. $\nu_{\mathrm{s}}$ below 
$10^{-10}$ Hz which is unacceptable since it would mean that, during nucleosynthesis, the Universe was dominated by the stiff fluid. 
In Fig. \ref{Figure10} (plot at the left) the region above the full line corresponds to a range of parameters  for 
which $\nu_{\mathrm{s}} > \nu_{\mathrm{bbn}}$: in such a range a decrease of $w_{\mathrm{t}}$ demands an increase of $\Sigma$.

The occurrence described in the previous paragraph is illustrated in Fig. \ref{Figure14} where, at the left, $w_{\mathrm{t}} =0.5$ and the values of $\Sigma$ are the same ones illustrated in Fig. \ref{Figure13}.
The full, dashed and dot-dashed curves illustrated in Fig. \ref{Figure14} (plot at 
the left) are incompatible with phenomenological considerations 
since the frequency of the elbow is systematically smaller than $\nu_{\mathrm{bbn}}$.
Once more, this choice of parameters would contradict the bounds of  Fig. \ref{Figure10} and would imply that the stiff ]
phase is not yet finished at the BBN time. In the left plot of Fig. \ref{Figure14} the diamonds denote a model which is compatible 
with BBN considerations but whose signal at the frequency of interferometers 
is rather small (always three orders of magnitude larger than in the case 
of conventional inflationary models).

The compatibility  with the phenomenological constraints demands that the parameters
of the T$\Lambda$CDM paradigm must lie above the full lines of Fig. \ref{Figure10}.
The requirements of Fig. \ref{Figure10} suggest, therefore, that 
$\Sigma$ should be raised a bit. In this case the frequency of the elbow gets 
shifted to the right but, at the same time, the overall amplitude of the spike 
diminishes.  The putative amplitude remains still much larger than 
the conventional inflationary signal reported in Fig. \ref{Figure9}.

In Fig. \ref{Figure14} (plot at the right) $\Sigma =0.2$ and $w_{\mathrm{t}} =0.6$. The tensor 
spectral index is allowed to depend upon frequency according to Eq. (\ref{int3}) (i.e. $\alpha_{\mathrm{T}} \neq 0$). 
Two different values of $n_{\mathrm{s}}$ are reported. In the example of Fig. \ref{Figure14} the phenomenological bounds 
are all satisfied.
\begin{figure}[!ht]
\centering
\includegraphics[height=6.2cm]{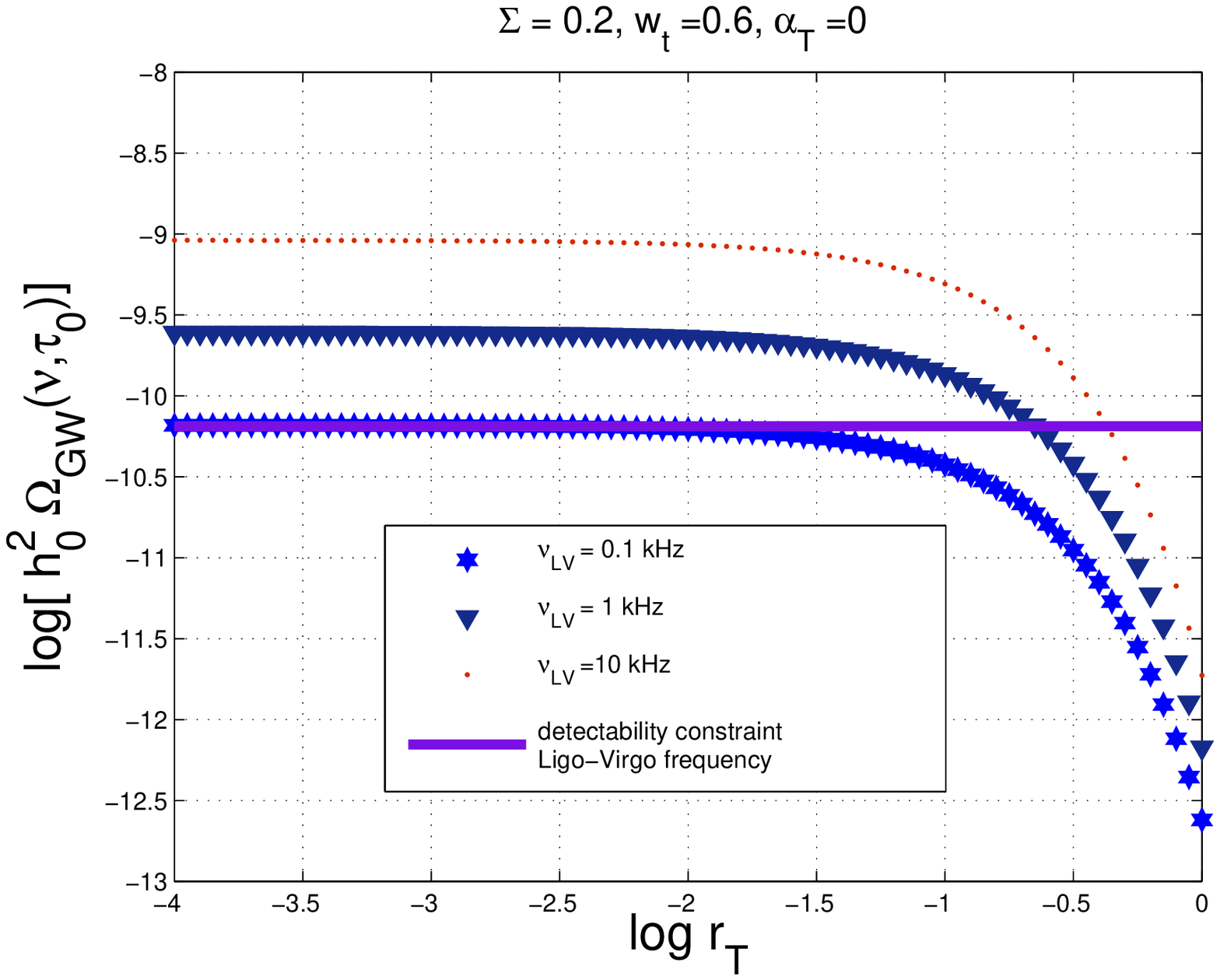}
\includegraphics[height=6.2cm]{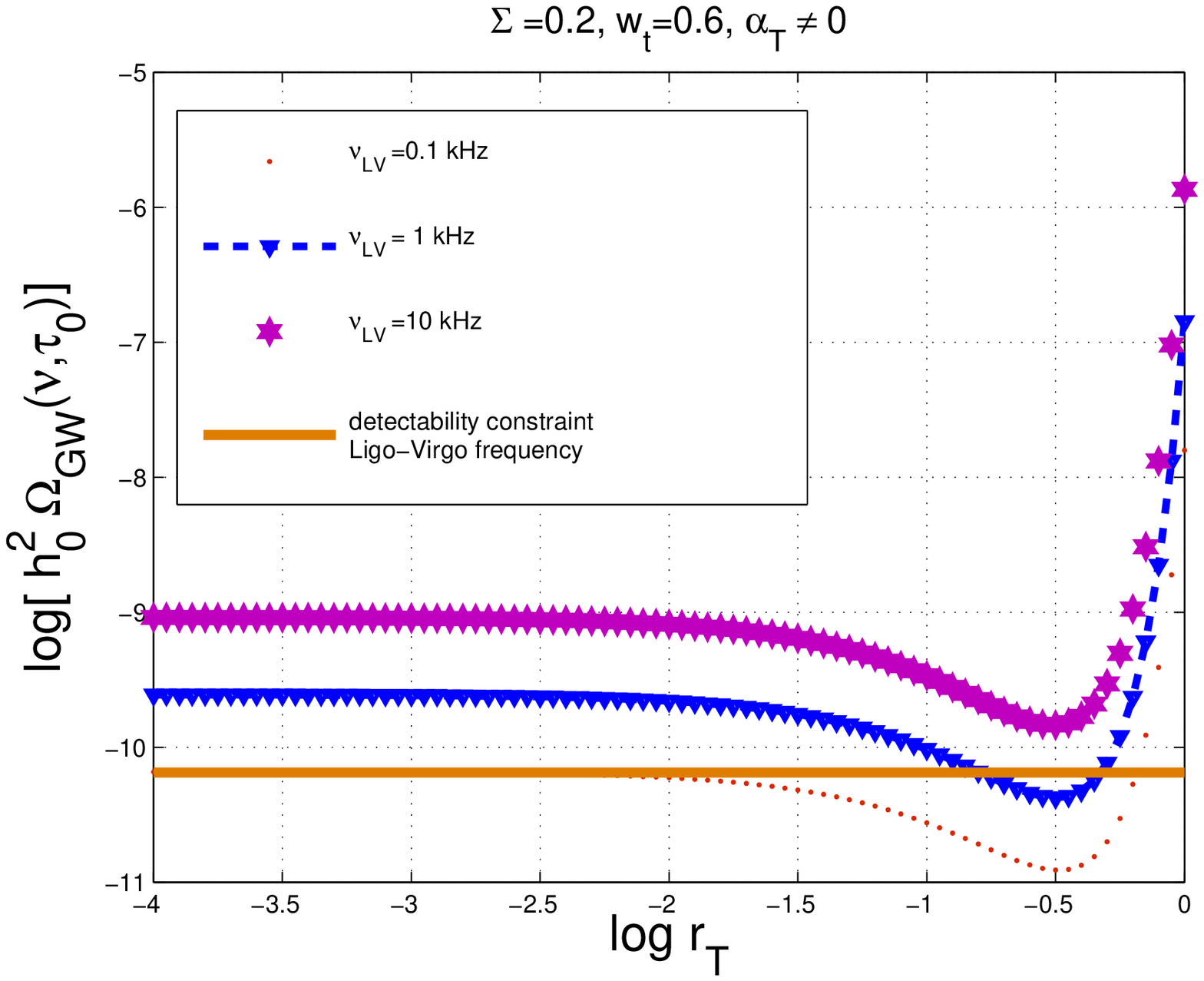}
\caption[a]{The graviton energy spectrum is illustrated, in the T$\Lambda$CDM scenario,  for $\nu= \nu_{\mathrm{LV}}$  and as a function of $r_{\mathrm{T}}$. As in Fig. \ref{Figure9} at the left $\alpha_{\mathrm{T}} =0$ while, at the right, $\alpha_{\mathrm{T}} \neq 0$.}
\label{Figure15}      
\end{figure}
In Fig. \ref{Figure15} the spectral energy density of the relic gravitons is illustrated as a function of $r_{\mathrm{T}}$ 
for a choice of parameters which is compatible with all the bounds applicable to the stochastic backgrounds of the 
relic gravitons. The three curves refer to three different frequencies, i.e. $0.1$ kHz, $1$ kHz and $10$ kHz. Indeed, if the spectrum is 
nearly scale-invariant (as in the case o Fig. \ref{Figure9}) we can compare the potential signal with the central frequency of the 
window. If the signal increases with frequency it is interesting to plot the same curve for some significant 
frequencies inside the window of wide-band interferometers. Even if the frequency window extends from 
few Hz to $10$ kHz the maximal sensitivity is in the central region and depends upon various important factors which will 
now be briefly discussed. 

To illustrate more quantitatively this point we remind the 
expression of the signal-to-noise ratio (SNR)  in the 
context of optimal processing  required for the detection of stochastic backgrounds: 
\begin{equation}
{\rm SNR}^2 \,=\,\frac{3 H_0^2}{2 \sqrt{2}\,\pi^2}\,F\,\sqrt{T}\,
\left\{\,\int_0^{\infty}\,{\rm d} \nu\,\frac{\gamma^2 (\nu)\,\Omega^2_{{\rm GW}}(\nu,\tau_{0})}
{\nu^6\,S_n^{\,(1)} (\nu)\,S_n^{\,(2)} (\nu)}\,\right\}^{1/2}\; ,
\label{SNR1}
\end{equation}
 ($F$ depends upon 
the geometry of the two detectors and in the case of the correlation between 
two interferometers $F=2/5$; $T$ is the observation time). 
In Eq. (\ref{SNR1}), $S_n^{\,(k)} (f)$ is the (one-sided) noise power 
spectrum (NPS) of the $k$-th 
$(k = 1,2)$ detector. The NPS contains the important informations concerning the 
noise sources (in broad terms seismic, thermal and shot noises)
 while $\gamma(\nu)$ is the overlap reduction function 
which is determined by the relative locations and orientations 
of the two detectors. In \cite{mg4} Eq. (\ref{SNR1}) has been used to assess the 
detectability prospects of gravitons coming from a specific model of stiff evolution with $w_{\mathrm{t}} =1$.
At that time the various suppressions of the low-frequency amplitude as well as the free-streaming effects 
were not taken into account. Furthermore, the evaluation of the energy transfer function was obtained, in 
\cite{mg6}, not numerically but by matching of the relevant solutions. We do know, by direct comparison, that 
such a procedure is justified but intrinsically less accurate than the one proposed here. 
It would be interesting to apply Eq. (\ref{SNR1})  for the (more accurate) assessment of the sensitivities 
of different instruments to a potential signal stemming from the stiff age \footnote{
For intermediate frequencies the integral of Eq. (\ref{SNR1}) is sensitive to the form of the overlap reduction 
function which depends upon the mutual position and relative orientations of the interferometers. The 
function $\gamma(\nu)$ effectively cuts-off the integral which defines the signal to noise ratio for a typical 
frequency $\nu\simeq 1/(2 d)$ where $d$ is the separation between the two detectors.
Since $\Omega_{\mathrm{GW}}$ increases 
with frequency (at least in the case of relic gravitons from stiff ages) at most as $\nu$ and since there is a $\nu^{-6}$ in the 
denominator, the main contribution to the integral should occur for $\nu < 0.1$ kHz.  This argument can be 
explicit verified in the case of the calculations carried on in \cite{mg4} and it would be interesting 
to check it also in our improved framework.}.

Equation (\ref{SNR1}) assumes that the intrinsic noises of the detectors are stationary, Gaussian, 
uncorrelated, much larger in amplitude than the gravitational strain, and 
statistically independent on the strain itself \cite{int1,int2,int3}.  The integral appearing in Eq. (\ref{SNR1})
extends over all the frequencies. However, the noise power spectra of the detectors are defined 
in a frequency interval ranging from few Hz to $10$ kHz. In the latter window, for very small frequencies 
the seismic disturbances are the dominant source of noise. For intermediate and high frequencies 
the dominant sources of noise are, respectively,  thermal and electronic (i.e. shot) noises.  
The wideness of the band is very important when cross-correlating two detectors: typically 
the minimal detectable $h_{0}^2\Omega_{\mathrm{GW}}$
 will become smaller (i.e. the sensitivity will increase) by a factor $1/\sqrt{\Delta \nu T}$ where 
 $\Delta \nu$ is the bandwidth and $T$, as already mentioned, is the observation time.
 Naively, if the minimal detectable signal (by one detector ) is 
 $h_{0}^2\Omega_{\mathrm{GW}} \simeq 10^{-5}$, then the cross-correlation of two 
 identical detector with overlap reduction $\gamma(\nu) =1$ will detect  
 $h_{0}^2\Omega_{\mathrm{GW}} \simeq 10^{-10}$ provided $\Delta \nu \simeq 100$ Hz and 
 $T\simeq {\mathcal O}(1\mathrm{yr})$ (recall that $1\mathrm{yr} = 3.15 \times 10^{7} \mathrm{Hz}^{-1}$). 
\begin{figure}[!ht]
\centering
\includegraphics[height=6.2cm]{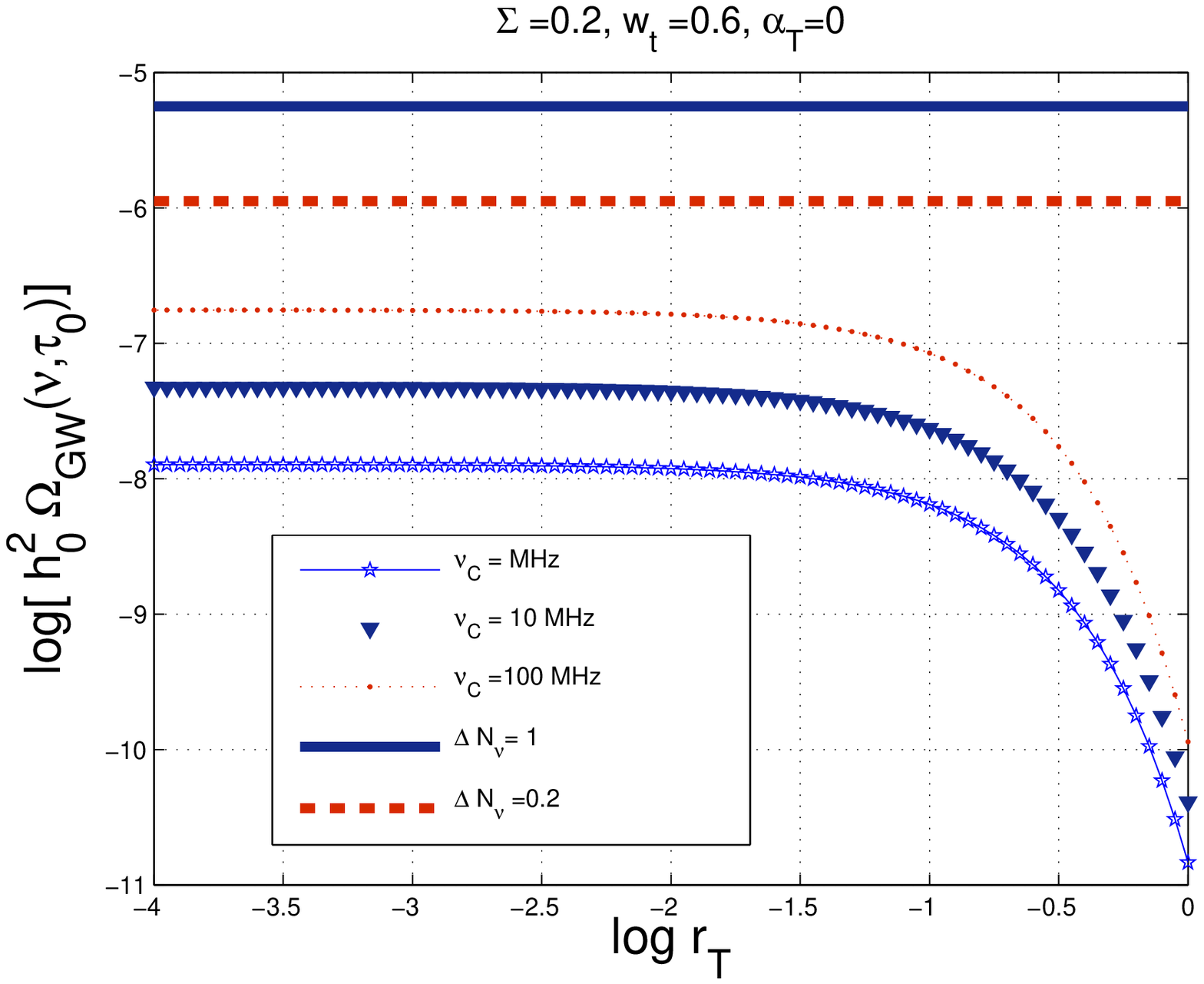}
\includegraphics[height=6.2cm]{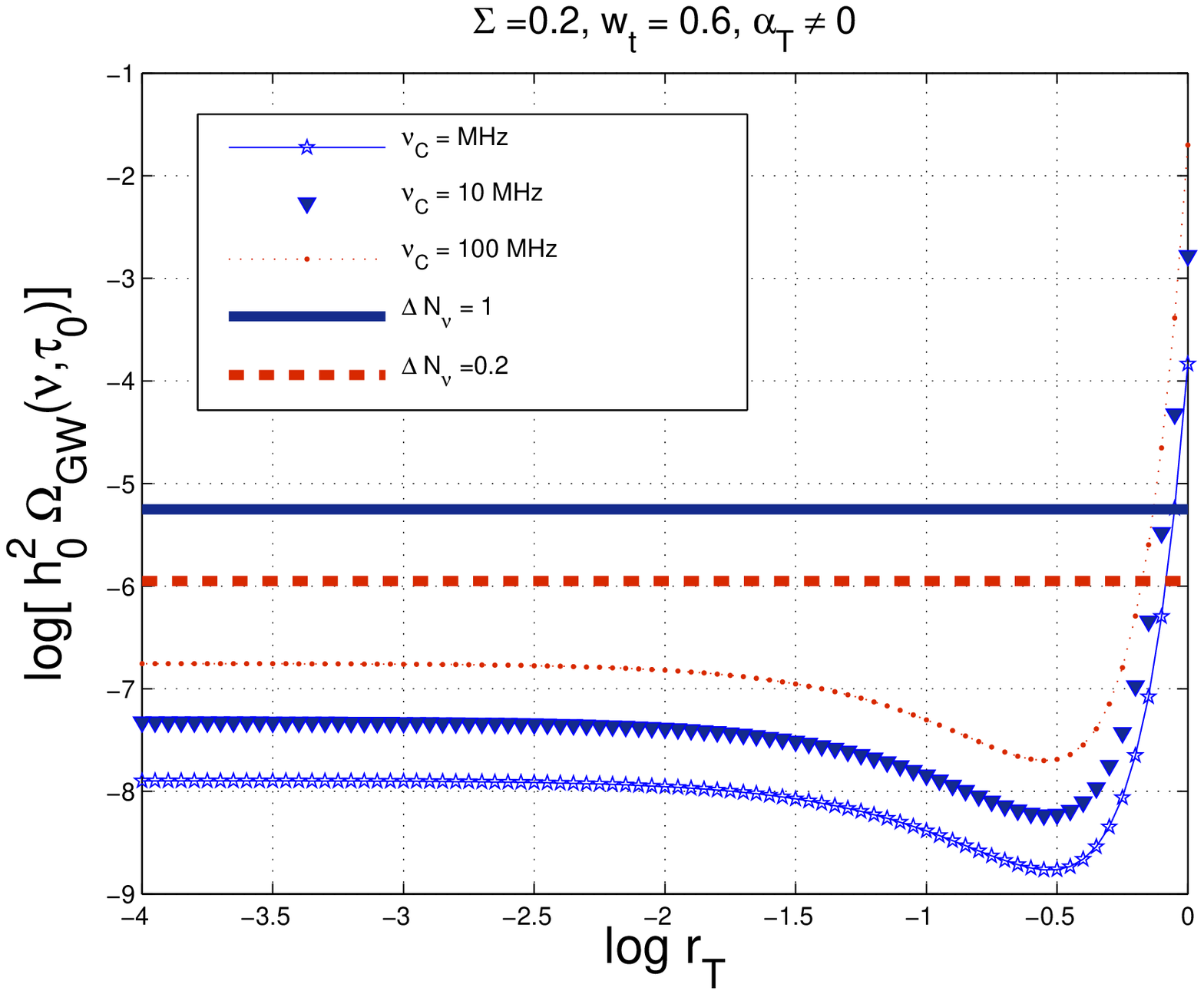}
\caption[a]{The graviton energy spectrum is illustrated, in the T$\Lambda$CDM scenario,  for $\nu= \nu_{\mathrm{C}}$  and as a function 
of $r_{\mathrm{T}}$. As in Figs. \ref{Figure9} and \ref{Figure15} at the left $\alpha_{\mathrm{T}} =0$ while, at the right, $\alpha_{\mathrm{T}} \neq 0$.}
\label{Figure16}      
\end{figure}
The achievable sensitivity of a pair of wide band interferometers crucially 
depends upon the spectral slope of the theoretical energy spectrum in the 
operating window of the detectors. So, a flat spectrum will lead 
to an experimental sensitivity which might not be similar to the 
sensitivity achievable in the case of a blue or violet spectra.   
Previous calculations \cite{mg4,mg5,mg6} showed that, however, 
to get a reasonable idea of the potential signal it is sufficient to compare 
the signal with the sensitivity to flat spectrum which has been 
reported in Eq. (\ref{SENS}). Of course any experimental 
improvement in comparison with the values of Eq. (\ref{SENS}) 
will widen the detectability region by making the prospects
of the whole discussion more rosy. 

In the T$\Lambda$CDM paradigm the maximal signal 
occurs in a frequency region between the MHz and the GHz. 
This intriguing aspect led to the suggestion \cite{mg4,mg5} that 
microwave cavities  \cite{HF1a} can be used as GW detectors precisely in the 
mentioned frequency range. Prototypes of these detectors \cite{HF1b} have been 
described  and the possibility of further improvements in their sensitivity received 
recently attention \cite{HF1c,HF1d,HF2,HF3,HF4,HF5}. 
Different groups are now concerned with high-frequency 
gravitons. In \cite{HF1d} the ideas put forward in \cite{HF1a,HF1b,HF1c} 
have been developed by using electromagnetic cavities (i.e. static electromagnetic fields).
In  \cite{HF2,HF3,HF4} dynamical electromagnetic fields (i.e. wave guides) have been studied always 
for the purpose of detecting relic gravitons.  In \cite{HF4} an interesting prototype 
detector was described with frequency of operation of the order of $100$ MHz (see also \cite{HF6}). 
In Fig. \ref{Figure16} the value of the spectral energy density is reported for $\nu = \nu_{\mathrm{C}}$ where 
$\nu_{\mathrm{C}}$ defines the frequency of operation of a given electromagnetic detector. In Fig. 
\ref{Figure16} $\nu_{\mathrm{C}}$ is taken in the MHz range.
In both plots the horizontal lines denote the bounds of Eqs. (\ref{BBN1}) and 
(\ref{BBN2}) for two typical values of $\Delta N_{\nu}$ (i.e., more specifically, 
$\Delta N_{\nu} =1$ and $\Delta N_{\nu} = 0.2$).  To be compatible with the bounds the 
values of the spectral energy density must be smaller than the horizontal lines. 
The region of large $r_{\mathrm{T}}$ (i.e. $r_{\mathrm{T}} \simeq {\mathcal O}(1)$) is 
already excluded from CMB upper limits: the plots have been extended also in that 
region for comparison with the analog plots illustrated in Fig. \ref{Figure9}.

Absent direct tests on the thermal history of the plasma 
prior to neutrino decoupling, the current bounds 
on a tensor component affecting the initial conditions 
of the CMB anisotropies (and polarization) do not forbid a potentially detectable 
signal for typical frequencies compatible with the window of wide-band interferometers.
The numerical approach described in the present paper 
allows for a sufficiently accurate estimate of the spectral energy density 
of the relic gravitons. In the context of the class of models analyzed here
 it is plausible to imagine, in the years to come,  
a rather intriguing synergy between large-scale observations (e.g. CMB physics, 
measurements of the matter power spectrum and supernovae)  and small 
scale observations such as the ones conducted by wide-band interferometers 
in the range between few Hz and $10$ kHz.

\end{document}